\documentclass{aa}  
\usepackage{arydshln}
\usepackage{scrextend}
\usepackage[dvipsnames]{xcolor}
\usepackage{rotating}
\usepackage{multirow}
\usepackage{ulem}
\usepackage{natbib,twoopt}
\usepackage{anyfontsize}
\usepackage{ccaption}
\usepackage[breaklinks=true]{hyperref} 
\bibpunct{(}{)}{;}{a}{}{,}             
\makeatletter
  \newcommandtwoopt{\citeads}[3][][]{\href{http://adsabs.harvard.edu/abs/#3}%
    {\def\hyper@linkstart##1##2{}%
     \let\hyper@linkend\@empty\citealp[#1][#2]{#3}}}
  \newcommandtwoopt{\citepads}[3][][]{\href{http://adsabs.harvard.edu/abs/#3}%
    {\def\hyper@linkstart##1##2{}%
     \let\hyper@linkend\@empty\citep[#1][#2]{#3}}}
  \newcommandtwoopt{\citetads}[3][][]{\href{http://adsabs.harvard.edu/abs/#3}%
    {\def\hyper@linkstart##1##2{}%
     \let\hyper@linkend\@empty\citet[#1][#2]{#3}}}
  \newcommandtwoopt{\citeyearads}[3][][]%
    {\href{http://adsabs.harvard.edu/abs/#3}
    {\def\hyper@linkstart##1##2{}%
     \let\hyper@linkend\@empty\citeyear[#1][#2]{#3}}}
\makeatother

\usepackage{xcolor}
\usepackage{pdflscape}
\usepackage{multirow}
\usepackage{hyperref}
\usepackage{amsmath}
\usepackage{txfonts}

\usepackage{xintcore}

\usepackage{xintfrac}

\usepackage{cleveref}
\crefdefaultlabelformat{\textsubscript{\textbf{(#2#1#3.)}}}
\usepackage[inline]{enumitem}

\usepackage{tikz}
\usetikzlibrary{shapes, arrows.meta, positioning}
\usepackage{longtable}
\usepackage{orcidlink}

\newcommand{\nicer}{\textit{NICER}}
\newcommand{\swift}{\textit{Swift}}

\newcommand{\xrism}{\textit{XRISM}}
\newcommand{\ep}{\textit{Einstein Probe}}

\newcommand\T{\rule{0pt}{2.6ex}}       
\newcommand\B{\rule[-1.2ex]{0pt}{0pt}}

\newcommand{\Fexxv}{Fe~{\sc xxv}}
\newcommand{\Fexxvi}{Fe~{\sc xxvi}}
\newcommand{\Caxx}{Ca~{\sc xx}}
\newcommand{\Sxvi}{S~{\sc xvi}}

\newcommand{\Crxxiii}{Cr~{\sc xxiii}}
\newcommand{\Nex}{Ne~{\sc x}}
\newcommand{\Nixxvii}{Ni~{\sc xxvii}}

\newcommand{\Fei}{Fe~{\sc i}}
\newcommand{\Feii}{Fe~{\sc ii}}

\begin{document} 

   \title{\vspace{-2em}XRISM reveals a variable, multi-phase outflow-inflow structure\\
during the X-ray obscured 2024 outburst of the black hole transient V4641 Sgr}
\titlerunning{XRISM reveals a variable, multi-phase outflow-inflow structure
during the 2024 outburst of V4641 Sgr}
   \author{
   M. Parra\inst{1}\orcidlink{0009-0003-8610-853X}\thanks{Corresponding authors:\\ \href{mailto:maxime.parrastro@gmail.com}{maxime.parrastro@gmail.com},  \href{mailto:shidatsu.megumi.wr@ehime-u.ac.jp}{shidatsu.megumi.wr@ehime-u.ac.jp}},
   M. Shidatsu \inst{1}\orcidlink{0000-0001-8195-6546}$^\star$,
   R. Tomaru \inst{2}\orcidlink{0000-0002-6797-2539},
   C. Done \inst{3}\orcidlink{0000-0002-1065-7239},
   T. Muñoz-Darias\inst{4,5}\orcidlink{0000-0002-3348-4035},
   M. Armas Padilla\inst{4,5}\orcidlink{0000-0002-4344-7334},
   S. Ogawa\inst{6}\orcidlink{0000-0002-5701-0811},
   A. Marino\inst{7,8,9}\orcidlink{0000-0001-5674-4664},
   N. Grollimund\inst{10},
   S. Corbel\inst{10}\orcidlink{0000-0001-5538-5831},
   E. De la Fuente\inst{11,12}\orcidlink{0000-0001-9643-4134}\thanks{Invited Project Professor at Institute of Cosmic Ray Research, The University of Tokyo on June-July 2025},
   H. Cheng\inst{13}\orcidlink{0000-0003-4200-9954},
   M. D\'iaz Trigo\inst{14}\orcidlink{0000-0001-7796-4279},
   R. Fender\inst{15,16}\orcidlink{0000-0002-5654-2744},
   K. Isogai\inst{17,18},
   S. B. Kobayashi\inst{19}\orcidlink{0000-0001-7773-9266},
   S. Motta\inst{15,20}\orcidlink{0000-0002-6154-5843},
   K. Murata\inst{18},
   H. Negoro\inst{21},
   S. Safi-Harb\inst{22},
   H. Suzuki\inst{6}\orcidlink{0000-0002-8152-6172},
   N. Tsuji\inst{23,24}\orcidlink{0000-0001-7209-9204},
   Y. Ueda\inst{25}\orcidlink{0000-0001-7821-6715},
   C. Zhang\inst{13},
   Y. Zhang\inst{26,27}\orcidlink{0000-0002-2268-9318},
   Z. Zhang\inst{28}\orcidlink{0000-0003-0847-1299}}
    \authorrunning{M. Parra et al.} 

   \institute{
   Department of Physics, Ehime University, 2-5, Bunkyocho, Matsuyama, Ehime 790-8577, Japan
   \and
   Department of Earth and Space Science, Graduate School of Science, Osaka University, 1-1 Machikaneyama, Toyonaka, Osaka 560-0043, Japan
   \and
   Centre for Extragalactic Astronomy, Department of Physics, University of Durham, South Road, Durham DH1 3LE, UK
   \and 
    Instituto de Astrofísica de Canarias, E-38205 La Laguna, Tenerife, Spain
    \and
    Departamento de Astrofísica, Universidad de La Laguna, E-38206 La Laguna, Tenerife, Spain
   \and
   Institute of Space and Astronautical Science (ISAS), Japan Aerospace Exploration Agency (JAXA), Kanagawa 252-5210, Japan
    \and 
    Institute of Space Sciences (ICE, CSIC), Campus UAB, Carrer de Can Magrans s/n, E-08193 Barcelona, Spain
    \and
    Institut d'Estudis Espacials de Catalunya (IEEC), 08860 Castelldefels (Barcelona), Spain 
    \and
    INAF/IASF Palermo, via Ugo La Malfa 153, I-90146 - Palermo, Italy 
   \and
    Université Paris Cité and Université Paris Saclay, CEA, CNRS, AIM, F-91190 Gif-sur-Yvette, France
    \and
    Departamento de F\'{i}sica, CUCEI, Universidad de Guadalajara, 44430, Guadalajara,Jalisco, M\'exico
    \and
    Doctorado en Ciencias Fisico-Matematicas, CUValles, Universidad de Guadalajara, Carretera Guadalajara--Ameca km. 45.5, 46600, Ameca, Jalisco, M\'exico
    \and
    National Astronomical Observatories, Chinese Academy of Sciences, 20A Datun Road, Beijing 100101, China
    \and
    ESO, Karl-Schwarzschild-Strasse 2, 85748, Garching bei M\"unchen, Germany
    \and
    Department of Physics, University of Oxford, Denys Wilkinson Building, Keble Road, Oxford OX1 3RH, UK
    \and
    Department of Astronomy, University of Cape Town, Private Bag X3, Rondebosch 7701, South Africa
    \and
    Okayama Observatory, Kyoto University, 3037-5 Honjo, Kamogatacho, Asakuchi, Okayama 719-0232, Japan
    \and 
    Department of Multi-Disciplinary Sciences, Graduate School of Arts and Sciences, The University of Tokyo, 3-8-1 Komaba, Meguro, Tokyo 153-8902, Japan
    \and 
    Department of Physics, Tokyo University of Science, 1-3 Kagurazaka, Shinjuku-ku, Tokyo 162-8601, Japan
    \and
    Istituto Nazionale di Astrofisica, Osservatorio Astronomico di Brera, Via E. Bianchi 46, 23807 Merate, (LC), Italy
    \and
    Department of Physics, Nihon University, 1-8 Kanda-Surugadai, Chiyoda-ku, Tokyo 101-8308, Japan
    \and
    University of Manitoba, Winnipeg, MB, R3T 2N2, Canada
    \and
    Institute for Cosmic Ray Research, University of Tokyo, 5-1-5, Kashiwa-no-ha, Kashiwa, Chiba 277-8582, Japan
    \and
    Interdisciplinary Theoretical \& Mathematical Science Center (iTHEMS), RIKEN, 2-1, Hirosawa, Wako, Saitama 351-0198, Japan
    \and
    Department of Astronomy, Kyoto University, Kitashirakawa-Oiwake-cho, Sakyo-ku, Kyoto, Kyoto 606-8502, Japan
    \and
Center for Astrophysics, Harvard \& Smithsonian, 60 Garden St, Cambridge, MA 02138, USA
    \and
    Kapteyn Astronomical Institute, University of Groningen, P.O.\ BOX 800, 9700 AV Groningen, The Netherlands
    \and
    Astrophysics, Department of Physics, University of Oxford, Oxford OX1 3RH, UK
   }

   \date{}
 
  \abstract
   {We report the results of a simultaneous X-ray and optical spectroscopy campaign on the Galactic black hole X-ray binary V4641 Sgr, carried out with XRISM and the Seimei telescope during a low-luminosity phase towards the end of its 2024 outburst. Despite a very low X-ray luminosity of $10^{34}$ erg s$^{-1}$, the continuum spectrum is well reproduced by a disk blackbody model with a high inner disk temperature ($1.8$ keV). XRISM/Resolve provides the highest-resolution X-ray spectrum ever obtained from the source, and several strong, narrow emission lines were detected, resolved and characterized at a high significance level. The continuum shape and narrow emission lines both indicate that the inner disk region is obscured by the surrounding high-density gas, and the intrinsic luminosity is several orders of magnitude higher. 
In the simultaneous optical observation from the Seimei telescope, the line features are largely dominated by the optical companion. Although we detect a clear emission component in H$\alpha$ that could originate from a cold outflow or the disk atmosphere, there are no signs of the strong outflow signatures historically detected in this source. 
In X-rays, the combination of significantly redshifted ($\sim 700$ km s$^{-1}$) and weakly blueshifted ($\sim-250$ km s$^{-1}$) components, all varying strongly on ks timescales, along with a marginally significant (99.2\%) highly blueshifted ($\sim-1200$ km s$^{-1}$) component, indicates a complex, inhomogeneous outflow geometry. This is corroborated by the erratic long-term evolution of the source seen in the complementary X-ray monitoring, and radio detections spanning 3 orders of magnitude.}

   \maketitle

\setcounter{page}{2}
\section{Introduction} \label{sec:intro}

X-ray Binaries (XRBs) are a subcategory of X-ray sources, corresponding to one of the many evolutionary pathways of binary stellar systems, during which mass transfer from a main sequence ``companion'' star onto a stellar mass compact object leads to intense emission along the entire electromagnetic spectrum, and particularly in X-rays. The dynamics of these systems, and notably those hosting a black hole (BH) or a neutron star (NS), strongly depend on the mass and behavior of the companion. In low-mass X-ray binaries (LMXBs; \citealt{Bahramian2023_LMXB_review}), the evolution of a low-mass ($\lesssim 1$ M$_\odot$) star and the binary system itself will trigger mass transfer through Roche Lobe overflow, leading to the formation of an accretion disk around the compact object. On the other hand, high-mass X-ray binaries (HMXB; \citealt{Fornasini2023_HMXB_review}) host a B or O-type supergiant of few or more solar masses, and tend to feed the compact object directly with intense stellar winds. The majority of BH LMXBs, along with part of the HMXB population, is made of transients, and shows rare, erratically repeating periods where accretion and X-ray emission intensify during few months to few years and become detectable, the so-called outbursts. The remaining time is spent in quiescence, at low accretion levels and negligible luminosities. 

These outburst patterns show a coherent evolution through distinct spectral-timing states (see, e.g., \citealt{Remillard2006_BHXRB_properties,Done2007_BHXRB_accretion} in X-rays, \citealt{fender2003_jet_review} in the radio band). The start is characterized by a rise in luminosity of more than 5 orders of magnitude through the hard state (HS), defined by a spectral cut-off powerlaw  peaking at $\sim$100 keV, originating in a hot, horizontally extended \citep{Krawczynski2022_CygX-1_pola} plasma region near the BH called the corona, and a steady radio emission at GHz radio frequencies with a flat synchrotron spectrum, 
likely produced by a compact, optically thick jet. Strong and rapid X-ray variability (up to 50\% rms) is observed on time scales between 10ms and 100s, with characteristic frequencies (often manifesting as Quasi-Periodic Oscillations or QPOs, \citealt{Ingram2019_QPO_review}) increasing with luminosity. 
Then, once above a critical luminosity between $\sim1-10\% L_{Edd}$, the spectrum begins to soften. A hitherto secondary component, tracing an optically thick disk, becomes dominant, and extends close to or at the innermost stable circular orbit (ISCO), while the power law steepens and fades away, as the corona mostly disappears. During this hard-to-soft transition, the source shows major radio flares linked to transient, optically thin ejecta, before the radio emission gets quenched, indicating a cessation of core jet activity. The source then remains in this soft, disk dominated state (SS), with inner temperatures of $kT_{in}\sim$1-1.5 keV and low variability, and fades to lower luminosities over the course of few weeks to months. Once below a precise Eddington ratio threshold \citep{VahdatMotlagh2019_soft-hard_transition_lum,Wang2023_soft-hard_transition_lum}, the source transits back to the HS, and the jet reappears as the emission continues to fade down to quiescence.

In addition to the continuum evolution, the last two decades have seen a series of discoveries of blueshifted, narrow, highly ionized absorption features in X-rays (mainly ionized iron lines from Fe{\sc xxv} and/or Fe{\sc xxvi}) in transient XRBs \citep{Ueda1998_GROJ1655-40_wind_ASCA,Kotani2000_GRS1915+105_winds_ASCA,Lee2002_GRS1915+105_winds_first}. 
These features trace hot and massive 
 outflows, dynamically important for the evolution of these systems, due to the removal of a substantial fraction of the disk mass, or heating of the disk surface by scattering X-ray flux down onto the disc  \citep{tet18b,dub19,tet20,pon16,tet18a}. %
 Only high-inclination (dipping) sources  show such powerful ``hot'' winds, likely due to the equatorial nature of these outflows. Moreover, in X-rays, these winds are generally detected during the SS \cite[e.g.][]{Ponti2012_ubhw,Parra2024_winds_global_BHLMXBs} and remain undetected during the HS. The fact that the ``hot'' wind signatures disappear during the HS is likely related to photoionization instability of the wind in the ionization domain consistent with the presence of Fe{\sc xxv} and Fe{\sc xxvi} in that state (\citealt{cha13,bia17} but see \citealt{gat19}). The shape of these stability curves are however strongly dependent on the spectral shape of the ionizing Spectral Energy Distribution or SED \citep{cha13,bia17,Petrucci2021_outburst_wind_stability,Parra2025_4U}.

Yet the picture is more complicated than a simple on-off wind process between spectral states 
and recent results during the (hard and soft) intermediate states (HIS and SIS respectively), characteristic of the hard-soft state transition, show a large variety of X-ray wind properties and signatures \citep{mil08,kal09,nei16,shi16,mun17,gat19}. In parallel, recent observations in the optical and infrared (OIR) range \cite{Jimenez-Ibarra2019_optical,Munoz-Darias2019_1820_hard,MataSanchez2022,Sanchez-Sierras2023_optical}
have shown the presence of low ionization absorption features during the entirety of the outbursts. Whether these ``cold'' winds are different from the ``hot'' ones or are both the same component but seen through different ionization states is still highly debated, notably due to the lack of simultaneous wind detections in different wavelengths in standard systems \citep{Munoz-Darias2022_winds_simultaneous,Parra2024_winds_global_BHLMXBs}.

V4641 Sgr is a transient galactic X-ray binary located at
$6.2 \pm 0.7$ kpc. It consists of a BH with a mass 
of $6.4 \pm 0.6 M_\sun$ and a B9III companion with 
$2.9 \pm 0.4 M_\sun$ \citep{MacDonald2014} orbiting 
each other with a period of a 2.82 days \citep{orosz01,uemura02}.
The source was identified as an X-ray source due to a giant 
outburst in 1999 \citep{smith99,hjellming00,stubbings99}. 
Since then, it has shown
a dozen smaller outbursts, 
with a reccurence interval of 2--4 years \citep{tetarenko16}. The source 
is also one of the several BH X-ray binaries around which 
diffuse emission in the TeV gamma-ray band has been detected 
\citep{alfaro24, lhaaso24}, and its possible X-ray counterpart 
was recently discovered \citep{Suzuki2025a}.

During its outbursts, the spectral evolution of the source remains completely different from the classical Q-shape evolution and spectral states seen in standard outbursting BHXRBs (see, e.g., \citealt{Dunn2010_BHXRB_spectral}). 
 It either shows a hard, reflection dominated spectrum \citep{morningstarSEYFERT2LIKESPECTRUMHIGHMASS2014} or a disk blackbody spectrum \citep{pahari15, bahramian15, Shaw2022_SAXJ1819_wind_emission_2020}, both at very low luminosities, typically below $L/L_{Edd}\sim10^{-3}$. These unusual spectral states suggest that the outer disk or its atmosphere completely obscure the inner regions of the accretion flow,
and that the observed flux is only the tiny fraction of the central source 
revealed by scattering and photoionized emission from the X-ray heated disk atmosphere, or a wind. In the disk dominated state, several narrow, highly ionized emission lines, including \Fexxvi{} Ly$\alpha$ (also referred as \Fexxvi{} K$\alpha$) and 
\Fexxv{} He$\alpha$ (also referred as \Fexxv{} K$\alpha$), were first discovered with Chandra/HETG in 2020 \citep{Shaw2022_SAXJ1819_wind_emission_2020}. In these spectra, the iron emission lines showed negligible velocity shifts, but a lower ionized layer, with marginal (non-corrected) blueshifts in absorption and marginal redshifts in emission, was also detected from the lines of less massive elements. The former can retroactively explain the iron line feature seen at lower spectral resolution in soft state NuSTAR observations of previous outbursts \citep{pahari15}. 
The presence of a high-speed, ``cold'', low-ionized wind, first associated with the September 1999 X-ray outburst, and later with subsequent outbursts, was first suspected from the complex emission profiles seen in optical spectroscopy (Charles et al. 1999; \citep{chatyOpticalNearinfraredObservations2003,lindstromNewCluesOutburst2005}, before being confirmed unambiguously with P-Cygni profiles and blueshifted absorption lines \citep{munoz-darias18}, with terminal velocities of several 1000 km s$^{-1}$. Yet the association between these strong optical wind signatures and a potential ``hot'', highly-ionized wind phase, which has yet to be directly identified, remains unclear.  

A new outburst of V4641 Sgr was detected on 2024 September 6 \citep{negoro24}, three years after the previous outburst. To better determine the properties and profiles of the emission lines, which were only partly constrained by the previous Chandra observations, we observed the source with XRISM \citep{tashiro20}, as the first Director’s Discretionary Time (DDT) observation of the mission since the beginning of the guest observer phase. We also performed simultaneous optical spectroscopy with the 3.8-m Seimei telescope in Japan, to search for optical wind signatures. 

In this paper, we present the results of our spectral analysis of both data sets, complemented by publicly available X-ray monitoring of this outburst using MAXI, NICER, and Swift, as well as the coverage of Einstein Probe, and reports of radio monitoring using MeerKAT. We describe our observations and the data reduction procedures in Section~\ref{sec:xobs} and the results of the X-ray spectral analysis in Section~\ref{sec:xana}. Section~\ref{sec:optana} describes our analysis of the optical 
observation. In Section~\ref{sec:disc}, we discuss these results, in particular for the obscuration of the source and its possible outflow structure. Our conclusions are given in Section~\ref{sec:conclu}.

\section{Observation and Data Reduction} \label{sec:xobs}

\subsection{\xrism{}}

\begin{figure}
    \centering
    \includegraphics[clip,trim=0.1cm 0cm 0cm 0cm,width=0.5\textwidth]{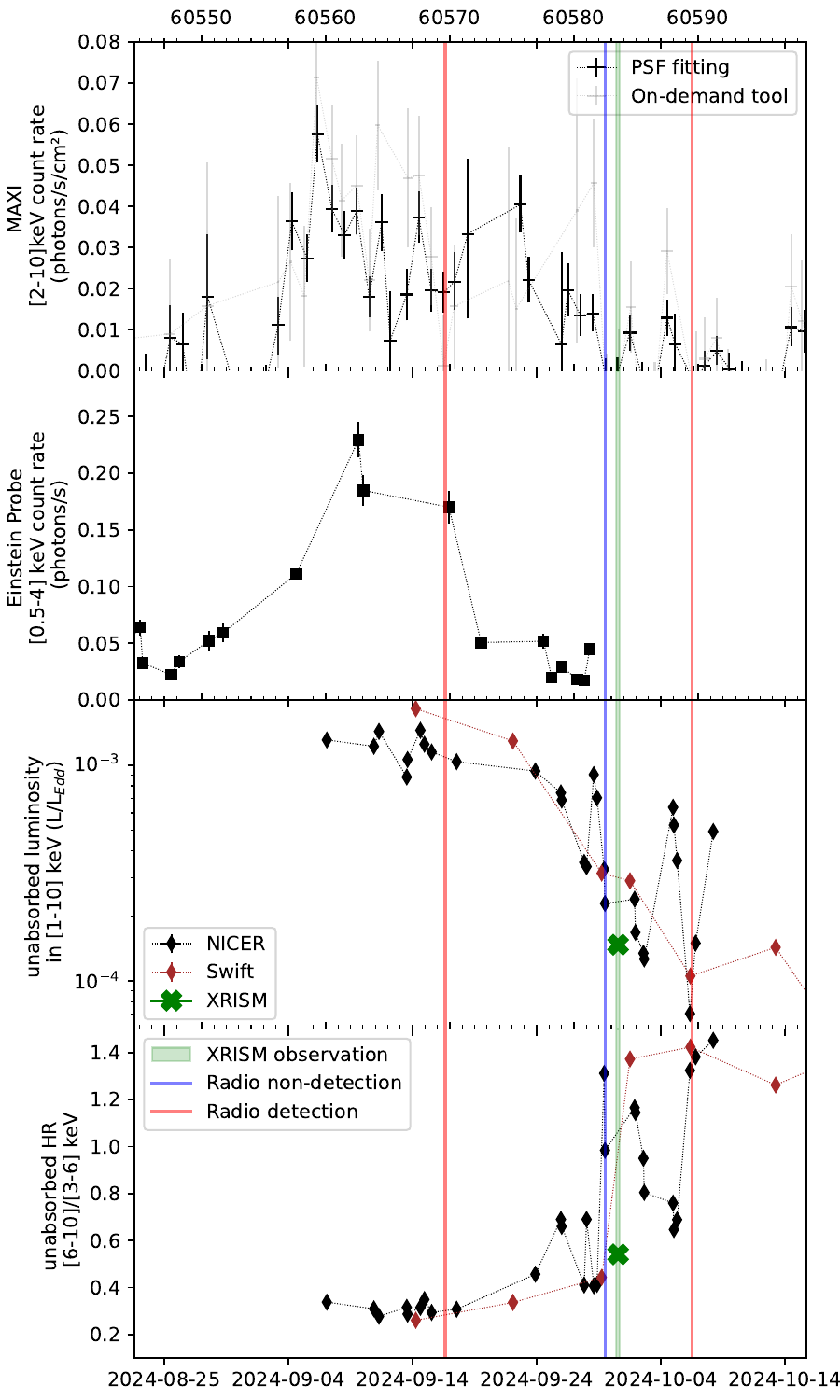}
    \vspace{-2em}
    \caption{Long-term monitoring of V4641 Sgr with different instruments during the 2024 outburst. From top to bottom:  \textbf{(First)} MAXI daily light curve in the 2--10 keV band, obtained via PSF fitting (black), and generated via the MAXI on-demand interface (gray). See Sect.~\ref{subsub:MAXI} for details.
    \textbf{(Second)} Einstein probe light curve in the 0.5--4 keV band. \textbf{(Third)} NICER / Swift / XRISM 1--10 keV luminosity light curves in Eddington units, after removal of the  ISM absorption component.\textbf{(Fourth)} NICER / Swift / XRISM 6--10/3--10 keV hardness ratio (HR) evolution. The results in the last two panels come from fitting the time-integrated spectra of the individual ObsIDs in this period (see Sect.~\ref{sub:x-ray_monit} for more details). The blue and red lines highlight the radio observations reported in \cite{grollimundMeerKATDetectionRadio2024,grollimundMeerKATSwiftConfirm2024}, and the green line the XRISM and simultaneous Seimei observations.}
    \label{fig:lc_monit}
\end{figure}

The XRISM observation of V4641 Sgr was carried out on 2024 September 30 UT 09:41--17:02 (OBSID 901001010), with a net exposure of $\sim$ 13 ks. During the observation, the microcalorimeter Resolve \citep{ishisaki22} was 
operated with no filters, and the soft X-ray imager Xtend \citep{mori22, noda25,uchidaInorbitPerformanceSoft2025} was operated in the full window mode. As shown in Figure~\ref{fig:lc_monit}, the source exhibited a flaring activity for a few days in early September and then a rapid flux drop. The observation was executed at the end of the outburst, close to quiescence. We note that while the MAXI evolution appears to follow the results of the \nicer{} and \swift{} monitoring, the flux values derived from MAXI, both with the on-demand tool and PSF-fitting, remain higher than those obtained with direct fitting from NICER and Swift, with the brightest Swift observation has a 1--10 keV flux of $\sim2\times10^{-10}$ erg s$^{-1}$ cm$^{-2}$, and are overestimated by a factor $\gtrsim3$. This is likely because the flux values in the MAXI data are converted from the count rates by assuming the spectrum of the Crab nebula. 

The data were reduced with Heasoft version 6.34 and the latest 
version of XRISM Calibration Database as of the observation date, 
following the XRISM Quick Start Guide v2.3\footnote{\url{https://heasarc.gsfc.nasa.gov/docs/xrism/analysis/quickstart/}}. 

\subsubsection{Resolve}
The energy scale was measured using a set of $^{55}$Fe sources on the filter wheel, 
which illuminated the entire array at fiducial time intervals selected 
to follow the characteristic pattern of gain changes caused by the cycles of the 50~mK cooler. 
The Mn K$\alpha$ line complex from $^{55}$Fe was fitted for each fiducial interval and pixel using 
an eight-component model with intrinsic Lorentzian profiles, 
which were convolved with Gaussian smoothing to account for the detector’s resolution. 
The energy scale was corrected by nonlinear interpolation between gain curves 
measured at different heat-sink temperatures and linear interpolation between the fiducial measurements. 
Out of the 36 pixels in the Resolve detector, 35 pixels are usable for scientific purposes
and the rest is a calibration pixel.

Since random gain jumps of unknown origin, lasting approximately 1--2 hours, were observed in Pixel 27
and the current fiducial strategy was unable to track these gain jumps, data from Pixel 27 were excluded. 
Gaussian modeling of line spread of the detector for the Mn K$\alpha$ line indicated 
a resolution of $4.45 \pm 0.04$~eV FWHM, with an energy shift of $0.0071 \pm 0.0182$~eV\footnote{\url{https://heasarc.gsfc.nasa.gov/FTP/xrism/postlaunch/gainreports/9/901001010_resolve_energy_scale_report.pdf}}. 
Early calibration measurements showed that the energy scale was accurate to within approximately $\pm$1 eV over the 2–8 keV range \citep{porterInflightPerformanceXRISM2024,eckartEnergyGainScale2024}. We applied additional screening of the standard energy-dependent rise time cut \citep{mochizuki24}. Only High-resolution primary (Hp) events are used for spectral analysis.

The response matrix file (RMF) was generated with \texttt{rslmkrmf}.
Its normalization is computed according to the fraction of events with the relevant grade in the cleaned event file. However, at low count rates, Resolved cleaned event files can include a signification fraction of low-resolution secondary (Ls) events that are not due to the source,  
but instead result from the secondary-pulse detection algorithm on clipped pulses. To assess the degree of Ls event contamination, we computed both the theoretical and observed branching ratios (proportion of events in each grade) of each pixels in the observation. We show the result in App.\ref{app:resolve_branch}. In pixel 18, the brightest pixel of the observation, the combined branching ratio of all Middle and Low-resolution events should be less than 0.8\%. The entire observation should thus have a Hp grade branching ratio of virtually 100\%. Yet, we see a non-negligible  contribution (between few percent and few tens of percent) of Ls events, with values between 1 and 3 order of magnitudes above theoretical expectations, which confirms the high degree of Ls contamination. 
We thus removed the Ls events from the cleaned event file before  calculating the RMF. On the other hand, the Mp/Ms/Lp contribution, while higher than normal, remains negligible and will not affect our analysis.

The ancillary response file (ARF) was generated with \texttt{xaarfgen}
assuming a point-like source at the aim point as input.
To produce the spectrum, we adapted an energy range of 
2--12 keV, below which the X-ray signals are mostly absorbed 
by the closed gate valve.

We generated the Non-X-ray Background (NXB) file using \texttt{rslnxbgen} task, 
excluding pixel 27 and using only the Hp events.
The NXB spectrum is fitted with the model provided by the XRISM Calibration team, 
which consists of a power-law and 17 instrumental emission lines presented by narrow positive Gaussians: Al K$\alpha$1/K$\alpha$2,
Au M$\alpha$1, Cr K$\alpha$1/K$\alpha$2, Mn K$\alpha$1/K$\alpha$2, Fe K$\alpha$1/K$\alpha$2, Ni K$\alpha$1/K$\alpha$2, 
Cu K$\alpha$1/K$\alpha$2, Au L$\alpha$1/L$\alpha$2, and Au L$\beta$1/L$\beta$2.

\subsubsection{Xtend}
The time-averaged source spectra were extracted from a circular region with a radius of 2.5 arcmin centered on the source peak and the background from a source-free circular region with the same radius. The RMF and ARF were generated with \texttt{xtdrmf} and \texttt{xaarfgen}, respectively. 
We used 0.4--12 keV, above and below which the source count rates 
become lower than the background rates. 

\subsection{X-ray monitoring}\label{sub:x-ray_monit}

\subsubsection{\textit{MAXI}}\label{subsub:MAXI}

Due to the very limited flux of the source during the outburst, data from the MAXI \citep{matsuoka2009} all-sky monitoring instruments must be analyzed carefully. We thus computed the MAXI lightcurve in the 2-10 keV band using 2 methods. First, we used an updated PSF fitting, following the method in \cite{morii2016}, which uses the well-calibrated cameras (Camera ID: 2, 4, 5, and 7) and considers the contribution of nearby sources including GS 1826$-$238, GX 5$-$1, and NGC 6624. Secondly, for comparison, we generated lightcurve products using the MAXI on-demand interface\footnote{available at \url{http://maxi.riken.jp/mxondem/}} (gray), with an improved methodology \citep{Nakahira2012_MAXI} compared to the standard products provided directly on the standard website\footnote{\url{http://maxi.riken.jp/star_data/J1819-254/J1819-254.html}}. The results are shown in the top panel of Fig.~\ref{fig:lc_monit}.
 
\subsubsection{\nicer{}}

The Neutron Star Interior Composition Explorer (NICER, \cite{Gendreau2016_NICER}) started observing the source soon after the first report of the 2024 outburst from \cite{negoro24}. We analyzed every observation of this last outburst, using an automated data reduction procedure based on the NICERDAS software\footnote{\href{https://heasarc.gsfc.nasa.gov/docs/nicer/nicer_analysis.html}{https://heasarc.gsfc.nasa.gov/docs/nicer/nicer\_analysis.html}} version 12, with the latest calibration files available (CADLB ver:xti20240206). In order to use the \texttt{scorpeon} background model, we downloaded geomagnetic data using the NICERDAS task \texttt{nigeodown}. We split the data into continuous observation periods for each day, filtered them to remove background flares following the procedures detailed in the NICER data analysis threads\footnote{\href{https://heasarc.gsfc.nasa.gov/docs/nicer/analysis_threads/}{https://heasarc.gsfc.nasa.gov/docs/nicer/analysis\_threads/}}, and extracted spectra using the \texttt{nicerl3-spect} task of NICERDAS. We used the xspec-model version of the \texttt{scorpeon} background model, which is important to better constrain the background contribution in such faint observations. All spectra was grouped following the optimized binning of \citet{Kaastra2016_binning_opt}.

We performed spectral analysis independently on each of the 25 days of observation, further splitting the individual observation periods of each day when their count rate in the 3--8 keV band, or their hardness ratio in the 5--8/3--5 keV band changed by more than 20\%, and removing observation periods where very low exposures of high background contamination prevented good spectral fitting. The resulting 30 epochs were fit in the 0.3--10 keV band, with an absorbed Comptonized disk model, using \texttt{phabs*Tbabs*thcomp*diskbb}. We set interstellar abundances to \texttt{wilm} \citep{Wilms2000_xspec_abundances}, and the interstellar column density (in \textbf{TBabs}) was fixed at $1.5\cdot10^{21}$ cm$^{-2}$, following our results for the XRISM analysis (see Sect.~\ref{sec:xana}). The second phabs absorption component was left free to account for  the additional absorber intrinsic to the source, and varied between between 0 and $\sim3\cdot10^{21}$ cm$^{-2}$ between observations. The electron temperature of the Comptonization component was fixed at 100 keV. 
For each epoch, we then derived the unabsorbed 1--10 keV flux, the 6--10 keV/3--10 keV hardness ratio, and their 90$\%$ uncertainties. In order to facilitate the comparison between observations, and due to the highly unknown nature of the intrinsic absorber, we only remove the fixed interstellar absorption component (\texttt{TBabs}), to 
effectively show the absorbed intrinsic luminosity of the source in the bottom panels of Fig.~\ref{fig:lc_monit}. The source remained very soft and relatively stable during $\sim 2$ weeks, between September 7 (date of the first NICER observation) and 24. Although a detailed spectral analysis of the NICER observations is out of scope of this paper, we note that the spectra are perfectly fit with a thermal disk with temperature $\sim1.1-1.2$ keV, and up to two highly variable narrow emission lines between 6.4 and 7 keV. Then, after an extremely erratic $\sim3$ week period during which both the hardness ratio and luminosity vary intensely down to single hour timescales, the X-ray spectrum remains dominated by a $\Gamma\sim1-1.5$ component from October 6 (date of the second radio jet detection) onward, with Swift detecting a decay in luminosity down to $\sim10^{-5} L_{Edd}$ in the following month.

\subsubsection{\swift{}}

The Neil Gehrels Swift Telescope (\swift{}) observed the source on multiple occasions during the decay of the outburst. We derived spectral products in all available XRT observations of the outburst using the online \swift{}/XRT products generator\footnote{\href{https://www.swift.ac.uk/user_objects/}{https://www.swift.ac.uk/user\_objects/}}. After regrouping the epochs with the optimized binning of \citet{Kaastra2016_binning_opt}, we performed spectral analysis following the same methodology as for the NICER data.

\subsubsection{\ep{}}

The Einstein Probe \citep[EP;][]{Yuan2022,Yuan2025} is a time-domain X-ray astronomy mission launched in January 2024. The instruments onboard EP are a wide-field X-ray telescope \citep[WXT][]{Cheng2025} to serve a soft X-ray (0.5--4\,keV) all-sky monitor and a follow-up X-ray telescope (FXT) enabling deeper spectral and timing studies in the 0.3--10\,keV band. 
V4641 Sgr has been observed by EP-WXT with $\sim$daily cadence from August 3rd to September 28th, 2024. In order to extract a WXT light curve, we have produced cleaned event files for each observation using the processing pipeline available within the WXT Data Analysis Software (\texttt{WXTDAS}). A circular region centred on the source position with a radius of 9\arcmin\ (corresponding to $\simeq$90\% of the encircled energy fraction of the instrument’s point spread function) was used to extract source photons. A region of the same size, but located away from the source, was used as background region. We finally used \texttt{XSELECT} to extract background-subtracted light curves from the cleaned event files in the 0.5--4, 0.5--2, and 2--4\,keV energy bands.

\subsection{Optical Spectroscopy} 
\label{sub:optobs}

\begin{table*}[t!]
\centering
\caption{Best-fit continuum parameters with empirical models. Both models also include the line components described in Tab.~\ref{tab:line_param}, and the statistical fit value refers to the final model with all line components. 
\label{tab:cont_mod}}
\begin{footnotesize}
\begin{tabular}{l c cccccccccc}

  & TBabs &  & diskbb &  & & thcomp & & constant & & &  \T\B\\
\hline 
Model & NH & $kT_{\rm in}$ & norm & $R_{\rm in}$$^{(a)}$ & $\Gamma$ & C$_{frac}$ & $kT_{\rm e}$ & $C_{\rm rsl}$${(b)}$ & $F_{\rm 1-10}$ & $L_{\rm 1-10}$$^{(c)}$ & Fit statistic \T\\

 & [$10^{22}$ & & & & & & & & [erg s$^{-1}$ & 10$^{-4}$ & C-stat\T\B \\
  &  cm$^{-2}$] & [keV] & & [km] & & & [keV] & & cm$^{-2}$] & [$L/L_{Edd}$]& /d.o.f.\B \\
\hline
pure & 
\multirow{1}{*}{$0.15$} & 
\multirow{1}{*}{$1.77$} &
\multirow{1}{*}{$0.14$} &
\multirow{1}{*}{$0.41$}&
\multirow{2}{*}{/} &
\multirow{2}{*}{/} &
\multirow{2}{*}{/} & 
\multirow{2}{*}{$1.0_{-0.04}^{+0.03}$} &
\multirow{2}{*}{$2.55 \cdot 10^{-11}$} &
\multirow{2}{*}{1.45} &
\multirow{2}{*}{3212/3698}
\T\\
disk&
\multirow{1}{*}{$\pm 0.01$} & 
\multirow{1}{*}{$\pm0.03$} &
\multirow{1}{*}{$\pm 0.01$} &
\multirow{1}{*}{$\pm 0.01$}&
\B\\
\hline
compt. & 
\multirow{1}{*}{$0.16$} & 
\multirow{2}{*}{$1.65_{-0.12}^{+0.11}$} &
\multirow{2}{*}{$0.18_{-0.04}^{+0.05}$} &
\multirow{2}{*}{$0.47_{-0.05}^{+0.06}$}&
\multirow{2}{*}{$3.5_{-1.5}^{+0.0}\dagger$} &
\multirow{2}{*}{$0.57_{-0.5}^{+0.16}$} &
\multirow{2}{*}{100$^{\dagger\dagger}$} & 
\multirow{2}{*}{$1.0_{-0.03}^{+0.04}$} &
\multirow{2}{*}{$2.55 \cdot 10^{-11}$} &
\multirow{2}{*}{1.45} &
\multirow{2}{*}{3207/3696}
\T\\
disk &
\multirow{1}{*}{$\pm 0.01$} & 

\B\\
\hline
\end{tabular}
\end{footnotesize}

\tablefoot{
The energy bands used for the flux and luminosity calculations are given in keV.
$\dagger$ Parameters unconstrained within the allowed parameter space (2-3.5). $\dagger\dagger$ Frozen parameter. 
(a) Computed from {\tt diskbb} normalization assuming a distance of 6.2 kpc and an inclination of $72^\circ$ \citep{MacDonald2014}. 
(b) Due to the broader energy band of Xtend and the potential uncertainty in the Ls events in Resolve, we froze the xtend constant factor to 1 and allowed that of Resolve to vary freely.(c) Computed using the orbital parameters of \cite{MacDonald2014} (see Sect.~\ref{sec:intro}).}
\end{table*}

\begin{figure*}[h!]
\vspace{-1em}
    \includegraphics[clip,trim=0.4cm 0.4cm 0.3cm 0cm,width=0.99\textwidth]{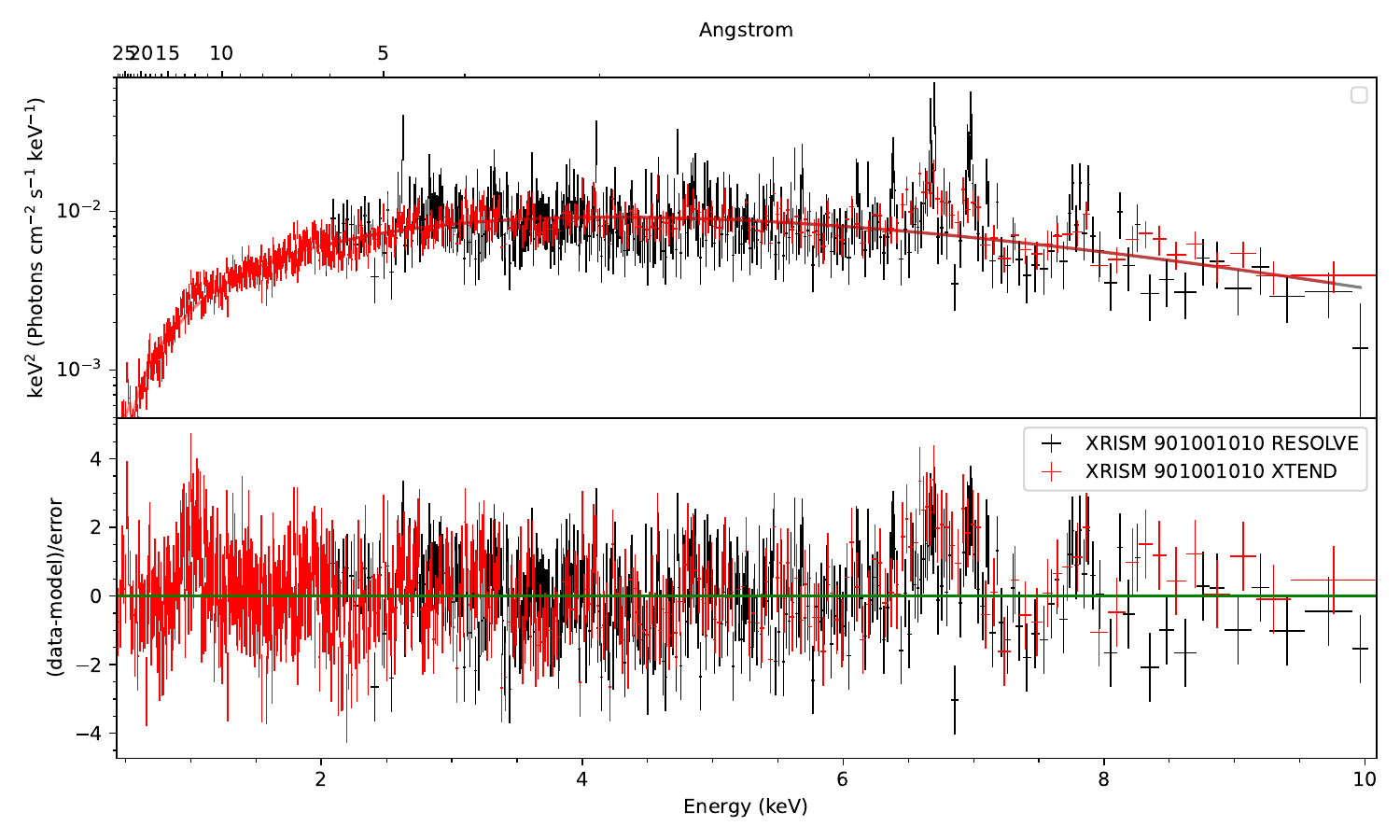}
        \vspace{-1.em}
    \caption{Unfolded spectra and residuals for the time-averaged products of Resolve (black) and Xtend (red). We use the best-fit absorbed disk blackbody model described in Tab.~\ref{tab:cont_mod}, with all line components removed. In all panels, the Resolve spectra are further binned at a 2$\sigma$ significance level for illustrative purposes.}
    \label{fig:timeavg_resid_nolines}
\end{figure*}

\begin{table*}[t!]
\centering
\caption{Significant spectral lines identified in the time-averaged Resolved+Xtend spectra.}
\label{tab:line_param}
\begin{tabular}{lccccccc}
\hline
Line ID & $E_{\rm rest}$ (keV) & $v_{raw}$ (km s$^{-1}$) & $v_{real}$ (km s$^{-1}$) & $\sigma$ (eV) & $\sigma_{raw}$ (km s$^{-1}$) & EW (eV) & $\Delta$ C-stat \T\B \\
\hline
\Nixxvii{} He$w$ & 7.806 & $-43^{+180}_{-390}$ & $-160^{+193}_{-403}$ & 0$^\dagger$ & 0$^\dagger$ & $39_{-30}^{+24}$ & 13\T\B \\\cdashline{1-8}
 Fast \Fexxvi{} Ly$\alpha_1$ & 6.973  & \multirow{2}{*}{$-1169^{+72}_{-79}$} & \multirow{2}{*}{$-1286^{+85}_{-92}$} &  \multirow{2}{*}{$<3.2$} & \multirow{2}{*}{$<150$} & $14^{+9}_{-10}$ & \multirow{2}{*}{12} \T\B \\
 Fast \Fexxvi{} Ly$\alpha_2$ & 6.952  &  & &  &  & $7^{+5}_{-5}$ &  \T\B \\\cdashline{1-8}
 Slow \Fexxvi{} Ly$\alpha_1$ & 6.973  & \multirow{2}{*}{$-131^{+372}_{-131}$} &  \multirow{2}{*}{$-248^{+385}_{-144}$} &  \multirow{2}{*}{$6.2^{+13.8^\dagger}_{-2.0}$} & \multirow{2}{*}{$270^{+590\dagger}_{-90}$} & $52\pm16$ & \multirow{2}{*}{89} \T\B \\
 Slow \Fexxvi{} Ly$\alpha_2$ & 6.952  &  & & &  &  $25^{+8}_{-7}$ & \T\B \\\cdashline{1-8}
 \Fexxv{} He$w$ & 6.700 & \multirow{2}{*}{$-41^{+170}_{-116}$} & \multirow{2}{*}{$-158^{+183}_{-129}$} &\multirow{2}{*}{$7.6^{+3.4}_{-2.1}$} & \multirow{2}{*}{$340^{+150}_{-90}$} & $69^{+23}_{-22}$ & \multirow{2}{*}{177} \T\B \\
 \Fexxv{} He$y$ & 6.668 &   &  & & & $78^{+24}_{-22}$ & \T\B \\\cdashline{1-8}
 %
  \Fei{} K$\alpha_1$ & 6.404  & \multirow{2}{*}{$847^{+292}_{-536}$} &  \multirow{2}{*}{$732^{+305}_{-549}$} &  \multirow{2}{*}{$4.9^{+14.4}_{-4.9}$} & \multirow{2}{*}{$230^{+680\dagger}_{-230}$} & $20\pm10$ & \multirow{2}{*}{18} \T\B \\
 \Fei{} K$\alpha_2$ & 6.391  &  & & &  &  $10\pm5$ & \T\B \\\cdashline{1-8}
  \Crxxiii{} He$w$ & 5.682  & $53^{+75}_{-82}$ & $-64^{+88}_{-95}$ & 0$^\dagger$ & 0$^\dagger$ & $13^{+8}_{-8}$ & 11\T\B \\
 \Caxx{} Ly$\alpha$ & 4.105  & $-232^{+170}_{-151}$ & $-349^{+183}_{-164}$ &$1.9^{+3.2}_{-1.9}$ & $140^{+230}_{-140}$ & $14\pm8$ & 15\T\B \\
 \Sxvi{} Ly$\alpha$ & 2.622 & $-356^{+243}_{-204}$ & $-473^{+256}_{-217}$ & $1.6^{+3.6}_{-1.6}$ & $ 180^{+410}_{-180}$ & $20^{+10}_{-10}$ & 20\T\B \\
 \hline
 \multicolumn{8}{c}{Xtend Only} \T\B \\
 \hline
  \Nex{} Ly$\alpha$ & 1.022 & 0$^\dagger$ & -  & 0$^\dagger$ & 0$^\dagger$ & 40$_{-9}^{+8}$ & 67\T\B \\
\hline
\end{tabular}
\tablefoot{Errors and upper limits are computed at a 90\% confidence level. $\dagger$ indicates parameters that were frozen during the fit or are at the limit of their allowable parameter space. The quoted rest energies are adopted from NIST, with the energies of all unresolved H-like lines set at the 2/1 mean of the two main individual transitions. Common widths and blueshifts between different lines indicate complexes for which these values were linked during the fit. The norms of the individual components of the H-like complexes were tied with a ratio of 1--2 during the fit. $v_{real}$ values include the total velocity correction of $-117\pm13$ km s$^{-1}$ derived for the observation. $\sigma_{raw}$ values are converted directly from the measured $\sigma$, and thus include an artificial contribution of $28$ km s$^{-1}$ from the evolution of the orbital velocity of the BH within the observation. See App.~\ref{app:vel_corr} for the detailed computation of the corrections.}
\end{table*}

\begin{figure*}[h!]
        \includegraphics[clip,trim=0.cm 0.cm 0.2cm 0cm,width=0.99\textwidth]{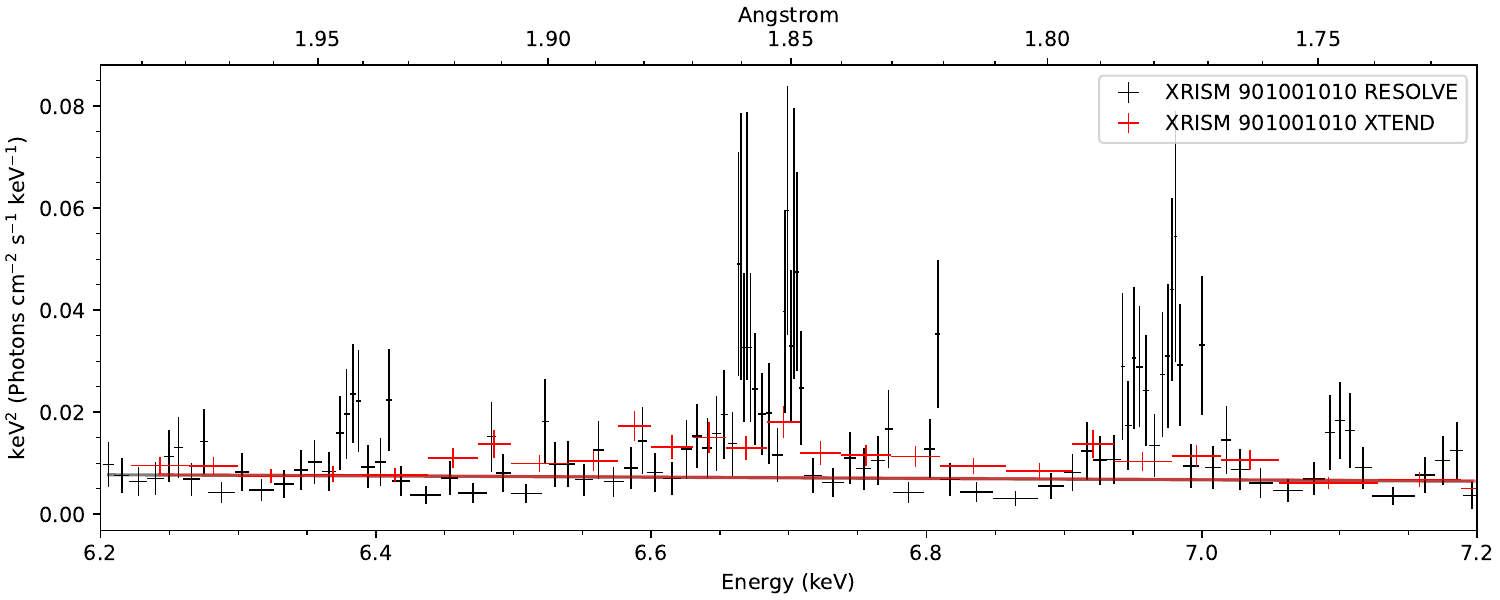}
    \vspace{-1.em}
    \caption{ Zoom of the unfolded spectra of Fig.~\ref{fig:timeavg_resid_nolines} in the 6.2-7.2 keV band, highlighting the emission features.}
    \label{fig:timeavg_resid_nolines_zoom}
\end{figure*}

We performed simultaneous optical spectroscopy with the 3.8 m Seimei telescope, located at Okayama Observatory of Kyoto University \citep{kurita10}, using the Kyoto Okayama Optical Low-dispersion Spectrograph with optical-fiber Integral Field Unit \citep[KOOLS-IFU][]{yoshida05,matsubayashi19}. We adopted the VPH-683 grism with the O56 filter, which covers 5800--8000 \AA~with a wavelength resolution of $R=\lambda/\Delta \lambda \sim$ 2000 and the VPH-495 grism, 
covering 4300--5900 \AA~with $R=\lambda/\Delta \lambda \sim 1500$. 
The observation was conducted simultaneously to the XRISM exposure, from 2024 September 30 UT 09:59--11:26 
with the total exposures of the target of 3600 s and 900 s for 
VPH-683 and VPH-495, respectively. 

We reduced the data following the standard procedures for optical 
spectra: overscan and bias-pattern subtraction, flat-fielding,
spectral extraction, wavelength calibration, sky subtraction, 
and then flux calibration, using the IRAF \citep{tody86}
and the pipeline tools\footnote{\url{http://www.kusastro.kyoto-u.ac.jp/~iwamuro/KOOLS/}; the package was downloaded 
on 2024 March 11.} dedicated to the KOOLS-IFU. The Hg, Xe, and Ne 
comparison lamp data were used in the wavelength calibration. 
We have confirmed from the calibrated comparison lamp data 
that the typical uncertainty in the wavelength determination 
was within $\sim 0.2$ \AA, and the wavelength resolution 
was typically $130$--$140$ km s$^{-1}$ (in full width at half maximum; FWHM) 
in both grisms.
The sky spectrum was created from the fibers on blank-sky areas
in the object frames. We obtained the HR 9087 data and used them 
as the standard star data for the flux calibration. 
We took the median of the spectra for the individual grisms 
to obtain high signal-to-noise-ratio spectra. 
The barycentric correction factor estimated from the IRAF tool {\tt bcvcorr} is $\Delta z = -1 \times 10^{-4}$ or $-27$ km s$^{-1}$ (corresponding to $-0.65$ \AA~at the H$\alpha$ line) 
during this observation, which was not corrected in the data and the results shown below. 

\section{Spectral Analysis}\label{sec:xana}

\subsection{Time-averaged X-ray Spectrum} \label{sub:xana_ave}

The X-ray spectral analysis was conducted using Xspec version 12.14.1 and abundances from \citet{Wilms2000_xspec_abundances}. In this analysis, we adopted the Cash statistic
and binned the Xtend and Resolve data such that each data bin
contains at least 1 count. All errors represent the 90\% confidence 
range of one parameter, unless otherwise specified.

Figure~\ref{fig:timeavg_resid_nolines} shows the time-averaged Resolve and X-tend 
spectra. The spectrum is characterized by a thermal continuum, 
consistent with a typical high/soft state spectra seen in black 
hole X-ray binaries, and emission line features around the Fe K$\alpha$ band (6--8 keV). Although we fitted both the continuum and lines together, we first focus on the description of the former, with the procedure around the later detailed in Sec.~\ref{sub:lines_empirical}. We note that the choice of the continuum (see below) has no influence on the lines, but a correct fitting of the lines is important to determine the continuum, due to the limited signal-to-noise (SNR) at high energies, and the presence of a strong feature at 1 keV that, if unaccounted for, affects the estimate of $N_{\rm H}$.

To determine the continuum profile, we first applied an absorbed disk blackbody model ({\tt TBabs*diskbb} in Xspec terminology). The model characterized the overall continuum shape well, yielding the 
best-fit parameters in Table~\ref{tab:cont_mod}, although with an unusually high disk temperature of $kT_{\rm in}=1.78_{-0.04}^{+0.03}$ keV.

Since the monitoring of other instruments (see Fig.~\ref{fig:lc_monit}) shows that the XRISM observation was taken during a decaying soft-to-hard transition,  we also tested the presence of a hard tail by adding a \texttt{thcomp} component, with a high-energy rollover fixed at a fiducial 100 keV due to the lack of high-energy coverage. We were careful to expand the model energy range from 0.01 to 1000 keV to ensure the validity of its computations. The resulting fit (see Table~\ref{tab:cont_mod}) is virtually identical to the pure disk in terms of statistic, and while the photon index of the Comptonization component is completely unconstrained, it retains a non negligible Comptonization fraction and significantly lowers the disk temperature. When limiting the possible photon index to a range of 2.0--3.5, which is what we would expect for a hard tail at the start of a transition state, we derived $kT_{\rm in}=1.65_{-0.02}^{+0.11}$ keV. In an attempt to provide additional constraints on the hard tail, we computed \swift{}/BAT survey spectra around the period of the XRISM observation, but the very low flux of the source and its erratic variability prevent any meaningful conclusions. 

\subsection{Empirical line fitting}\label{sub:lines_empirical}

Figure~\ref{fig:timeavg_resid_nolines} shows that clear narrow emission line are present in the Resolve spectrum. Visually, we can identify several features at $\sim 6.4$ keV, $\sim 6.7$ keV, and $\sim 7.0$ keV, likely the neutral Fe K$\alpha$, Fe He$\alpha$, and Fe Ly$\alpha$ complexes respectively. In parallel, similarly to what was found in \cite{Shaw2022_SAXJ1819_wind_emission_2020}, several other features are present at other energies. To identify the features that warrant spectral fitting, we performed a blind search for narrow emission/absorption line features, following the methodology highlighted in \cite{Parra2024_winds_global_BHLMXBs}. To account for the resolution of Resolve and the very prominent emission features, we fixed the line width to a fiducial 5 eV, varied the energy between 2 and 10 keV with steps of 5 eV, and the normalization range in 500 logarithmic steps between $10^{-2}$ and 50 times the local continuum flux. For the 0.3--2 keV range, covered exclusively by Xtend, we used a similar normalization range, a line width fixed at 0, and steps of 20 eV.

We show the result of the blind search in the left panels in Fig.~\ref{fig:blind_search_total_gauss} (App.~\ref{app:add_spectral}), with clearly identifiable features for \Caxx{}, \Crxxiii{}, \Fei{}, \Fexxv{}, \Fexxvi{} and \Nixxvii{} in resolve. In Xtend, we note a significant emission feature compatible with a \Nex{} K$\alpha$ transition. Not only are there no strong edges at this energy, but the feature strongly varies along the observation (see Sec.~\ref{sub:time_resolved}), and is clearly detected in the NICER observations of the previous days, thanks to a luminosity an order of magnitude higher. Most importantly, a \Nex{} K$\alpha$ feature was already unambiguously detected in one of the Chandra observations of a similar state in a previous outburst \citep{Shaw2022_SAXJ1819_wind_emission_2020}.

\begin{figure*}[t!]
    \centering
    \includegraphics[clip,trim=2.5cm 0.8cm 0.5cm 0.5cm,width=0.99\textwidth]{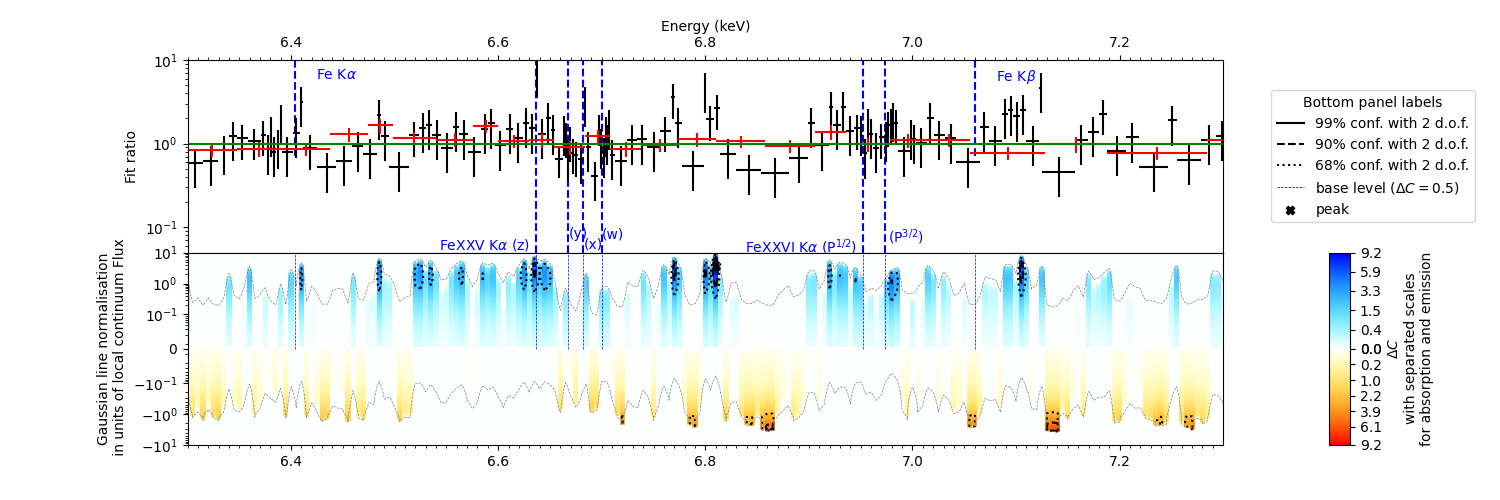}
    \includegraphics[clip,trim=2.5cm 0.8cm 0.5cm 1.4cm,width=0.99\textwidth]{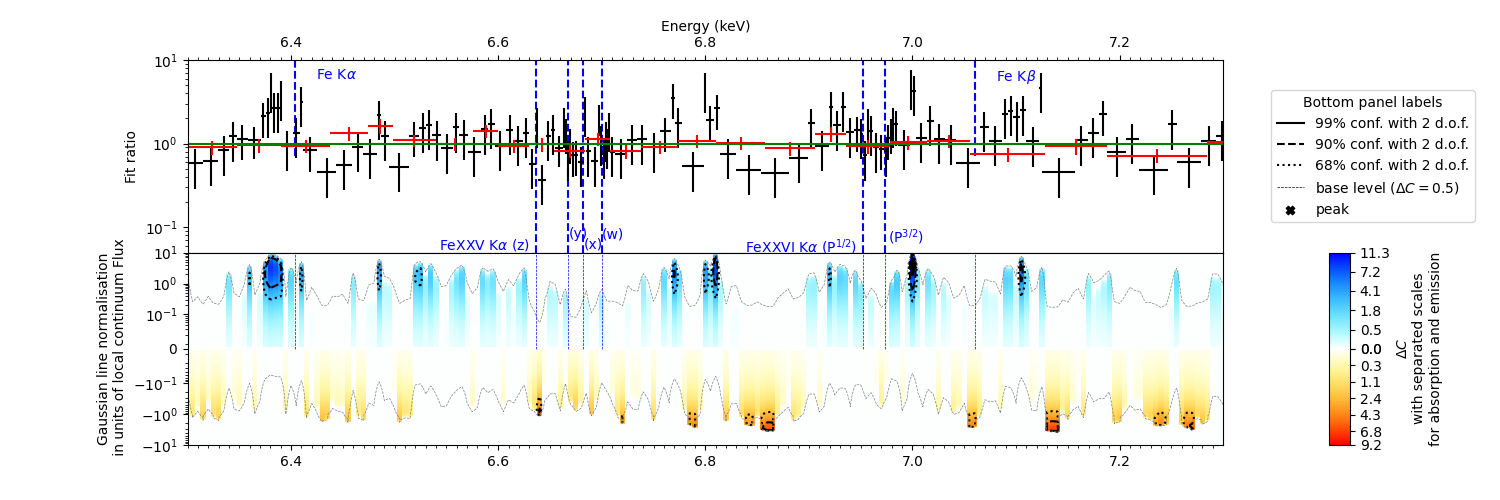}
    \vspace{-1em}
    \caption{Blind search showing the significance of the residuals after the Gaussian fit \textbf{(top)} and the pion fit \textbf{(bottom)} in the most critical region. Due to the very high ratios ($\gtrsim4)$ of the remaining emission features after the pion fit, the ratio panels are both shown in log scale.}
    \label{fig:blind_search_6_7}
\end{figure*}

We thus fit the lines in the Resolve and Xtend spectra, focusing on the emission lines with an individual component at more than 2$\sigma$ confidence, using the features in the blind search as a reference. With the limited SNR of the observations, \texttt{gaussian} components were sufficient to describe the line shapes. We fitted the individual components separately whenever possible, and otherwise used a single line when a single line was enough to describe the entire feature, namely for all features below 6.5 keV except for the \Crxxiii{} He complex, whose $w$ transition can clearly be identified.
We present the results of our fitting in Tab.~\ref{tab:line_param}, 
and show the significance of the lines using C-stat. We leave the widths free for all components except the 3 that remain unconstrained within the observation, namely \Nex{} (due to being only present in Resolve), and \Crxxiii{} and \Nixxvii{} (both with a very limited SNR). In all 3 cases, we freeze the line widths at 0 in the analysis. Most lines were previously identified in \cite{Shaw2022_SAXJ1819_wind_emission_2020} thanks to a much higher luminosity, except for the \Crxxiii{} complex and \Nixxvii{} complexes, whose $w$ transitions have EWs significant at a 95$\%$ and 97$\%$ significance level respectively. For \Fexxv{} K$\alpha$, we only fit components for the $w$ and $y$ transition, since only these two transitions are seen at a statistically significant level in the blind search. We show the line parameters derived from our fit in Tab.~\ref{tab:line_param}, including detailed orbital corrections considering the Earth movement in the solar system, the relative velocity of the binary relative to the solar system, and the orbital velocity of the BH within the binary. 

First, fitting with a single component for \Fexxvi{} K$\alpha$ leaves a clear residual at $\sim 7.00$ keV (which can be seen as an additional peak in Fig.~\ref{fig:blind_search_total_gauss}, and as a residual to physical models in Fig.~\ref{fig:blind_search_6_7}). We expect no strong line at this precise energy, and a \Fei{} K$\beta$ line would require a redshift of $\sim3000$ km s$^{-1}$ to reach 7.00 keV. We thus model this feature with a second resolved \Fexxvi{} K$\alpha$ component, resulting to an improvement of $\Delta C$ of 12. To assess more precisely the significance of this component, it is necessary to perform MC simulations \citep{Porquet2004_stat_MC}. We thus generated 1000 simulated Resolve+Xspectra using the \texttt{fakeit} command in xspec from the best fit continuum model, retaining all observational parameters of the initial spectra. In each case, we refit the continuum only model to derive a baseline C-statistic, before scanning the $[-3000;3000]$ km/s velocity shift band around the \Fexxvi{} K$\alpha$ line energy with a resolved \Fexxvi{} complex. To reproduce this complex, we use  2 gaussians at the rest energy of the individual transitions, with tied widths, a Ly$\alpha_1$ normalisation fixed to twice that of Ly$\alpha_2$, and both transitions fixed at a width of 1eV. Our choice of velocity shifts corresponds to an energy band of [6.90,7.04] keV and is a conservative estimate of the band where spurious detections would be associated to the \Fexxvi{} line. The distribution of the maximum $\Delta$C-stat of this artificial line in each spectrum can then be compared to that of the real spectrum. The significance of the real line is  defined as $P=1-N/1000$, with N the number of simulated spectra where the maximum $\Delta$C-stat of an artificial line is lower than the real $\Delta$C-stat value obtained in our observation. With this method, we derive a significance of 99.2\% (2.6$\sigma$) for this secondary component. While line searches in Active Galactic Nuclei (AGNs), with flux levels comparable to our observation, typically use thresholds of 95-99\% for their line detections \citep{Tombesi2010_wind_AGN_XMM,Matzeu2023_UFO_AGNs}, the commonly accepted detection threshold for X-ray Binaries (albeit suited for objects 1 to 2 orders of magnitude brighter) is of 3 $\sigma$. We thus only consider this line as a marginal or tentative detection. In addition, we note that this significance is derived for a width of the main \Fexxvi{} K$\alpha$ component below $\sim20$eV. At widths of $\gtrsim20$ eV, the main component covers a part of the secondary feature at higher energies, and thus the  $\Delta C$ of the secondary component is reduced to 8. This parameter space remains however very unlikely, since the corresponding $\Delta v$ for the main \Fexxvi{} K$\alpha$ component would end up $\gtrsim$ 3 times higher than that of every other line detected in the spectrum. 

\begin{figure*}[t!]
    \centering
    \includegraphics[width=0.95\hsize]{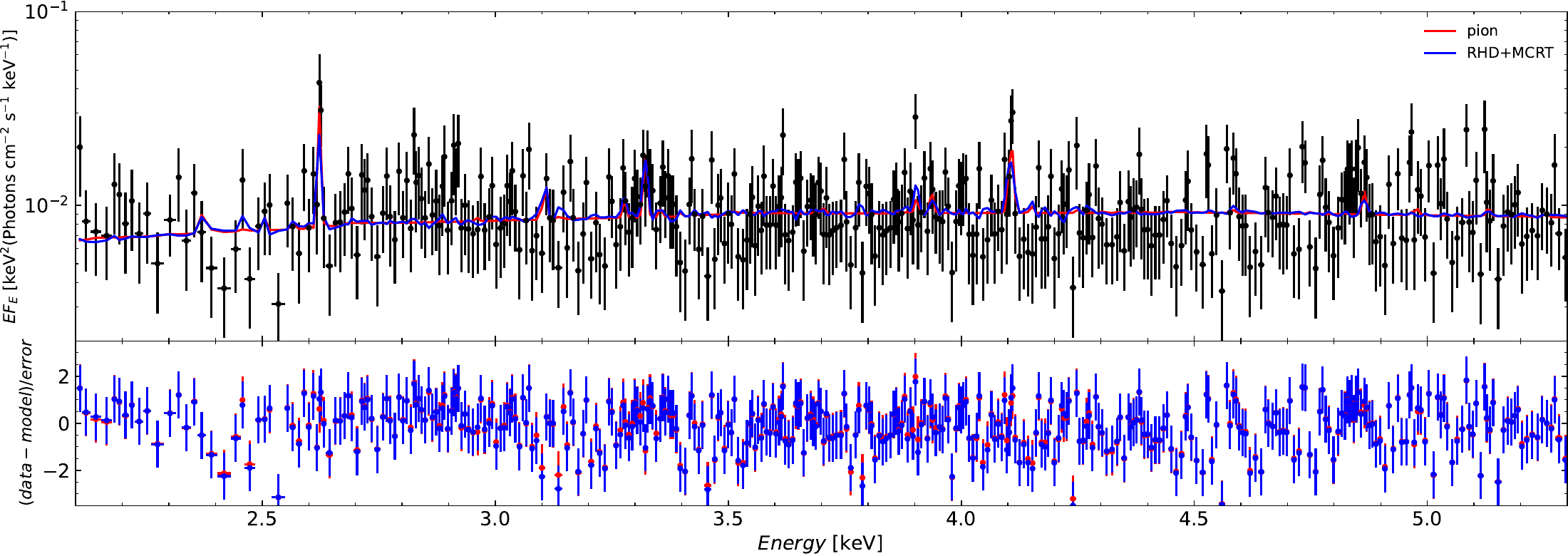}\\
    \includegraphics[width=0.95\hsize]{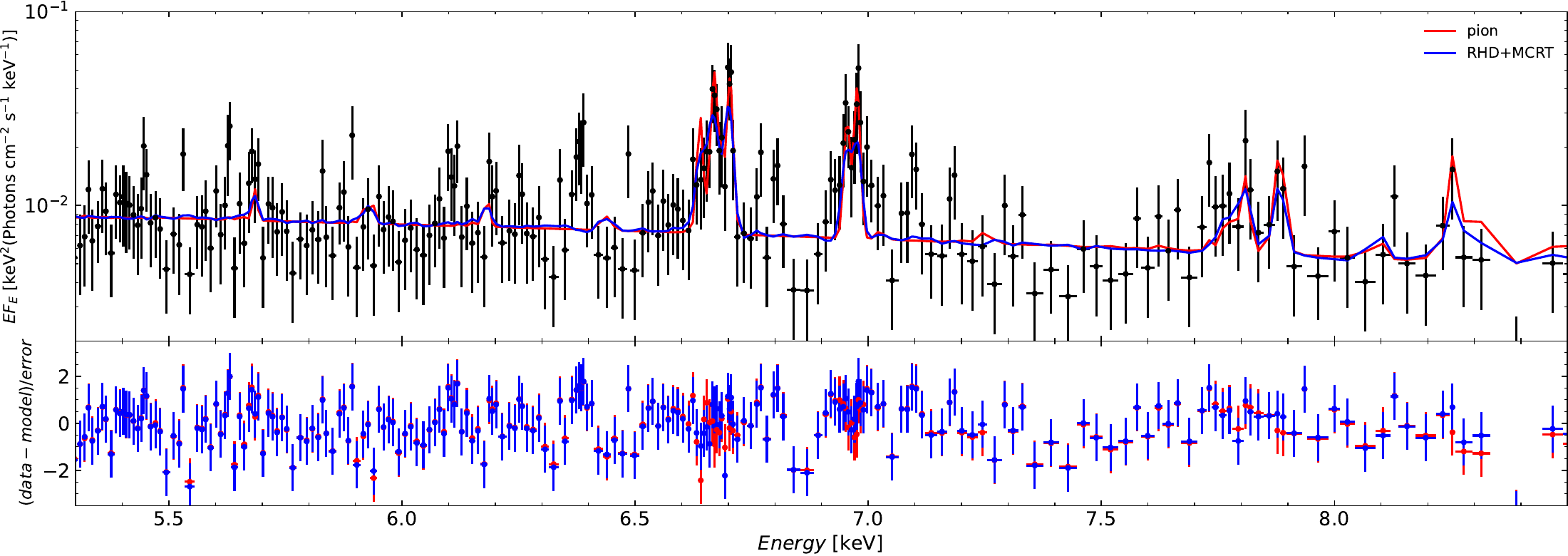}\\
    \vspace{-1em}
    \caption{Comparison of the Resolve spectrum with the model of photo-ionized plasma emission (red) and the numerical Monte-Carlo radiative transfer simulation (blue).
    These two models are fitted simultaneously to the Xtend and Resolve data, with the Xtend data omitted from the plot for clarity.
    }
    \label{fig:pion_Fe}
\end{figure*}

This tentative secondary component (which we will refer to as the ``fast'' component in the rest of the analysis, in contrast to the main ``slow''component), has a low width, which we constrain to below 3.2 eV at 90\% significance. This allows for very good constraints on its velocity, derived as $-1286^{+85}_{-92}$  km s$^{-1}$, and below $-800$ km s$^{-1}$ at a 3$\sigma$ confidence level. If this can be confirmed with subsequent observations, to the best of our knowledge, it would be the first time a narrow emission component with a blueshift above $1000$ km s$^{-1}$ would be identified in the X-ray band in a BH LMXB. The neutral Fe K$\alpha$ line, on the other hand, shows an important redshift, but with much weaker constrains, with a velocity derived at $722^{+305}_{-549}$ km s$^{-1}$ after correction, and above 0 at a $> 2 \sigma$ confidence level.

The rest of the lines show a trend for weak blueshifts, but none are statistically significant. Since their values are all compatible within errors, we tested the effect of linking the blueshift of the 6 components (SXVI, \Caxx{}, \Crxxiii{}, \Fexxv{}, \Fexxvi{}-slow, and \Nixxvii{}). This results in a very similar fit ($\Delta C=+9$ for 5 free d.o.f.), and a common velocity of $-124_{-105}^{+101}$ km s$^{-1}$ at 90\% confidence level after correction,  showing a trend of weak blueshifts for these lines, but is insufficient for a physical interpretation. We note that aside from being corroborated by other detections in previous observations \citep{Shaw2022_SAXJ1819_wind_emission_2020}, the very low velocity shift of these components means that odds of spurious detections due to photon noise are much lower compared to the fast \Fexxvi{} component, for which we purposely scanned a wide range of energies. Even the lines with a similarly low $\Delta$ C-stat value are thus all more than 3 $\sigma$ significant.

Following the fit of these components, we perform a new blind search to assess the significance of the residuals remaining in the spectra, whose results we show in the right panels of Fig.~\ref{fig:blind_search_total_gauss}. All lines are correctly fitted, and only very weak (and non-significant) residuals remain, notably from expected transitions around the \Fexxv{} K$\beta$ and \Fexxvi{} K$\beta$ complexes. We do note a duet of puzzling emission features around 6.8keV, which, while not individually significant at 3$\sigma$, match the separation between the (y) and (w) transitions of \Fexxv{} K$\alpha$, and could thus be the sign of an additional, highly blueshifted ($\sim-5000$km s$^{-1}$) component. Finally, the emission feature at $\sim7.1$ keV is unlikely to come from a \Fei{} K$\beta$ component, as it would require a very high blueshift ($\sim-1800$ km/s) incompatible with the \Fei{} K$\alpha$. Instead, it is more likely to be due to the start of the iron edge at $7.12$ keV, which can exhibit a very complicated profile depending on the composition of the absorbing material \citep{Rogantini2018_ironedge}, which we cannot resolve with our limited SNR.

\subsection{Photoionized plasma models}

\begin{table}[h!]
\centering
\caption{Best-fit parameters of the photoionized plasma model}
\label{tab:photo_model}
\begin{tabular}{lcc}
\hline
Model & Parameter [Unit] & Value \T\B\\
\hline
{\tt TBabs} & $N_{\rm H}$ [$10^{21}$ cm$^{-2}$] & $1.21 \pm 0.09$ \T\B\\
    {\tt pion }& $\log (\xi/[{\rm erg~cm~s^{-1}}])$ & $3.9\pm 0.1$\T\B\\         
    & $N_{\rm H}~[{\rm 10^{22} cm^{-2}}] $ & $50^{+70}_{-30}$\T\B\\
         & $v_{\rm rms}~[{\rm km~s^{-1}}] $ & $ 220^{+60}_{-50}$\T\B\\
         & $z_\textrm{raw}$ & $(-4\pm 2) \times 10^{-4}$ \T\B\\
         & $z_\textrm{real}$$^{(a)}$ & $(-8\pm 2) \times 10^{-4}$ \T\B\\
         & norm$^{(b)}$  & $1.0^{+0.7}_{-0.4}\times 10^{-10}$\T\B\\
        {\tt diskbb} & $kT_{\rm in}~[{\rm keV}] $&$ 1.80 \pm 0.04$\T\B\\
             & norm  & $0.13 \pm 0.01$ \T\B\\
  & $R_{\rm in}$$^{(c)}$ [km] &  $0.40 \pm 0.01$ \T\B\\ 
 {\tt constant} &  $C_{\rm xtd}$$^{(d)}$ &$1.0$ (fix)\T\B\\
       & $C_{\rm rsl}$$^{(d)}$ &  $1.04 \pm 0.03$ \T\B\\
  \hline 
   Fit statistic & C-stat$/$\rm dof & $4279/4778$ \T\B\\
\hline
\end{tabular}
\tablefoot{
\tablefoottext{a}{Before and after applying the velocity correction of $-117\pm13$ km/s derived for the observation (see App.~\ref{app:vel_corr} for details)}\\
\tablefoottext{b}{$=(L_{\rm irr}/4\pi D^2)\times \Omega/4\pi$}\\
\tablefoottext{c}{Calculated from {\tt diskbb} normalization assuming a distance of 6.2 kpc and an inclination of $72^\circ$ \citep{MacDonald2014}.}\\
\tablefoottext{d}{Cross normalization factors of Xtend and Resolve.}}
\end{table}

We modeled the photoionized emission lines with the plasma code {\tt pion} (v3.08.00), which is implemented within the X-ray spectroscopy analysis software {\sc spex}. We used the default cross section measurements, namely from the OPEN-ADAS database \footnote{see \href{https://open.adas.ac.uk/}{https://open.adas.ac.uk/}}. 
 Here, instead of fitting in {\sc spex},
 we extracted emission lines calculated by {\tt pion} and constructed an additive table model ({\tt atable}) that can be implemented into {\sc xspec}. This makes fitting much faster, as 
 {\sc spex} recalculates the photoionization state self consistently with every iteration of the continuum. 
 To address this issue, we fixed an illumination SED shape and precomputed the photoionization states and spectra, compiling them into a table model.
 
We calculate a grid of models for solar abundances, simulating emission with different values of column density $N_{H}$, ionization parameter $\log \xi$, 
and turbulent velocity $v_{\rm rms}$.
This simulation includes free-free emission, line emission, and recombination lines. 
Each of the simulations has 65536 logarithmically-spaced bins to cover the energy range from $10^{-4}$ to $10^{3}$ keV with a resolution of 1.5 eV around 6 keV, enough to fit Resolve data. 
In total, we perform 8736 simulations with values $21\leq \log N_H \leq 25$ spaced by 0.2 (21 points), 
$2 \leq \xi \leq 7$ spaced by 0.2 (26 points),
and $-5 \leq \log(v_{\rm rms}/c) \leq -2$ spaced by 0.2 (16 grid points). 
We assume a single number density $n_{p} = 10^{12}$ ${\rm cm^{-3}}$ to reduce the size of the tables. 
We calculate the population levels from radiative recombination, cascade, radiative, and collisional excitation correctly for metastable levels. 

Adding this photo-ionized emission to the continuum gives a good fit to the data (see 
Table \ref{tab:photo_model}), and successfully matches all the emission lines in the data (see Fig.~\ref{fig:pion_Fe}), except for the 6.4~keV Fe K$\alpha$ fluorescence line and the secondary \Fexxvi{} component, which are expected to arise from different material considering their distinct velocity shifts. The ionization state is well constrained at $\log\xi=3.9\pm 0.1$, but the column density is not as this mostly sets the overall normalization. In addition, the velocity, after adding the correction of $-117\pm13$ km/s, is derived as $-237\pm73$ km/s, which is not only twice higher than the common blueshift values derived from line fitting in the previous section, but also more than 3$\sigma$ significant. We note that the uncertainty in the systematics, unlike that of the fit, is a fixed value.
 Once again, we confirm the presence of residuals with a blind search in the entire Resolve band. The results for the iron band are shown in Fig.~\ref{fig:blind_search_6_7}, and the rest of the spectrum in App.~\ref{fig:blind_search_total_phys}.

\subsection{Time-resolved analysis}\label{sub:time_resolved}

\begin{figure*}[p!]
  \begin{minipage}[c]{0.4\textwidth}
    \includegraphics[width=1.0\textwidth]{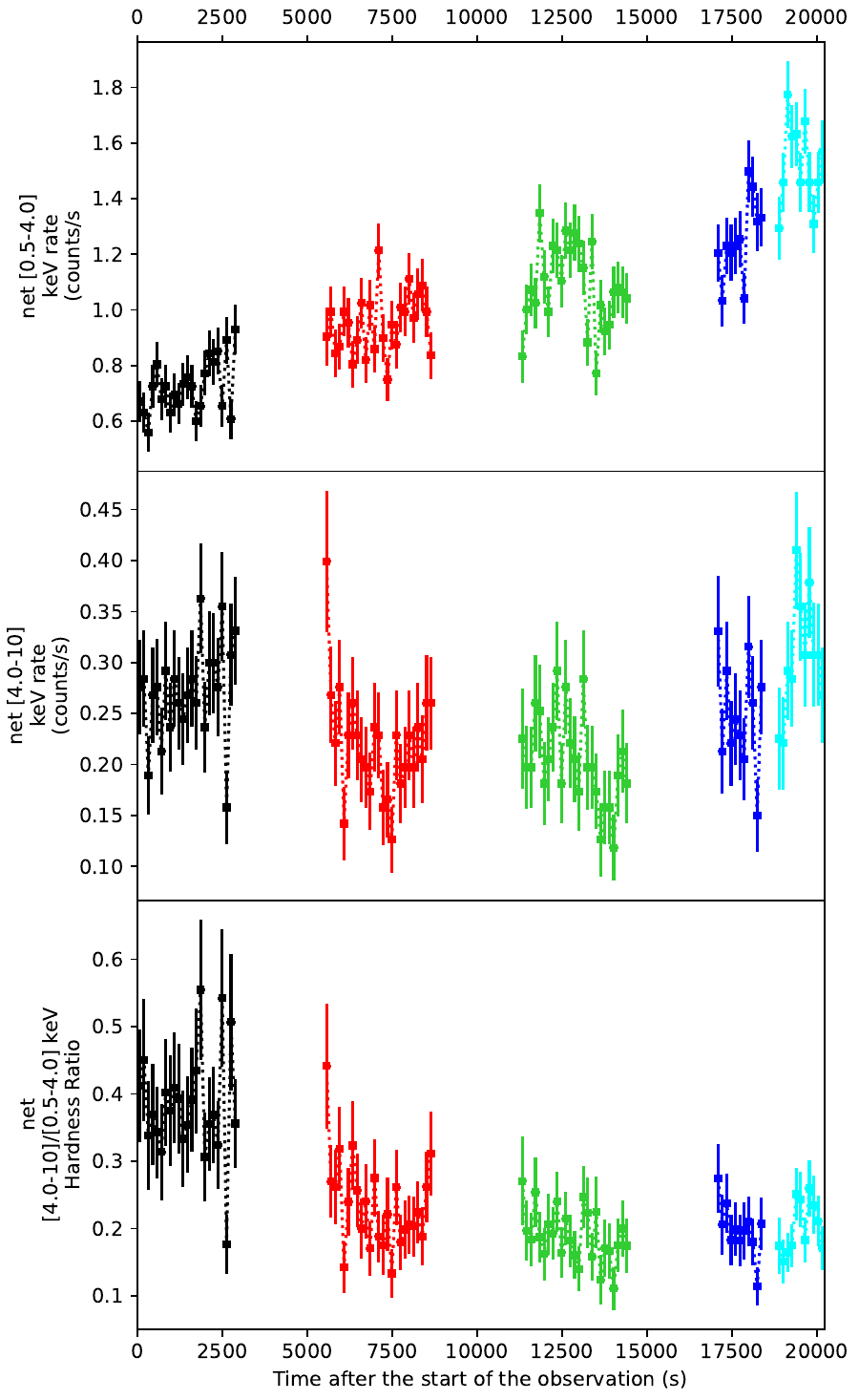}
  \end{minipage}\hfill
  \begin{minipage}[c]{0.6\textwidth}
    \includegraphics[clip,trim=0cm 0cm 0cm 0.2cm,width=1.0\textwidth]{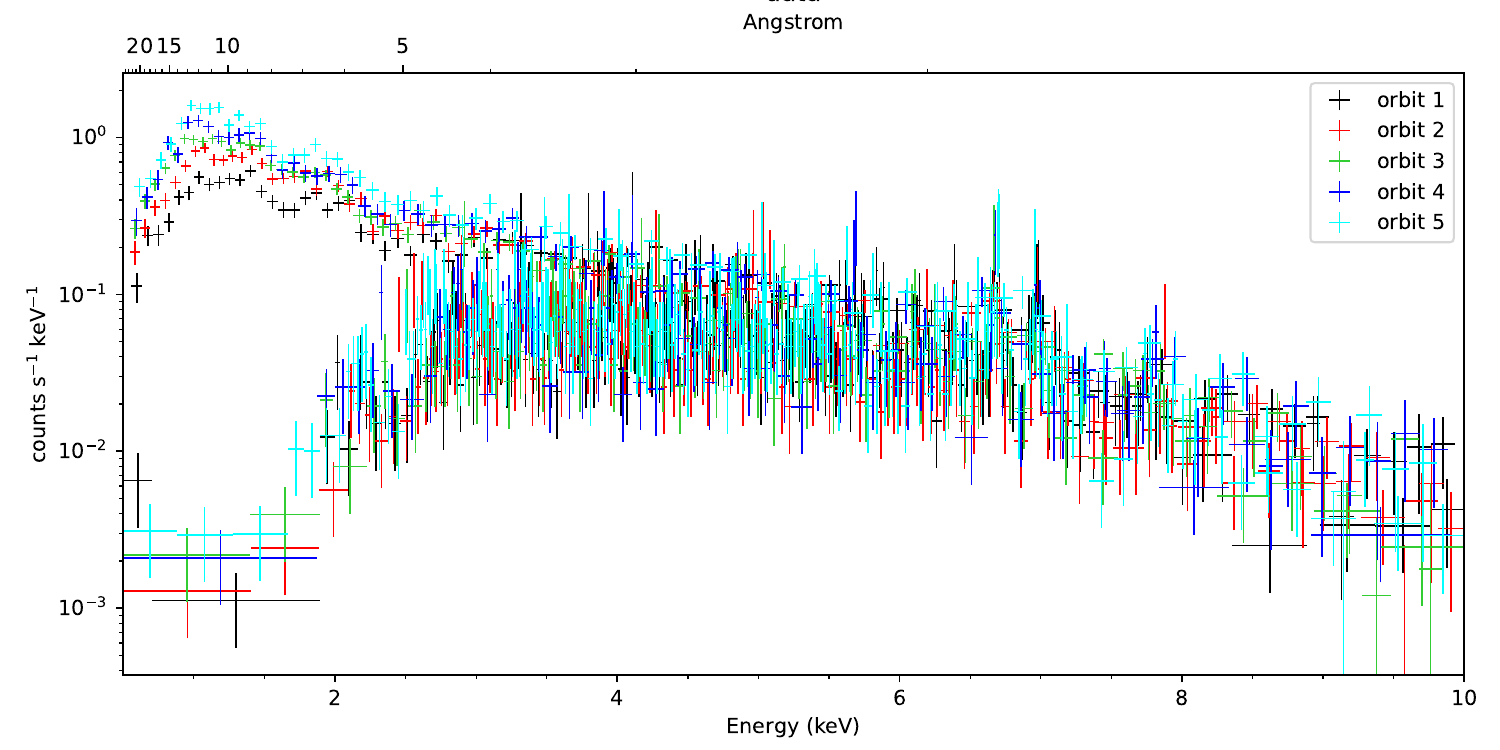}
    \includegraphics[clip,trim=0cm 0cm 0cm 0.2cm,width=1.0\textwidth]{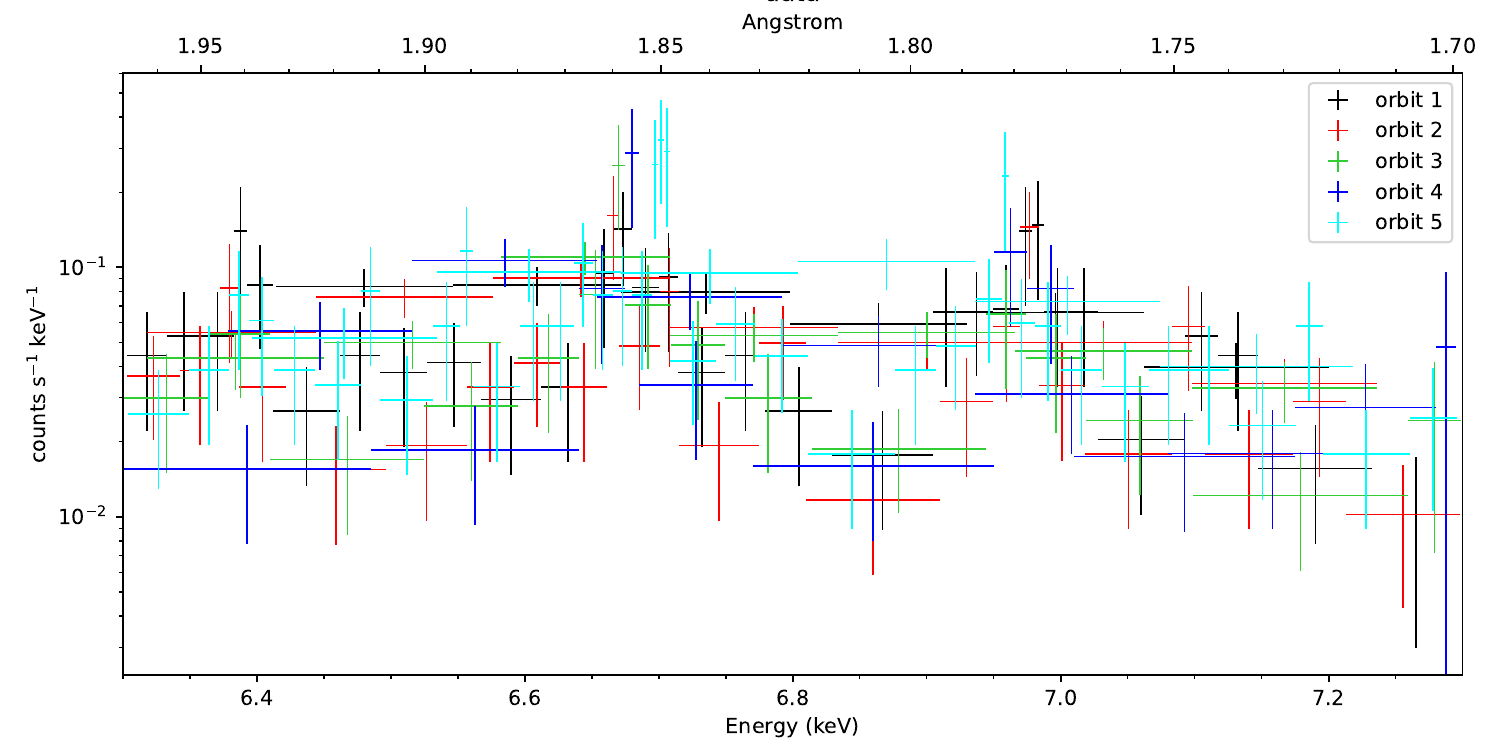}
  \end{minipage}
\caption{ \textbf{(Left)} Xtend light curve and HR of the observation with bins of 128 s, highlighting the different orbits used for time-resolved spectroscopy. \textbf{(Right)} folded Xtend and Resolve spectra of individual orbits of the observation \textbf{(top)} and zoom of Resolve to the 6.3--7.3 keV band \textbf{(bottom)} rebinned for visual clarity. In the top right panel, the Resolve effective area is near zero below 2 keV due to the presence of the Gate Valve}
    \label{fig:obs_timeres}
\end{figure*}

V4641 Sgr is know for its dramatic spectral evolution on very short time-scales \citep{hjellming00}. When computing the light curve of the source in several bands, as show in the left panels of  Fig.~\ref{fig:obs_timeres}, we see that although the higher energies remain relatively stable, at lower energies, the source brightens by a factor $\sim3$ between the beginning and the end of the observation. To investigate the changes in continuum and line properties, we thus derive Resolve and Xtend spectra for each of the individual orbits, which we show in right panels of Fig.~\ref{fig:obs_timeres}. We further divide the last orbit into two periods due to its strong evolution in flux, and refer to the two as different orbits due to the small observing gap between the two. Despite a very limited SNR, the source shows both a strong continuum evolution at low energy and changes in line properties.

We first assess the evolution of the continuum by fitting the Resolve and Xtend spectrum of each orbit together, while ignoring all the energy bands where lines were detected in the time-averaged spectrum. Although this progressive variation at low energy is characteristic of a change in the intrinsic absorber obscuring the source, we first attempted to fit each orbit with an independent \texttt{diskbb} and a common interstellar absorption component. The resulting fit, although globally satisfactory (C-stat/d.o.f.$\sim$1.2 in Xtend), is much worse for the first 3 orbits, and results in completely unphysical disk parameters, with notably an inner temperature in the first orbit of $T_{in}=2.5_{-0.2}^{+0.1}$ keV. We thus fitted all orbits together with a second model including a partial covering absorber, modeled with \texttt{cabs*pcfabs} within Xspec, whose column density and covering fraction can vary between orbits within the limits described below, and a disk that remained constant along the observation. To consider properly the influence of Compton Scattering in the partial covering absorber, as \texttt{cabs} doesn't include a covering fraction parameter, we defined our model in xspec as \texttt{TBabs(constant*diskbb+cabs*pcfabs*constant*diskbb)}, which each disk representing the uncovered and covered fractions. We thus link the parameters of both \texttt{diskbb}, fix the covering fraction in \texttt{pcfabs} to 1 (this represents the fully absorbed component), and fix the value of the second \texttt{constant} to 1 - the first \texttt{constant}, so as to act as a covering fraction parameter. Finally, we link the absorption of the \texttt{cabs} component to that of the \texttt{pcfabs}. To limit the degeneracy between the absorption components and the disk normalization, we fix the covering fraction in the brightest orbit at 0, and the interstellar column density at $N_{H}=0.15 \times 10^{22}$ cm$^{-2}$. With this choice, we focus on constraining the additional absorber between the last and first orbit.
Finally, since Resolve and Xtend were perfectly matching in the time-averaged spectrum, we do not use a constant factor between the two instruments.

\begin{table*}[p!]
\centering
\caption{Best-fit continuum parameters for the individual orbits using a partially covered disk model. 
$\dagger$ indicates that the parameters are frozen or at the limit of their allowable interval.}
\vspace{0.5em}
\label{tab:xparam_orbits}
\begin{tabular}{llccccc}
\hline
\hline
 \multirow{2}{*}{component} & \multirow{2}{*}{Parameter[Unit]} &  \multicolumn{5}{c}{disk-only model}  \T\B\\
 &  & Orbit 1 & Orbit 2 & Orbit 3 & Orbit 4 & Orbit 5 \T\B\\
  
   \hline

\tt TBabs & $N_{\rm H}$ [$10^{22}$ cm$^{-2}$] &\multicolumn{5}{c}{$0.15^\dagger$} \T\B\\
\multirow{2}{*}{\tt cabs*pcfabs} & $N_{\rm H}$ [$10^{22}$ cm$^{-2}$] & $8.5_{-1.7}^{+2.1}$ & $63_{-22}^{+50}$ & $94_{-38}^{+106\dagger}$ & $101_{-65}^{+99\dagger}$ & 0$^\dagger$\T\B\\
 & f & $0.56_{-0.03}^{+0.02}$ & $0.34_{-0.03}^{+0.02}$ & $0.27_{-0.03}^{+0.02}$ &$0.16\pm0.04$ & 0$^\dagger$ \T\B\\
\multirow{2}{*}{\tt diskbb} & $kT_{\rm in}$ [keV] &\multicolumn{5}{c}{$1.66\pm 0.03$} \T\B\\
&norm  & \multicolumn{5}{c}{$0.23\pm 0.02$} \T\B\\
  \hline
   $F_{\rm X}$$^{(a)}$ & [$10^{-11}$ erg s$^{-1}$ cm$^{-2}$] & 2.19 & 2.17 & 2.33 & 2.65 & 3.13 \T\B\\
  $L_{\rm X} $$^{(a)}$ & [$10^{-4} L_{\rm Edd}$] & 1.25 & 1.24 & 1.33 & 1.51 & 1.79 \T\B\\   \hline
    $\rm HR$$^{(b)}$ & / & 2.18 & 1.43 & 1.37 & 1.36 & 1.33 \T\B\\
   Fit statistic & Xtend & 97/71 & 98/70 & 93/70 & 53/66 & 72/66 \T\B\\ 
   Fit statistic & Resolve & 1180/1345 & 1208/1346 & 975/1342 & 864/1341 & 1251/1350 \T\B\\
   \hline
\end{tabular}
\tablefoot{
\tablefoottext{a}{1--10 keV flux and luminosity after removal of the interstellar absorption contribution (\texttt{TBabs}). The Eddington ratio is calculated according to the orbital parameters of \cite{MacDonald2014}}
\tablefoottext{b}{3-10/1-3 keV flux hardness ratio after removal of the interstellar absorption contribution.}}
\end{table*}

\begin{figure*}[h!]
    \centering
    \includegraphics[clip,trim=0.5cm 0.8cm 1.2cm 0cm,width=\textwidth]{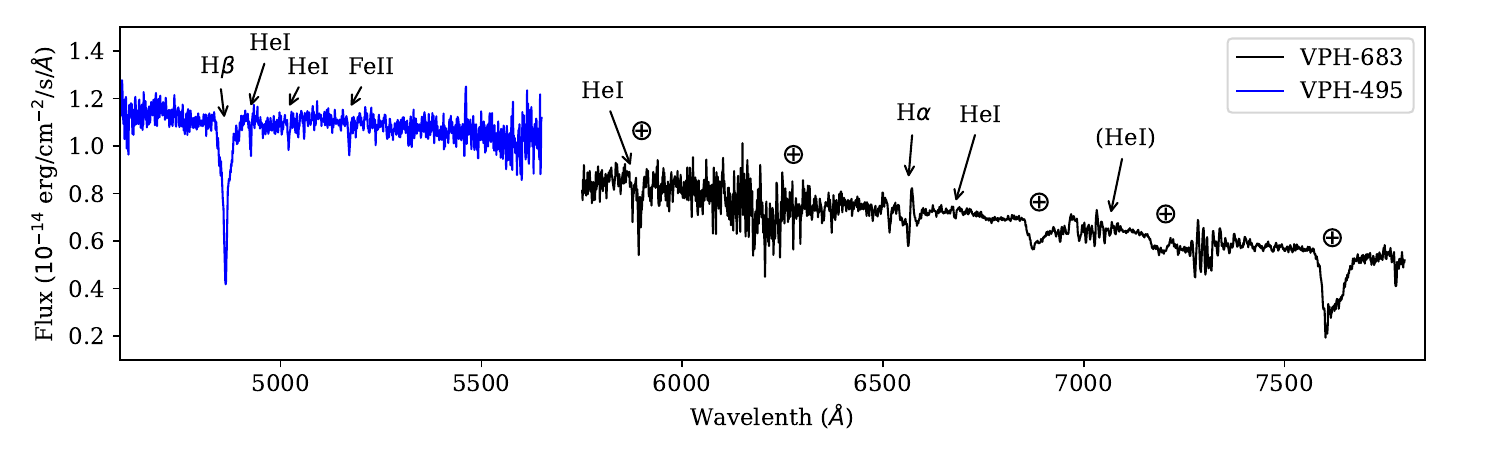}
    \vspace{-1.em}
    \tablefoottext{}{Wavelength(\AA)}
    \caption{Optical spectrum obtained with the Seimei telescope. The blue and black data show the spectra obtained with the VPH-495 and VPH-683 grisms, respectively. $\oplus$ indicate atmospheric lines.} 
    \label{fig:seimei}
\end{figure*}

We show the parameters of the fit in Tab.~\ref{tab:xparam_orbits}. The quality of the fit is similar to the empirical fit with varying disk parameters, while retaining physical parameters. We measure an increase of $50\%$ in flux between the first and last orbits, and a diminution of $\sim40\%$ of the 3--10/1--3 keV hardness ratio. However, in part due to our limited statistics, the fit retains very limited, as despite a progressive decrease in covering fraction between orbits, orbits 2 to 4 barely constrain the column density of the absorber. Moreover, the Xtend statistics of the first 3 orbits remain significantly higher than the last two orbits, hinting at a much more nuanced absorber structure than what our simply model can assess. In hope of reaching a better agreement between the first and last orbits, we tested several different variations of the model, such as thawing the interstellar column density, using a variable ionized partial covering (with \texttt{zxipcf}), and a physical comptonized disk model (\texttt{thcomp(kerrbb)}) inside the partial covering. None led to a meaningful improvement in the fit. 

In parallel, we can see the iron lines vary in both energy and strength in the zoomed iron spectrum see in the bottom panel of Fig.~\ref{fig:obs_timeres} (we stress that these spectra are visually rebinned but retain a much higher resolution). This hints at a much more complex evolution in the absorber obscuring the source than what can be assessed with continuum fitting, but the SNR of each line remains very marginal with only $\sim2-2.5$ks of exposure per orbit. We thus limit ourselves to the lines detected in at least one individual orbit. For this, we start by performing a blind search for narrow features in each observation period, of which we show a summary in Fig.~\ref{fig:blind_search_timeres}. We then fit lines in each orbit based on the result of the blind search, restricting ourselves to those with a 99\% significant feature in at least one of the five orbits. Our orbit-resolved line parameters are presented in Tab.~\ref{tab:line_param_orbit}.

Despite a very limited SNR, we confirm the trends seen visually in the spectrum, with a remarkable amount of variability in several different complexes, some of which unrelated to the changes in the continuum absorber. First, our fits show a surprising switch from an intercombination-dominated \Fexxv{} He profile in the first four orbits, to a resonance-combination dominated profile in the final orbit. In addition, although we detect a rise in the EW of the strongest component along the orbits, the significance is too low for any definitive conclusion. Meanwhile, the \Fexxvi{} Ly$\alpha$ profile, aside from a strong but marginally significant increase in EW in orbit 2, shows a remarkable velocity evolution, from being compatible with 0 in the first 4 orbits, to being very significantly redshifted in the last observation, at $790_{-110}^{+100}$ km s$^{-1}$ before the orbital corrections. Although this may be serendipitous, we note that this redshift matches that of the weak \Fei{} K$\alpha$ feature detected in the time-averaged spectrum. However, it remains very different from the well constrained, static ($-10_{-109}^{+136}$ km s$^{-1}$) \Fexxvi{} K$\alpha$ line seen in the same orbit, and thus the two lines must originate from two distinct absorbers (at least in that orbit).
In the weaker lines, the \Crxxiii{} He$w$ transition is only detected in a single orbit, but with a very high EW, and ends up diluted in the time-averaged spectrum. On the other hand, the \Fexxv{} Ly$\beta$ line, who was not significant in the time-averaged spectrum, is detected at more than 3 $\sigma$ in orbit 2. However, unlike the Cr transition, the respective lack of detection in other orbits are not significant enough to conclude on its evolution in the rest of the observation. Finally, although the evolution of the \Nex{} Ly$\alpha$ is difficult to disentangle from that of the poorly fitted continuum at low energies, there is a progressive change from a complete lack of line in the first orbit, to a very strong and significant feature in orbit 5.

\section{Optical Analysis}
\label{sec:optana}

\begin{figure*}[h!]
    \centering
    \includegraphics[clip,trim=1cm 0.4cm 4.5cm 2.5cm,width=0.32\textwidth]{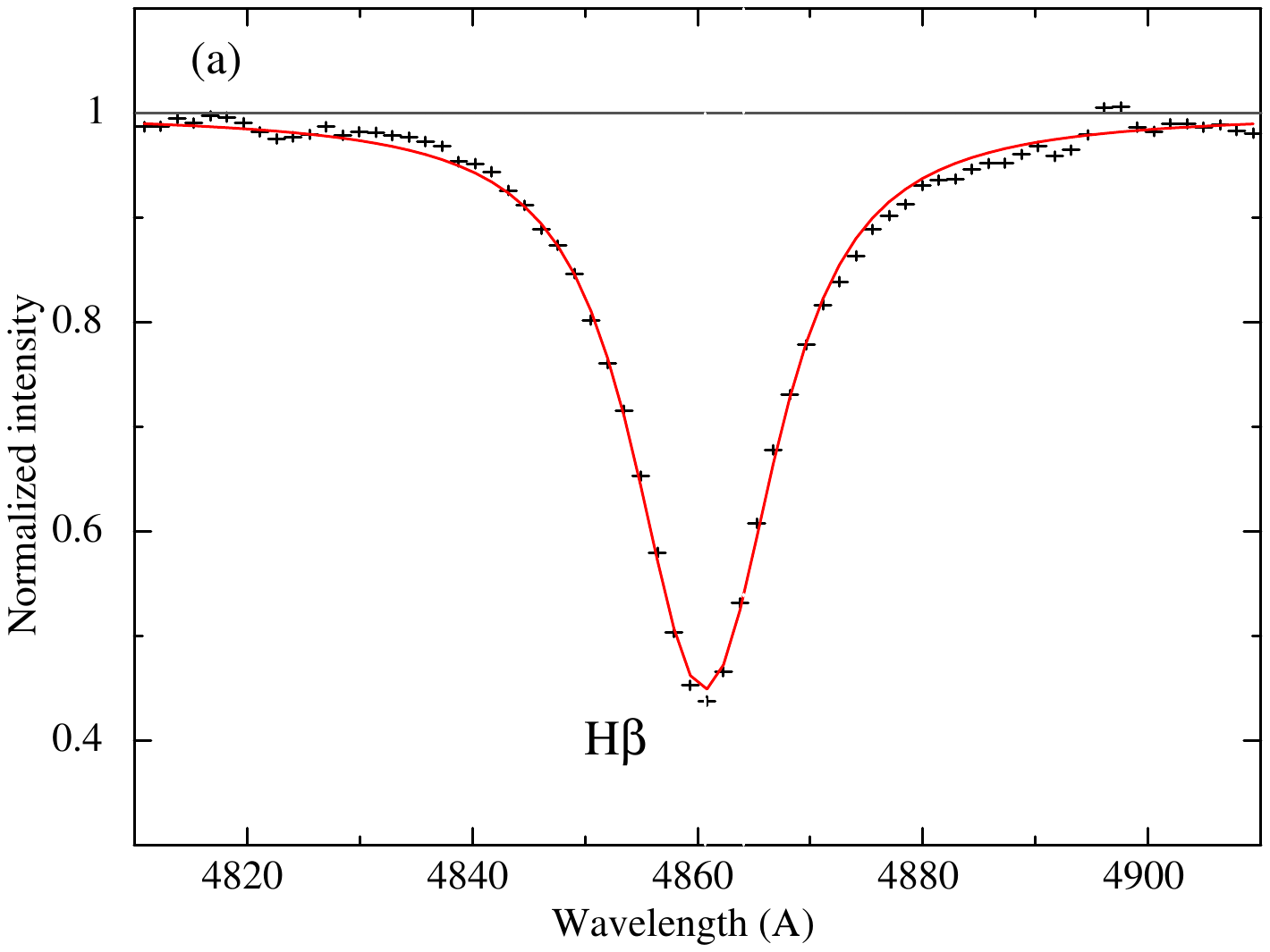}
    \includegraphics[clip,trim=1cm 0.4cm 4.5cm 2.5cm,width=0.32\textwidth]{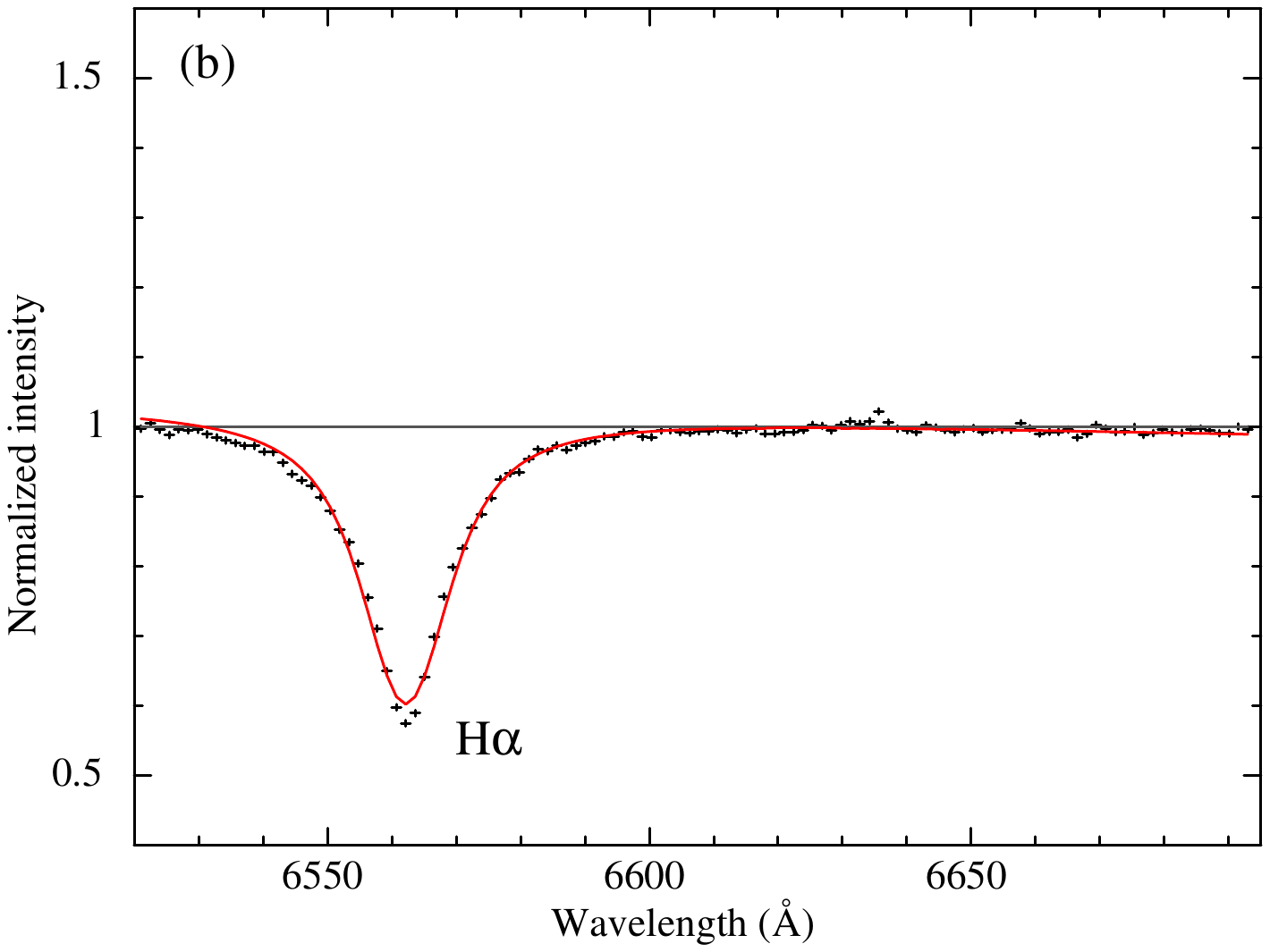}
    \includegraphics[clip,trim=1cm 0.4cm 4.5cm 2.5cm,width=0.32\textwidth]{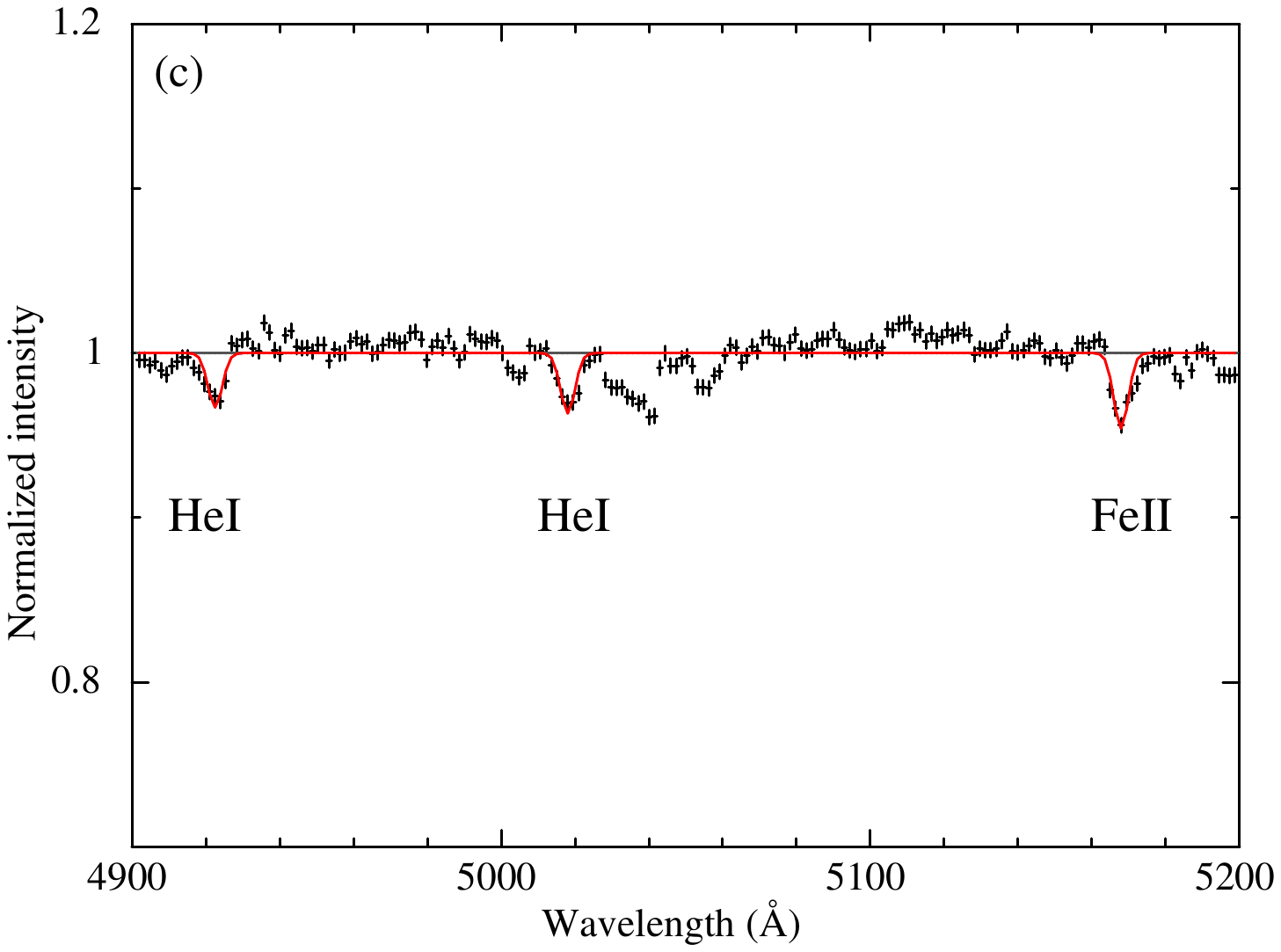}\\
    \includegraphics[clip,trim=1cm 0.4cm 4.5cm 2.5cm,width=0.32\textwidth]{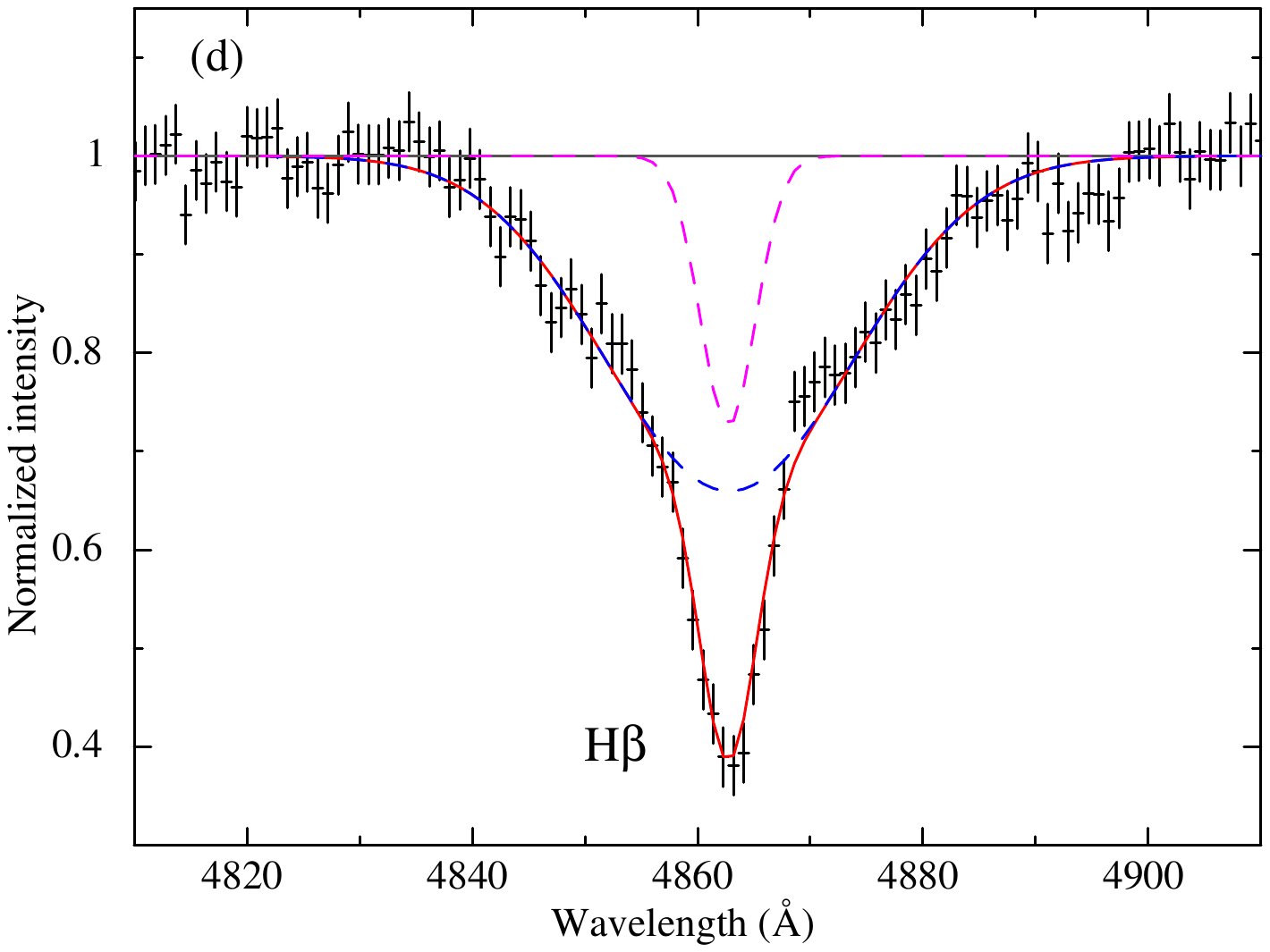}   
    \includegraphics[clip,trim=1cm 0.4cm 4.5cm 2.5cm,width=0.32\textwidth]{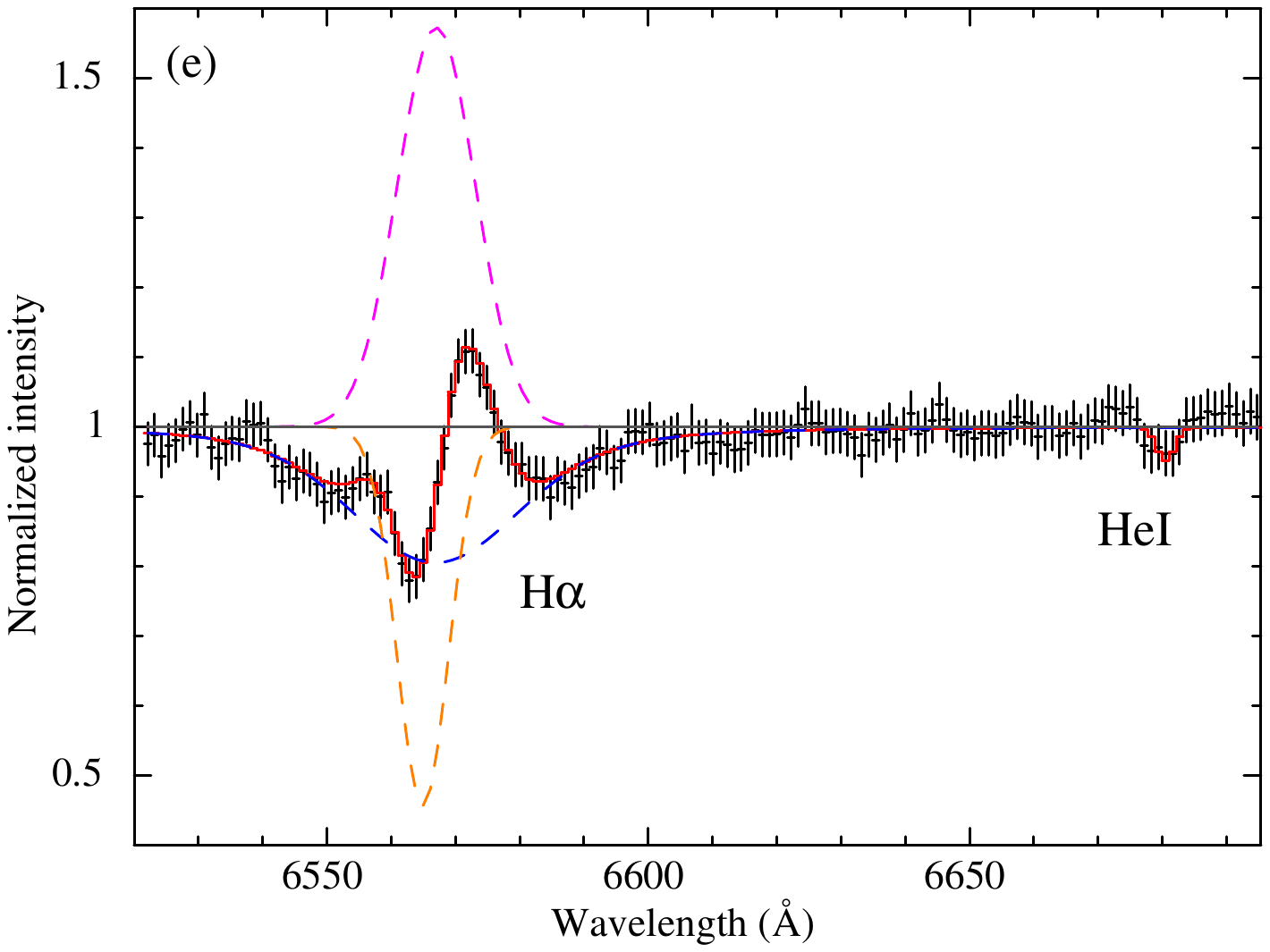}
    \includegraphics[clip,trim=1cm 0.4cm 4.5cm 2.5cm,width=0.32\textwidth]{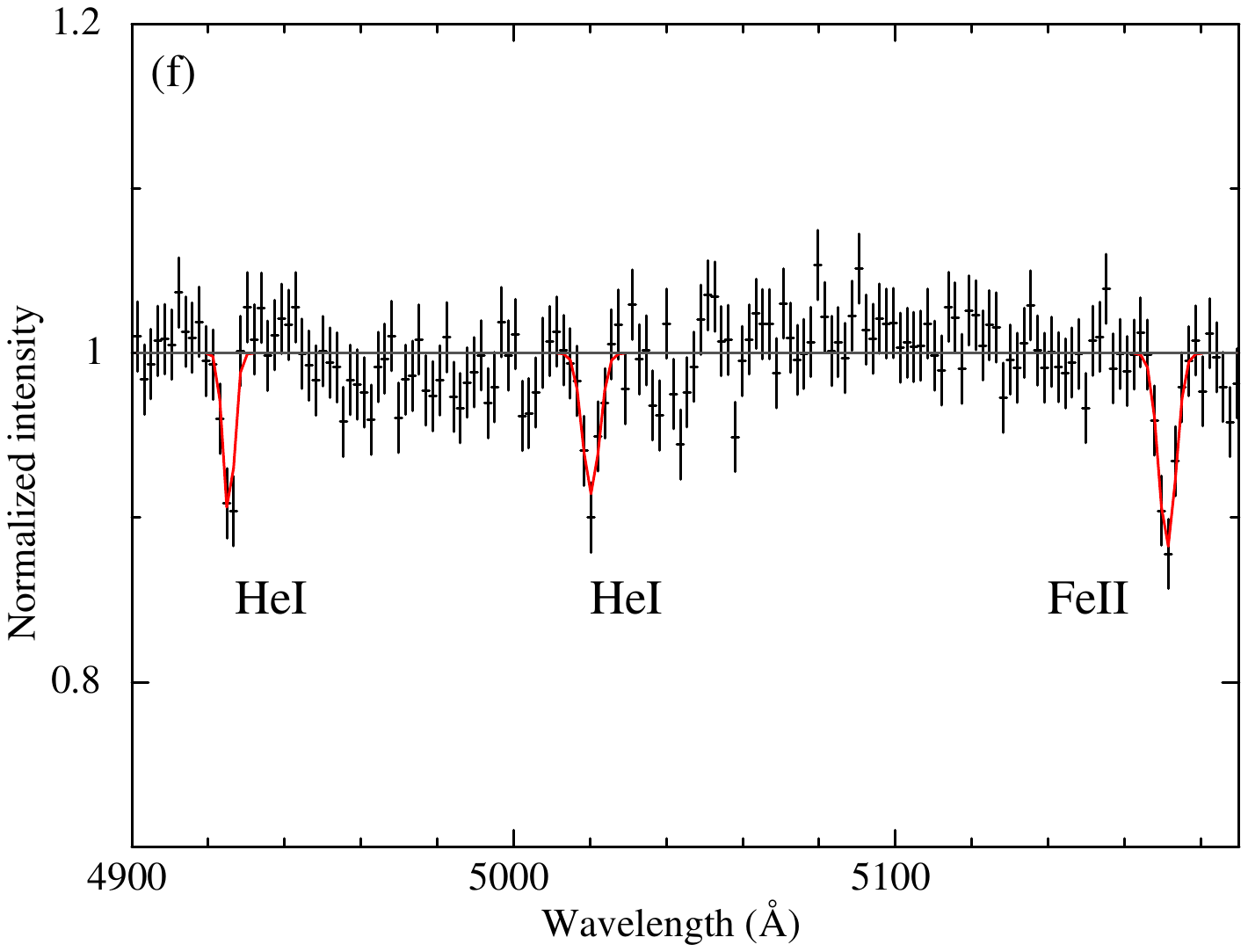}    
    \vspace{-1em}
    \caption{The most relevant spectral lines in the donor spectrum \textbf{(top)} and in the Seimei spectrum \textbf{(bottom)} and empirical best-fit models. In the Seimei spectra, the contributions of the individual components are indicated by dashed lines.
    } 
    \label{fig:seimei_fit}
\end{figure*}

We now focus on the optical Seimei spectra simultaneous to the first half of the XRISM observation, and whose data reduction procedure is detailed in Sec.~\ref{sub:optobs}.
Figure~\ref{fig:seimei} shows the overall spectra of the two grisms. The H Balmer lines (H$\alpha$ and H$\beta$), He lines, and \Feii{} line are detected in absorption, as normally seen from the intermediate mass companion star (B9III: \citealt{munoz-darias18}), but the H$\alpha$ profile shows an additional emission component. 
Although we can derive widths and velocities of the order of $\sim100$ km s$^{-1}$,  
the limited spectral resolution of the Seimei telescope is insufficient to properly characterize these spectral components, and disentangle quantitatively the contribution of the donor star. However, we still can compare the BH and donor spectrum qualitatively. For this, we perform empirical fitting of the components, both for our spectrum and the averaged donor spectrum obtained by \cite{munoz-darias18} on 1999 October during quiescence, in which the contribution of the accretion disk and the disk wind is expected to be negligible. We show the resulting comparison for the main lines of interest in Fig.~\ref{fig:seimei_fit}.
Although the H$\beta$ line profiles are broadly consistent between the two spectra, the H$\alpha$ line profiles are different, with a clear additional emission component in the BH spectrum. In addition, several additional lines, notably from HeI and \Feii{}, are seen in the Seimei spectrum in panel (c), and may include a component instrinsic to the BH, but the majority are also seen in the donor spectrum in panel (f), albeit with weaker normalization. The only exception is a very weak He I-6678 seen in panel (e), with a negligible velocity shift, and no counterpart in the donor spectrum.
The different H$\alpha$ profile could be due to an accretion disk atmosphere, or a disk wind with marginal velocities of few $\sim100$ km s$^{-1}$, to match the measurements considering the systematics of the Seimei observation. 
In addition, the Seimei observations were performed as ten 10-minute segments for VPH 683, and 3 5-minute segments for VPH 495, both covering the first half of the XRISM observation. In light of the highly variable line features seen in X-ray, and considering the fast variability seen in recent OIR spectroscopy results \citep{Miceli2024}, we also extracted the spectra of both grisms for each individual snapshot. Although we detect some variability in the profile of the H$\alpha$ component, the limited SNR and strong evolution in atmospheric transmission along the observation prevent any interpretation.

\section{Discussion} \label{sec:disc}

\subsection{continuum and global spectral evolution}\label{sub:continuum_evol}

\begin{figure}[t!]
    \centering
    \includegraphics[width=0.5\textwidth]{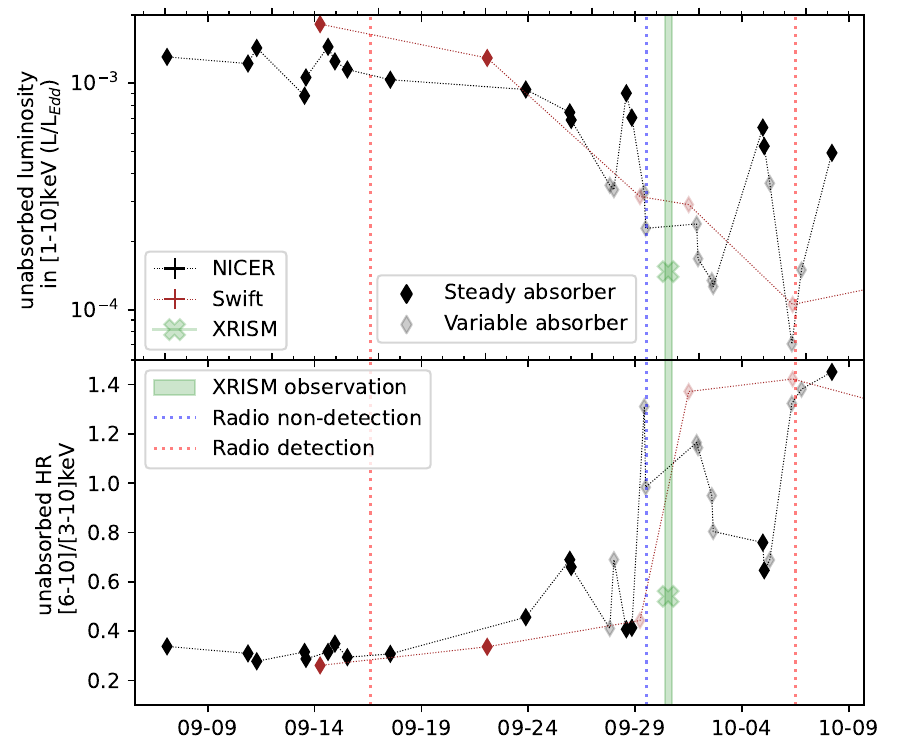}
    \vspace{-2em}
    \caption{Zoom on the possible state transition during the decay of V4641 Sgr, with the two panels as in Fig.~\ref{fig:lc_monit}. Fully colored markers highlight the standard evolution obscured by a steady absorber, and transparent markers highlight the erratic and typically harder flux dipping periods interpreted as the sign of a second, variable absorber. We note that the 6-10/3-6 keV hardness ratio does not fully encompass the spectral evolution of the dipping observations.}
    \label{fig:lc_monit_transi}
\end{figure}

Although the intrinsic luminosity and SED of the source are very difficult to constrain directly, we can still draw parallels to standard outburst patterns. The most obvious similarity is the apparent state change during the second part of the outburst: as seen in the lowest panels of Fig.~\ref{fig:lc_monit}, and according to the spectra analyzed in Sect.~\ref{sub:x-ray_monit}, this long-term evolution is reminiscent of a standard soft-hard transition at $\sim10^{-2}$ Eddington (see, e.g., \citealt{VahdatMotlagh2019_soft-hard_transition_lum,Wang2023_soft-hard_transition_lum}), obscured by a factor 10 by a steady, fully ionized absorber. Meanwhile, a second, highly variable layer would be responsible for the erratic dips and HR evolution during the transition, be it due to outflows or an inhomogeneous geometry.  We highlight this scenario in Fig.~\ref{fig:lc_monit_transi}, with the proposed distinction between the steady evolution (one absorber) and the more variable dips signaling an additional absorber. These dipping periods are typically harder, hinting at a partially ionized layer that would have a stronger absorption at low energies, and compatible with the NICER and Swift spectra at first order. The evolution of this layer could be what is seen in our time-resolved analysis in Sect.~\ref{sub:time_resolved}, and correspond to the main absorbing layer seen with the XRISM emission lines. The evolution of the lines as seen in NICER, and notably the lack of lines in most of the non-dip spectra, could be then directly due to that of the partially ionized layer, but a quantitative description is limited by the weak NICER SNR at these flux levels and requires a separate analysis.

However, this scenario remains insufficient to describe the entire outburst evolution. First, assuming that the steady absorber also dominated the long-term behavior of the source for the first weeks of the outburst, a standard outburst pattern would also include a noticeably brighter period (typically by one order of magnitude), matching the bright hard state, hard-to-soft transition, and bright soft state. However, the soft state plateau at $10^{-3} L_{Edd}$ (seen at the beginning of September and preceding the decay) seems to be the peak of the outburst luminosity, as seen in the EP and MAXI monitoring light curves with a longer coverage. 
This is also compatible with the soft state spectra seen across several weeks in the previous outburst \citep{Shaw2022_SAXJ1819_wind_emission_2020}, although they showed significantly lower disk temperature.
Secondly, the multi-wavelength behavior during the state transition hints at a much more complex evolution than a simple varying absorber. While the first (very weak, 0.17 mJy) radio detection, and subsequent non detection ($<$87$ \mu$Jy) at the start of the transition \citep{grollimundMeerKATSwiftConfirm2024}, are compatible with a low-luminosity soft state, the detection matching the very end of the transition \citep{grollimundMeerKATDetectionRadio2024} is at odds with our scenario. First, its low spectral index ($\alpha=-0.4$, Source: N. Grollimund, priv. communication) is more compatible with a transient ejecta than with a compact jet. Secondly, that detection (15.5 mJy) is not only an impressive 1000 times brighter than the weak soft state detection, but its absolute value itself remains uncharacteristically bright relative to the unabsorbed X-ray luminosity for a soft-to-hard transition.

Indeed, while it is difficult to draw conclusions on radio detections at such a low $\alpha$, the radio luminosity during soft-to-hard transitions remains comparable to or lower than the previous soft states (see e.g. Fig. 3 in \citealt{Hughes2025} for a recent study with extensive radio outburst coverage). We can thus use this detection to derive an approximate lower limit of the X-ray Eddington ratio during the soft state preceding the transition. 

Our radio detection translates to $L_{R-5GHz}=3.4\times 10^{30}$ erg s$^{-1}$ for a a distance of 6.2 kpc. Using the radio X-ray correlation (see \citealt{hannikainenMOSTRadioMonitoring1998b,corbelRadioXrayCorrelation2003,Gallo2003_radio_x_correl,corbelUniversalRadioXray2013a} and \citealt{hughesPeculiarHardState2025} for a more recent overview), even a radio bright source would require $L_X\gtrsim5\cdot10^{37}$ erg/s, or $L_{X}\gtrsim6\times10^{-2} L_{Edd}$, which would be at least an order of magnitude above our assumed Eddington ratio during the same period. If this extrapolation is valid, this could mean that the source is more obscured than expected, and that the state transition started at $L_{X}\gtrsim 10^{-1} L_{Edd}$ instead of the expected $L_{X}\sim10^{-2}L_{Edd}$. However, the previous radio measurements would nonetheless remain puzzling. The other, "simpler" possibility is that this event signals a more complicated accretion-ejection reconfiguration, that doesn't fit the standard radio outburst evolution, nor the Radio X-ray correlation. This could explain the very important X-ray luminosity and hardness-ratio evolution just before and after that event, if for instance this event is linked to an ejection of some of the obscuring layer.

 Finally, we highlight that the neutral column density we derived in our analysis is significantly lower than that of   \cite{Shaw2022_SAXJ1819_wind_emission_2020}, who fixed their interstellar column density to 2.5$\times 10^{21}$ cm$^{-2}$, following a previous Swift-NuSTAR fitting \citep{pahari15}. Our own analysis of Swift and NICER spectra of this outburst (see Sect.\ref{sub:x-ray_monit}) required an additional absorber of up to 3$\times 10^{21}$ cm$^{-2}$ to match the continuum below $\sim2$ keV, most notably at higher luminosities. We stress that without a good understanding of the intrinsic absorber, all values of NH remains highly degenerate, and the estimate in our analysis is still likely to be higher than the ISM absorption, which, in the direction of the source, only amounts to $\sim1.6\times 10^{21}$ cm$^{-2}$ when integrated over the entire galaxy \citep{hi4pi2016_NHsurvey}, and should thus be significantly lower for a source only at 6.4 kpc.

\subsection{physical interpretation of the X-ray line emitters}

In the X-ray spectrum, several narrow emission lines were successfully detected, despite the short 
exposure of $\sim$ 10 ks and the low source intensity of $\sim 1$ mCrab. 
All of these, except for the \Fei{} K$\alpha$ and the secondary blueshifted \Fexxvi{} K$\alpha$ component, are well modeled by emission from a single layer of photoionized material above the disc.
This model and the ion-by-ion line fitting show significant blueshift, meaning that this X-ray line emission is likely to come from a disk wind.
We can estimate the radius of the material by assuming that the velocity width comes from the Keplerian velocity around the launching radius, since the high-density, line-emitting gas is expected to be close to the disk surface.
This gives that $R/R_{g} =0.5 (c\sin i/v_{\rm rms})^2=(8\pm4)\times 10^5 $, where $i= 72\pm4^{\circ}$ is the inclination angle of this source \citep{MacDonald2014}.
The outer disk radius should be at $ 70\%$ of the Roche-lobe radius, giving $R_{\rm out}/R_{g} = 6\times 10^5 $. 
Thus the estimated radius is at the outer edge of the disk.

The fit parameters also enable a tentative estimate of the intrinsic luminosity $L_0$. The wind material is directly irradiated, so its ionization parameter is set by $\xi=L_0/(nR^2)$. With the column density as $N_H=n\Delta R\approx nR$, we derive 
$L_0\sim \xi N_{H} R = 3.5_{-3}^{+13}\times 10^{39}~{\rm erg~s^{-1}}$ .
This is about $10^{4}$ times higher than observed, and implies that the source is likely to be intrinsically super-Eddington, at  $\sim4_{-3.5}^{+16} L_{\rm Edd}$.
This large luminosity is also consistent with the observed high temperature of the disk blackbody continuum, but it is quite surprising that the wind is so slow at such super-Eddington fluxes, and that the scattered fraction is so small. However, we note that this is similar to the wind parameters inferred in the neutron star super-Eddington source GX 13+1 (Xrism collaboration, accepted). 

As a qualitative comparison to a more physical wind launching model, we compared this observation to a single simulation of scattering from a thermal wind (blue in Fig.\ref{fig:pion_Fe}).
This simulation was tailored for the neutron star binary system GX 13+1, with a Compton temperature of $1.3\times 10^7$~K, corresponding to a {\sc diskbb} component of 2.3~keV \citep{Tomaru2023,Tomaru2020}. 
This is a little hotter but still comparable to that seen here. 
The intrinsic luminosity was set to $L/L_{Edd,NS}=0.5$, or $8.8\times10^{37}~{\rm erg~s^{-1}}$, and would translate to $L/L_{Edd,BH}=0.11$ for our source. As seen in Fig.\ref{fig:pion_Fe}, this model gives a fairly good overall description of the photoionized emission, showing that emission from a thermal-radiative disc wind can explain the main ionized scattering layer. However, this model does not include obscuration, and we would thus require an additional, fully ionized absorber to obscure the spectrum by a factor 1000, down to the observed luminosity of $L/L_{Edd,BH}=1.45\times10^{-4}$. With pure photoelectric absorption, we can estimate this would require an additional column density of $N_{H}\sim8.5\times 10^{24} \textrm{cm}^{-2}$.

In parallel, the origin of the secondary components remain puzzling. The redshifted \Fei{} K$\alpha$ line, whose velocity is barely compatible with the main component at a $3\sigma$ confidence level,  
can be produced where X-rays illuminate mostly neutral material.
Potential sites are the accretion disk itself, but the inner disk is highly ionized, and the outer disc should be covered by the highly ionized wind. 
The outer disc surface under the wind (below the thermal instability) could also produce lines,
but they would likely still be mildly ionized, and give an apparent blueshift rather than a redshift. Meanwhile, the tentative fast blueshifted component, being detected in \Fexxvi{} only and not in \Fexxv{}, points towards a very high $\xi$ and thus an entirely different emitting region, closer to the central region than the main obscuring layer, which is important for e.g. radiation pressure to accelerate it to $\sim-1200$ km s$^{-1}$.

Within the context of the recent discoveries of systematically clumpy outflows from highly accreting AGN, not only for Ultra-Fast-Outflows but also for Warm Absorbers (see, e.g., \citealt{XRISM2025_PDS456,Xu2025_PDS456_XRISM,Mehdipour2025_XRISM_NGC3783,Xiang2025_XRISM_NGC4151}), it can be tempting to adapt this scenario to V4641 Sgr. Although this could explain very well both the fast variability and diversity in absorbers, ionization parameters and velocity shifts, the mix of redshifted and blueshifted components remains puzzling. One possibility would be for the \Fei{} K$\alpha$ line to originate in an outflowing material in the backside of the accretion disk, while the frontside would be suppressed by one of the main absorbing layers. This can however not explain the switch to a purely redshifted \Fexxvi{} K$\alpha$ line in one of the orbits. Another possibility would be that clumps of material from a ``failed'' wind progressing towards the BH occasionally cross our line of sight. As different components may evolve in different timescales, and we lack the signal to noise to distinguish meaningful changes in the Fe K$\alpha$ line in single orbits, this emission could be created by a similar scenario. Future observations, especially during brighter states, will allow to distinguish transient in-falling absorbers from and a more steady backside outflow. 

Finally, a comparison to the results of the photoionized computation derived in \citep{Shaw2022_SAXJ1819_wind_emission_2020} is not straightforward, due to the differences in illuminating SEDs, photoionization codes, and methodology. It would be worthwhile to analyze all spectra using the same approach, and assess how much the main ionized layer at the origin of the emission lines evolved between the previous and current outburst. Moreover, the Chandra observations, with their higher luminosity, have sufficient SNR to perform time-resolved spectroscopy and search for the high diversity of line variability found in our observation, at least qualitatively, something which was not investigated in previous works.

\subsection{Contextualization of the optical emission}

Although quantitative measurements are limited by the high contamination by the companion star, the simultaneous optical data offers a complementary view, with 
an H$\alpha$ emission line with negligible redshift yet low enough ionization parameter to match the \Fei{} K$\alpha$ emission. This H$\alpha$ emission is likely formed in a dense, clumpy optical wind \citep{munoz-darias18}, but we note very significant differences compared to the more active spectra in the previous outbursts, with a lack of significant blueshifted P-Cygni profiles or blue wings in H$\alpha$ and H$\beta$, or significant He-5876 absorption. This is the aspect most at odds with a highly accreting and outflowing scenario: in all outbursts previously observed and studied in \cite{munoz-darias18}, the optical magnitudes drop extremely rapidly after the initial bright events, and strong wind signatures can be seen up to several weeks after the outburst peak. We do note however that the wind signatures become much weaker in some observations, with the H$\alpha$ emission line being the most clear remaining component, and could then be compatible with what is seen at lower SNR in the Seimei spectrum. In 2024, no increase of optical magnitudes above $\sim0.5$ was detected \citep{goranskijOpticalObservationsBlack2024}. However, we note that the optical monitoring was limited around the approximate peak of the X-ray outburst, with a notable observing gap between September 4 and September 14 (V. Goranskij, priv. communication).

Several possibilities can thus explain the lack of strong wing signatures in our optical spectrum. The most straightforward result is that the optical monitoring spuriously coincided with a less pronounced wind phase, but this remains at odds with the high diversity of absorbers seen in X-rays and their variability, which indicates actively evolving and outflowing phases. Another is that this outburst truly never reached the high Eddington ratios of $\gtrsim0.1 L_{Edd}$, which preceded the presence of the strong optical wind signatures in previous outbursts. However, many other BHXRBs have shown strong OIR wind signatures down to very low Eddington ratios (e.g. $\sim 10^{-4} L_{Edd}$ in \citealt{Sanchez-Sierras2020_MAXIJ1820+070_wind_emission_infrared_soft_hard}). Finally, the obscured soft state X-ray SED may signal an accretion configuration which, despite its expected very high intrinsic luminosity and peculiar geometry, suppresses optical wind signatures independently of the state of the outflow, similarly to what is seen in standard outbursting BH LMXBs soft states \citep{MataSanchez2022}, or even suppresses the ``cold'' outflows completely . This would strongly contrast with the spectra discussed in \cite{Munoz-Darias2018_SAXJ1819.3-2525_winds_hard_optical}, which were typically simultaneous to hard X-ray SEDs. 

Confirming that this recent (also seen in e.g \citealt{Shaw2022_SAXJ1819_wind_emission_2020}) ``soft'' obscured state coincides  with an invisible, more ionized, or completely absent ``cold'' wind phase will require a denser multi-wavelength monitoring during future outbursts. Still, as long as this source remains the only obscured BHXRB with both ``soft'' and ``hard'' X-spectral states, it will remain one of the best candidates to study the link between the illuminating X-ray SED, highly ionized absorbers, and the presence (or detection) of optical wind signatures. Although this link is already largely suspected, as the hot wind phase is detected in X-rays in typical soft states \citep{Ponti2012_ubhw,Parra2024_winds_global_BHLMXBs}, and the cold wind phase can be seen in infrared along the entire outburst \citep{Munoz-Darias2019_1820_hard}, we currently lack in depth studies modeling the effect of soft X-ray SEDs on the presence of OIR wind signatures. In parallel, a more exhaustive study of the correlation between X-ray SED and OIR wind signatures (which has yet to be done for this source) could reveal whether the configuration of 2024 is an outlier or the standard behavior for such atypical outbursts.

\section{Conclusion} \label{sec:conclu}

V4641 Sgr is a unique BHXRB, whose X-ray spectra show an obscured central engine and narrow emission lines (instead of absorption lines) from a disk wind. Although similar features were observed in V404 Cyg and GRS 1915+105, these sources showed both hard spectra and extreme flux variability when obscured. 
In contrast, V4641 Sgr shows 
a soft thermal spectrum and can exhibit a stable flux over several weeks.

In this study, we present the results of our high-resolution multi-wavelength monitoring campaign of V4641 Sgr during the 2024 outburst. Our 10 ks XRISM spectrum, taken during an obscured dip at 1 mCrab at the end of a transition to an obscured hard state, shows a soft, thermal-like spectrum dominated by a high temperature thermal disk component of $1.77 \pm 0.03$ keV (or $1.65_{-0.11}^{+0.12}$ keV with a comptonized disk). The Resolve and Xtend spectra show many strong, narrow emission lines, most of them with marginal blueshifts, which can be well represented by a single ionized ($\log \xi =$ 3.9) photoionization layer in emission with a very high column density ($N_\mathrm{H} = 50_{-30}^{+70} \times 10^{22}$ cm$^{-2}$) and a small but significant blueshift of $-237\pm73$ km/s after corrections. The very low Eddington fraction of the observation ($\sim1.5 \times 10^{-4}$) and the high disk temperature strongly hint at an additional, strongly ionized and thus largely unconstrained obscuring layer, that would absorb the spectrum by at least one order of magnitude. 

We detected signs of two additional layers in the spectrum: the narrow, redshifted Fe K$\alpha$ line ($732_{-549}^{+305}$ km s$^{-1}$) indicates an infalling clump with low ionization or scattering from an outflow in the back of the disk, and a marginally significant (99.2\% with MC simulations) strongly blueshifted secondary \Fexxvi{} K$\alpha$ emission component ($-1286_{-85}^{+92}$ km s$^{-1}$) may be caused by a weaker inner wind component at high ionization. Moreover, the time-resolved spectrum shows a rapidly changing continuum, which softens by a factor of two and becomes 50\% brighter in the 10 ks of observation. The latter is accompanied by significant changes in the line properties and the ratios of individual transitions between the individual orbits. Although the weak SNR limits the analysis for such small exposures, the most remarkable changes are a very significant switch from a slight blueshift for the \Fexxvi{} K$\alpha$ line to a strong redshift ($790_{-110}^{+100}$ km s$^{-1}$) in the last orbit, and from intercombination dominated \Fexxv{} K$\alpha$ complex in the first orbits to a resonance dominated line in the last.

Although the optical Seimei spectrum is strongly dominated by the companion star and associated absorption features, a clear additional emission component is evident in 
H$\alpha$. This single detection starkly contrasts with the very strong P-Cygni and blueshifted asymmetric emission/absorption line profiles often observed in this source following the brightest portion of previous outbursts \citep{munoz-darias18}.

While the picture provided by these datasets is still incomplete, the context of the available X-ray monitoring shows that this observation was made at the end of the transition from a faint, weakly variable ``soft'' state -which may have marked the peak of luminosity of the outburst-, to a pure, decaying hard state, which later dimmed down to near-quiescent levels. Although the entire outburst is likely obscured by at least an order of magnitude, if this transition is comparable to a soft-hard transition in normal outbursts, erratic variability during the transition and the radio measurements, which are three orders of magnitude stronger at the end of the transition (15 mJy) than in the soft state (0.17 mJy) strongly suggest a second, more variable absorbing layer that may have been partially ejected during the radio flare at the end of the transition.

Meanwhile, a simple extrapolation from the absorber parameters, as well as comparisons with physical, radiation-driven wind models tailored to the near-Eddington source GX 13$+$1 show that the obscuration of the main ionized line-emitting layer of this source is consistent with a near- or Super-Eddington intrinsic luminosity. However, this scenario would require an extreme obscuration of three to four orders of magnitude.

This observing campaign perfectly demonstrates the power of XRISM Target of Opportunity observations, even with low exposures. Thanks to the exquisite spectral resolution of Resolve, the instrument can reveal entirely new information about the X-ray features in accreting sources, investigate line variability on new timescales, and small observing campaigns can serve as a probe to assess the scientific potential of longer observations. In our case, realistic future observations of V4641 Sgr in a less extremely obscured state (e.g. 20 mCrab) and with a higher exposure time (100 ks) would allow an order of magnitude better constraints on all line complexes. Aside from a definitive answer on the existence of the fast wind component, and more precise measurements of the velocities of the redshifted K$\alpha$ line and the weakly blueshifted main component, new observations would certainly allow the identification of many additional transitions and secondary components that we have not discussed here due to their weak significance. In parallel, it would be possible to constrain the evolution of these strong lines and absorbers down to timescales of one minute, and along an entire orbital period, which would open new opportunities to understand the influence of the (high-mass) companion on the absorber dynamics. A higher SNR would also enable more advanced line diagnostics for several line complexes \citep{porquetXrayPhotoionizedPlasma2000}, and notably \Fexxv{} K$\alpha$ \citep{Shaw2022_SAXJ1819_wind_emission_2020}. However, these results should be computed carefully due to degeneracy between several effects, among which continuum pumping \citep{Chakraborty2021_CLOUDY_3}, resonant Auger Destruction \citep{Chakraborty2020_CLOUDY_1}, and resonant scattering,  which cannot be well disentangled due to the very marginal detection of higher-order complexes \citep{meweCHANDRALETGSXrayObservations2001}. We also note the direct effects of high optical depths on low-ionization lines such as \Nex{} \citep{Chakraborty2022_CLOUDY_4}, that could be quantifiable with a higher SNR, most notably with the Gate Valve in its open configuration.

For now, a more comprehensive view of the global evolution of this outburst, using the extensive simultaneous coverage of the INTEGRAL observatory, would provide much stronger constraints on the presence and evolution of the comptonized component, and the XMM-Newton/RGS observation performed earlier in the outburst could shed more light on the nature and evolution of the absorbers. Finally, a more holistic study of the NICER coverage of this source during recent outbursts and comparisons with the existing radio monitoring will be critical to relate its outburst evolution to that of standard sources and extrapolate its intrinsic Eddington ratio.\\

\begin{acknowledgements}
We thank the XRISM operation team for accepting our DDT proposal and conducting the observation. This research has made use of software provided by the High Energy Astrophysics Science Archive Research 
Center (HEASARC), which is a service of the Astrophysics Science Division at NASA/GSFC. The Seimei telescope at Okayama Observatory is jointly operated by Kyoto University and the National Astronomical Observatory of Japan (NAOJ), with assistance provided by the Optical and Infrared Synergetic Telescopes for Education and Research (OISTER) program. 
This study makes use of the Seimei data obtained through the program 24B-N-CT03 in the open use of the observing time provided by NAOJ. 
Part of this work was financially supported by Grants-in-Aid for Scientific Research 19K14762, 23K03459, 24H01812 (MS) from the Ministry of Education, Culture, Sports, Science and Technology (MEXT) of Japan. MP acknowledges support from the JSPS Postdoctoral Fellowship for Research in Japan, grant number P24712, as well as the JSPS Grants-in-Aid for Scientific Research-KAKENHI, grant number J24KF0244. 
T.M.-D. acknowledges support by the Spanish \textit{Agencia estatal de investigaci\'on} via PID2021-124879NB-I00.
EdF gratefully acknowledges support from the Inter-University Research Programme (Grant2025i–F–05) and the Joint Research Program (international) of the Institute for Cosmic Ray Research (ICRR) at the University of Tokyo. He also thanks Marco P\'erez from the Centro Universitario de Ciencias Exactas e Ingenier\'ias (CUCEI), Universidad de Guadalajara (UdeG), the Coordinacion de Personal and the corresponding administrative offices of the UdG, as well as the administrative and scientific staff of the ICRR for the financial and logistical support during his invited research stays at the ICRR in 2025.
M.A.P. acknowledges support through the Ramón y Cajal grant RYC2022-035388-I, funded by MCIU/AEI/10.13039/501100011033 and FSE+
YZ acknowledges support from the Dutch Research Council (NWO) Rubicon Fellowship, file no.\ 019.231EN.021.

Facilities: XRISM, Seimei: 3.8m, NICER, Swift, Einstein Probe, INTEGRAL, Meerkat. Software: XSPEC  \citep{arnaud96},  IRAF  \citep{tody86}.
\end{acknowledgements}

\bibliographystyle{aa}
\bibliography{v4641_ms,files_parra/biblio_compl,files_parra/ref,files_parra/biblio}

\begin{onecolumn}
\appendix

\section{Velocity corrections}\label{app:vel_corr}

The velocity correction must consider several distinct components: the velocity of the Earth within the solar system $\Delta v_\mathrm{e-ss}$, the velocity correction between the solar system and the binary system V4641 Sgr $\Delta v_\mathrm{ss-b}$, and the velocity of the BH within the binary system $\Delta v_\mathrm{b-BH}$.
Considering the SNR of the observation, we do not consider the velocity of the satellite around the Earth, which adds an uncertainty of $\pm 7.6$ km s$^{-1}$.

To derive $\Delta v_\mathrm{e-ss}$, we calculate the projected velocity due to the Earth's motion in the direction of V4641 Sgr directly using \texttt{SkyCoord}, part of the astropy package. For the date of the observation, 2024 September 30, we get $\Delta v_\mathrm{e-ss}=-29.7$ km s$^{-1}$.

$\Delta v_\mathrm{ss-b}$ can be estimated from the velocity of the Local Standard of Rest, and the expected velocity of V4641 Sgr, considering its position in the galaxy (see, e.g., \citealt{russeilMilkyWayRotation2017,sofueRotationMassMilky2017}). However, in our case, this value is already part of the the orbital parameters derived by \cite{orosz01} from the analysis of the optical light curves. They derive a radial velocity of $107.4\pm2.9$ km s$^{-1}$, which means that the correction to our measurement is $\Delta v_\mathrm{ss-b}=-107.4\pm2.9$ km s$^{-1}$. 

$\Delta v_\mathrm{b-BH}$ can be calculated from the orbital parameters, by weighting the rotational velocity of the binary system $v_\mathrm{rot}$ sin$i$ by the mass ratio of the system $Q$ and applied w.r.t. the orbital phase of the BH during the XRISM observation, which is obtained from the orbital period $P_\mathrm{orb}$ and $T_{0}$ of the system. We take $v_\mathrm{rot}$ sin$i=101\pm1$ and $Q=2.2 \pm 0.2$, from the latest dynamical computation of the orbital parameters in \cite{MacDonald2014}. Since there is some small variability on the orbital parameters in the literature, we use the values from the most recent optical monitoring, namely $P_\mathrm{orb}=2.81727$ d and $T_{0,\mathrm{photo}}=2459410.4136$ from \cite{goranskijOpticalObservationsBlack2024}, with $T_0$ calculated at the middle of the eclipse when the BH is closest to the observer. 
Halfway through the observation (September 30 at 12:11 UT), we derive a phase of $\varphi_\mathrm{obs} =$ 0.57, resulting in $\Delta v_\mathrm{b-BH} = 20\pm2$ km s$^{-1}$ for the time-averaged spectrum. The correction of the companion star would be $\Delta v_{\mathrm{b-}\star}=-44 \pm 4$ km s$^{-1}$.

The combination of these different corrections results in a total of $\Delta v_\mathrm{cor}=-117 \pm 13$ km s$^{-1}$ for the BH.
We note that the phase evolution within the observation is $\Delta\varphi=0.11$, which leads to an increase of the linewidths by  $\sigma_{\Delta\varphi}\sim28$ km s$^{-1}$. However, this value is negligible compared to the errors in the linewidths in our analysis. Also, in the time-resolved spectra, the orbital velocity of the BH varies by 28 km s${-1}$ between orbits 1 and 5, but this is also insignificant compared to the fitting uncertainties of the line velocities within the individual orbits.

\vspace{-1em}
\section{Resolve branching ratios}\label{app:resolve_branch}
\vspace{-1em}

\begin{figure*}[h!]
    \centering

    \includegraphics[width=\textwidth]{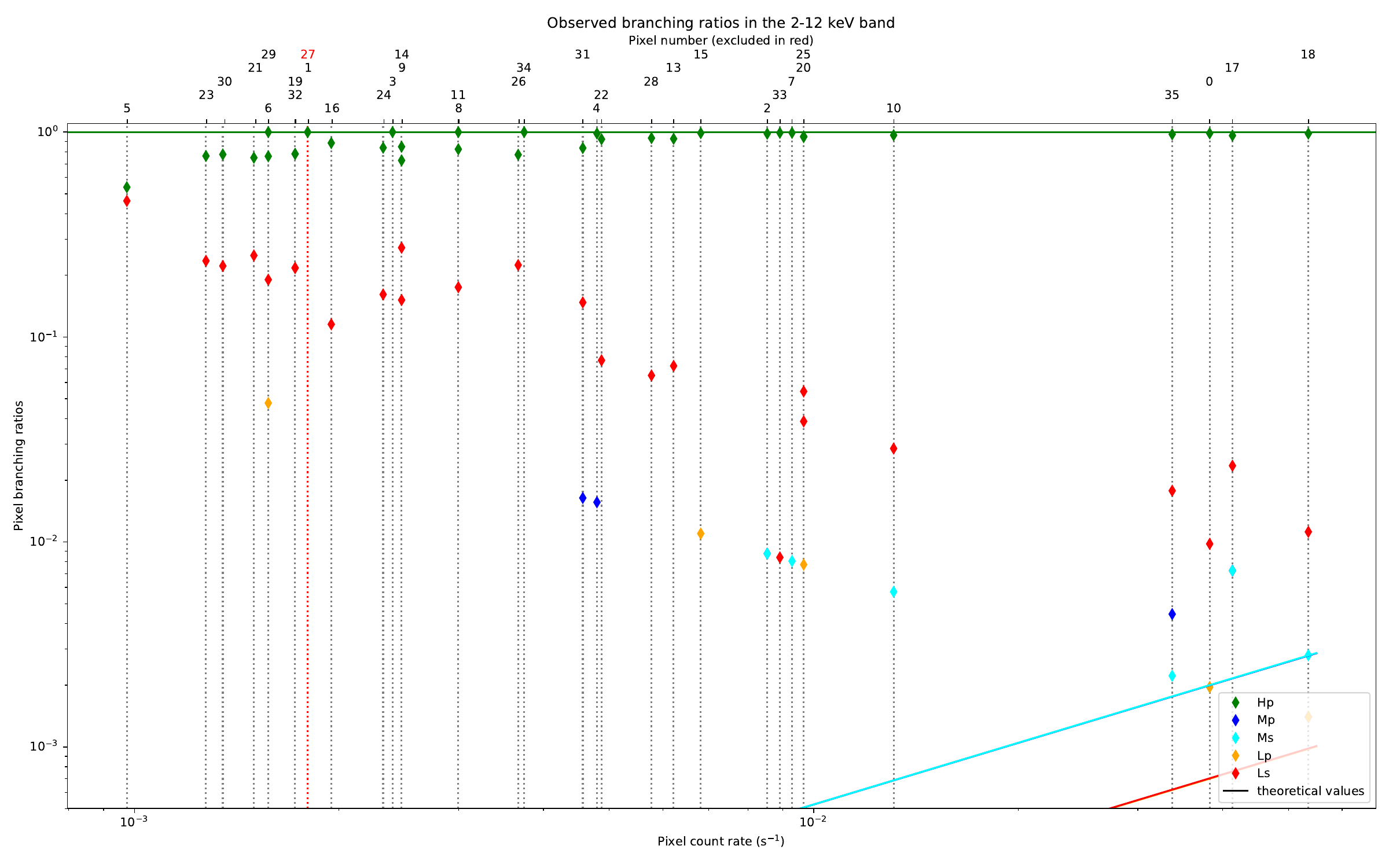}
    \caption{Branching ratios of all Resolve pixels in the 2--12 keV band, identified by count rate (bottom) and the identification number of each pixel (top), as in \href{https://heasarc.gsfc.nasa.gov/docs/xrism/proposals/POG/Resolve.html}{https://heasarc.gsfc.nasa.gov/docs/xrism/proposals/POG/Resolve.html}. The diamonds show the values and the lines the theoretical prediction. The theoretical values for Mp/Ms (blue and cyan) and Lp/Ls (orange and red) are identical in this range of count rates. We highlight pixel number 27, which was not included in our analysis, in red.} 
    \label{fig:resolve_branch}
\end{figure*}
\pagebreak

\section{Details of the blind searches and line characterization}\label{app:add_spectral}

\begin{figure*}[h!]
    \centering
\includegraphics[clip,trim=19cm 0.8cm 15cm 0.8cm,scale=0.315]{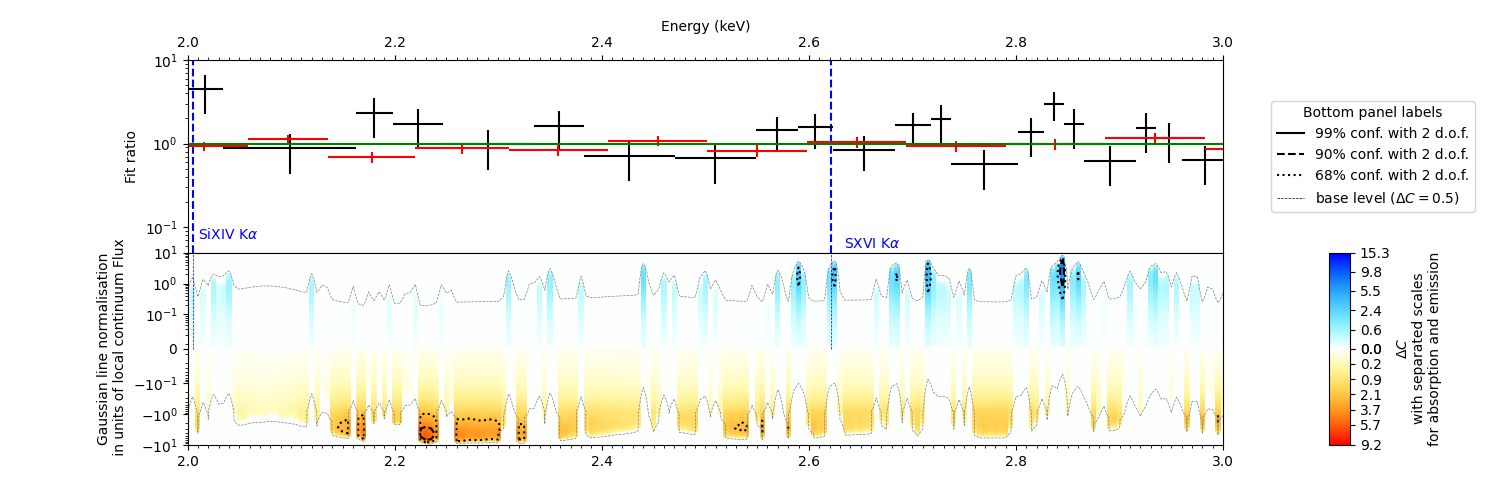}
\includegraphics[clip,trim=10.5cm 0.8cm 22.5cm 0.5cm,scale=0.315]{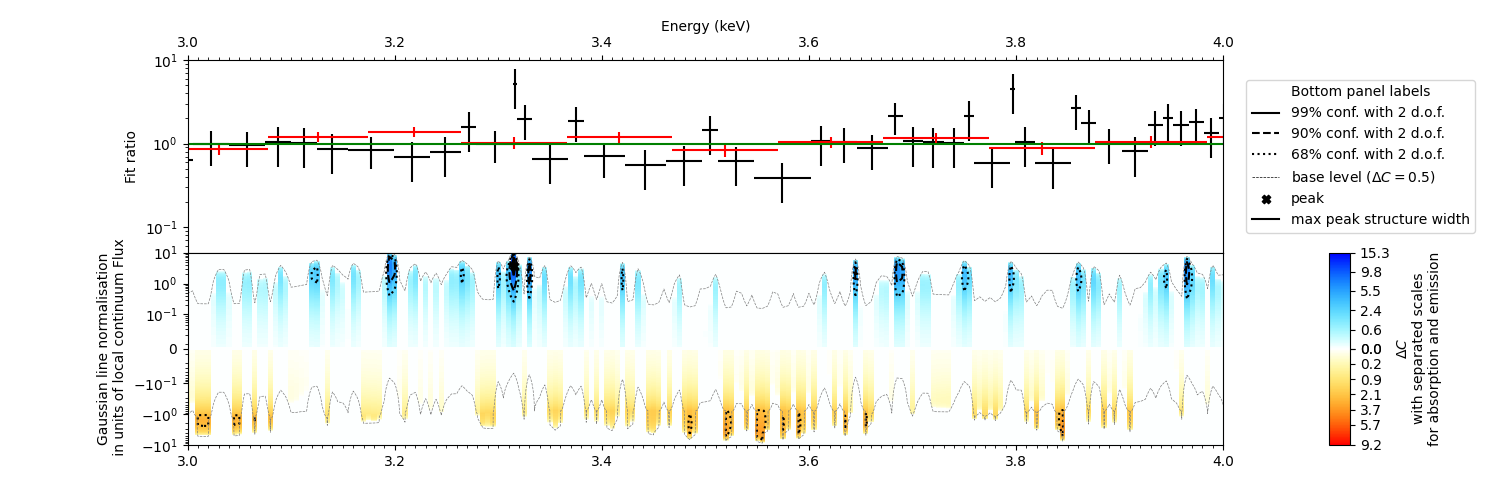}
\includegraphics[clip,trim=6.5cm 0.8cm 27.5cm 0.5cm,scale=0.315]{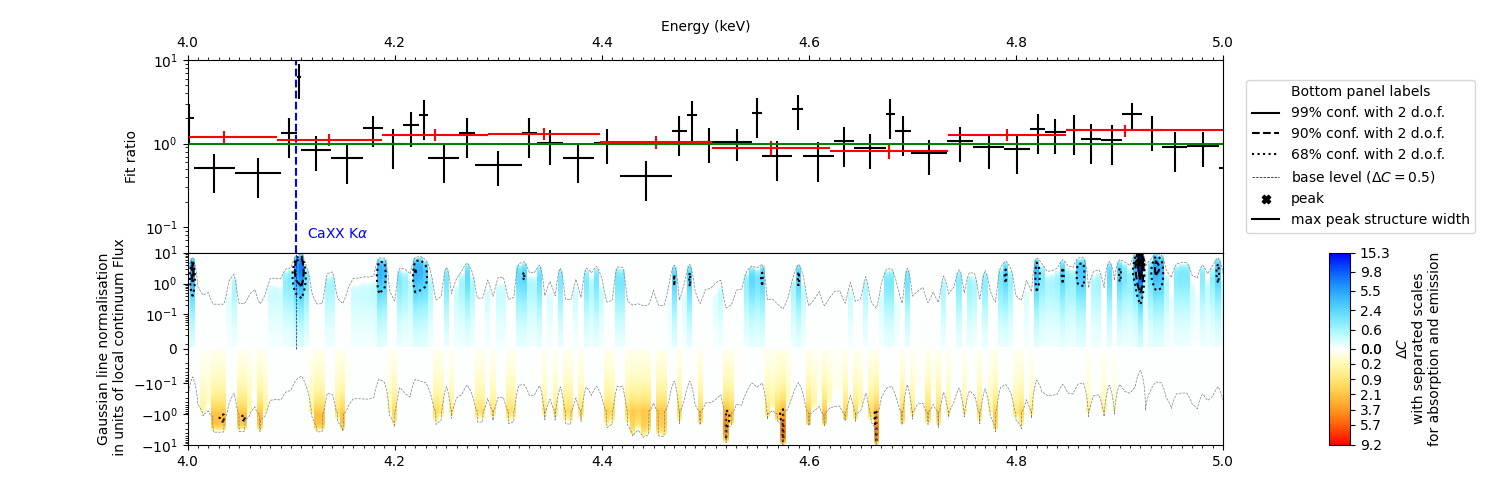}
\includegraphics[clip,trim=27.5cm 0.8cm 6.7cm 0.5cm,scale=0.315]{files_parra/blind_search_appendix/timeres/save_01_zoom_4.0_5.0.png}
\includegraphics[clip,trim=18.5cm 0.8cm 14.5cm 0.9cm,scale=0.315]{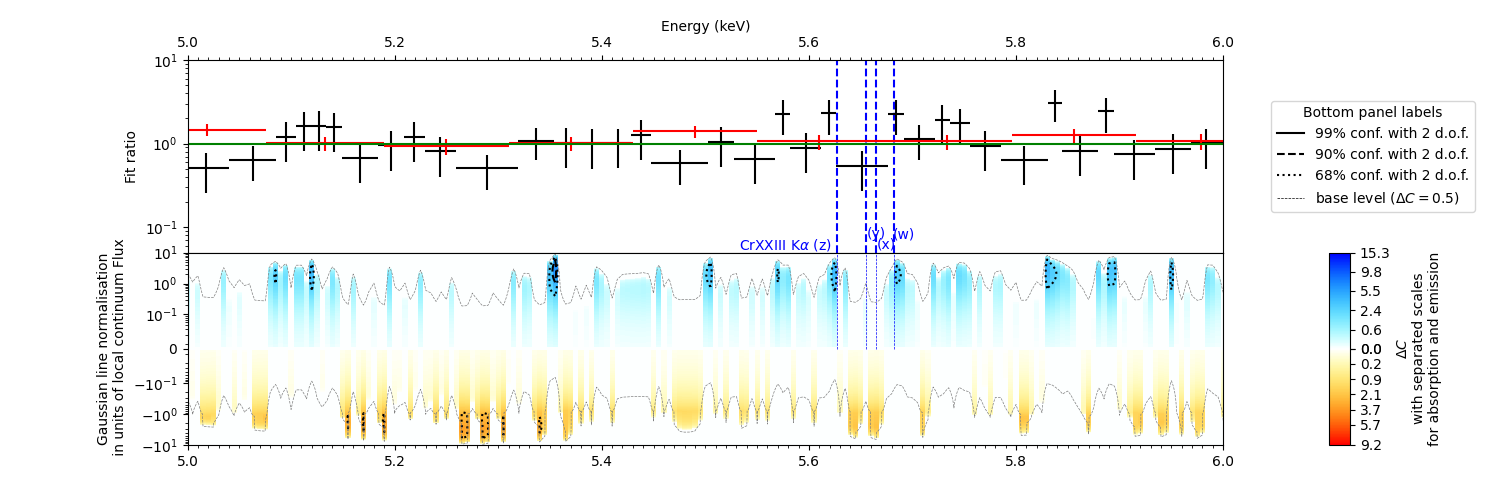}
\includegraphics[clip,trim=12.5cm 0.8cm 8.5cm 0.5cm,scale=0.315]{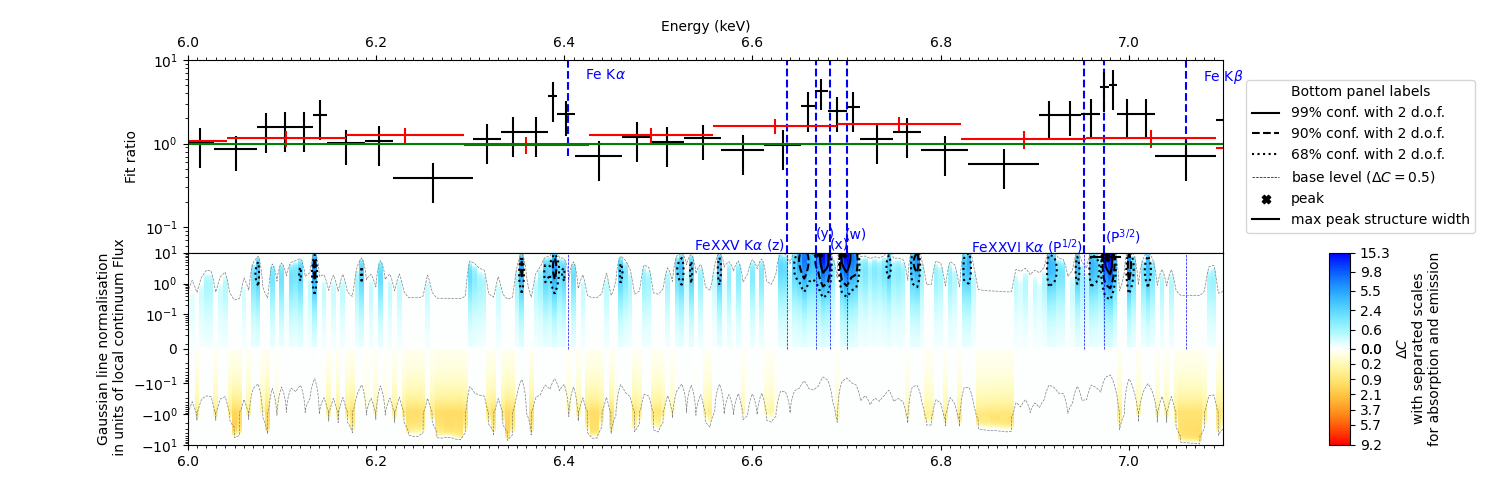}
\includegraphics[clip,trim=20.5cm 0.8cm 7.5cm 0.5cm,scale=0.315]{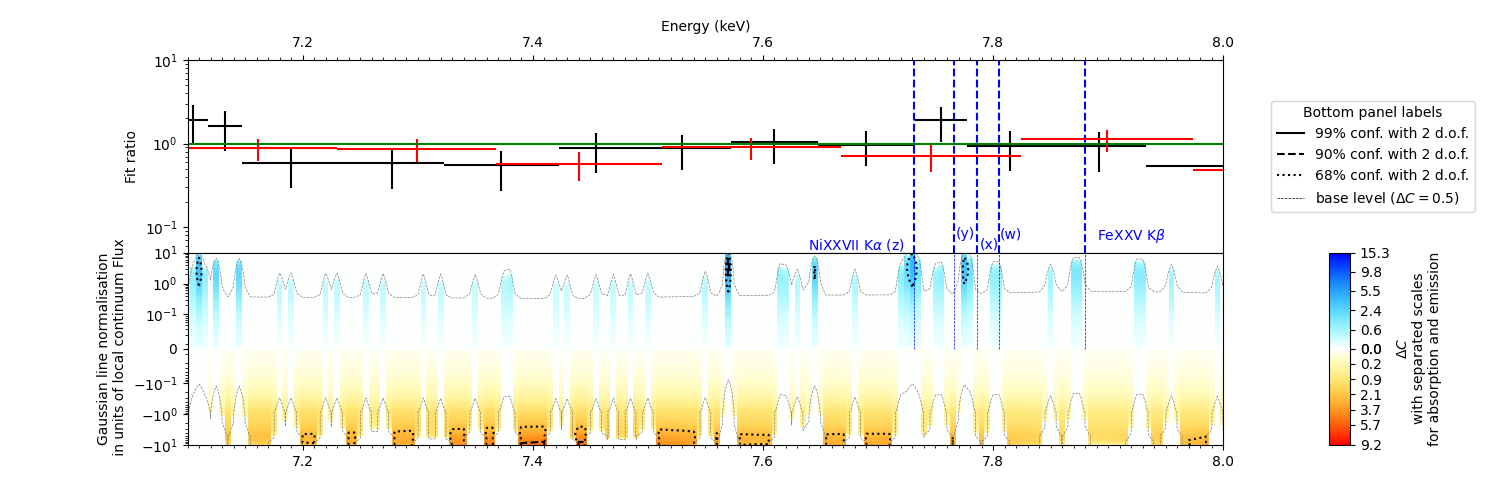}
\includegraphics[clip,trim=32.2cm 0.8cm 0.5cm 0.5cm,scale=0.315]{files_parra/blind_search_appendix/timeres/save_01_zoom_7.1_8.0.png}

\includegraphics[clip,trim=19cm 0.8cm 15cm 0.8cm,scale=0.315]{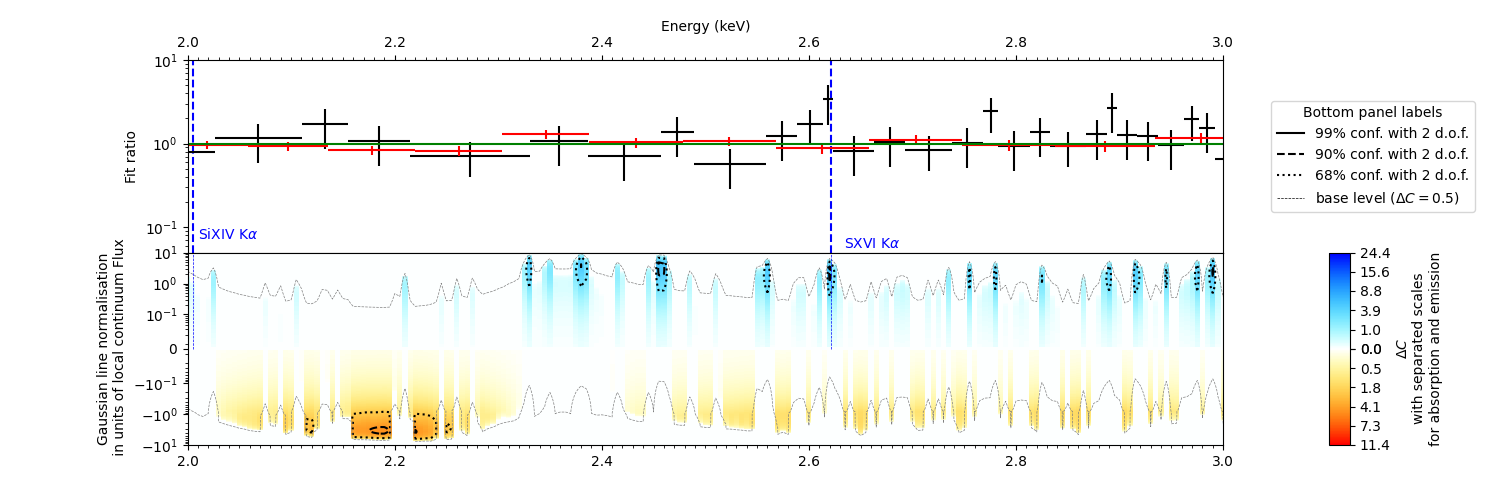}
\includegraphics[clip,trim=10.5cm 0.8cm 22.5cm 0.5cm,scale=0.315]{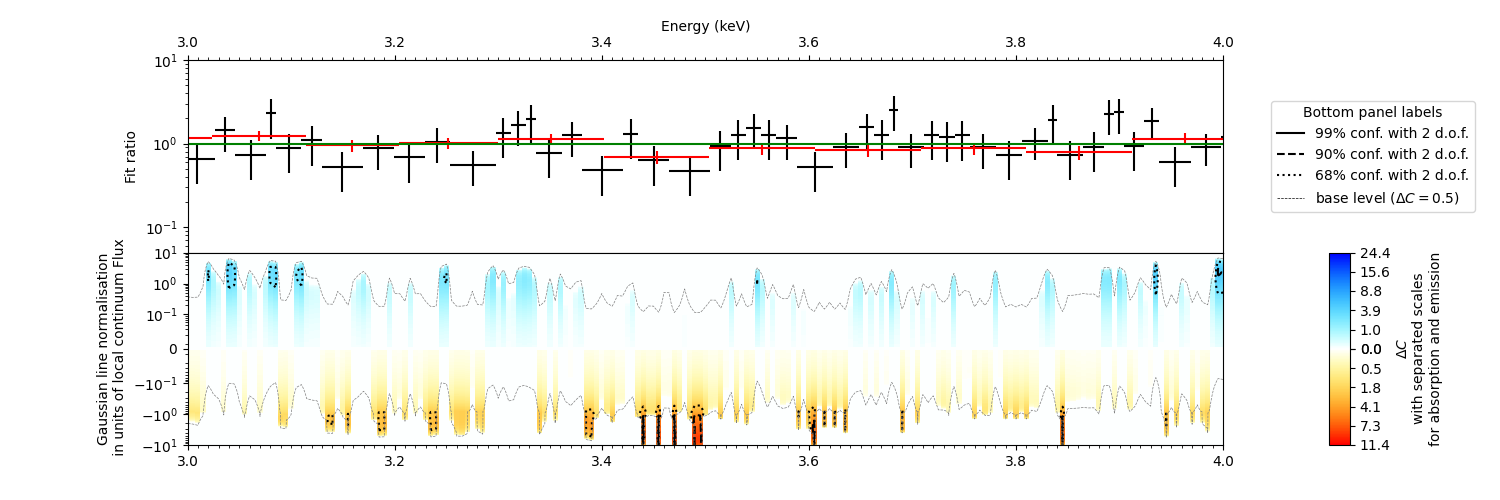}
\includegraphics[clip,trim=6.5cm 0.8cm 27.5cm 0.5cm,scale=0.315]{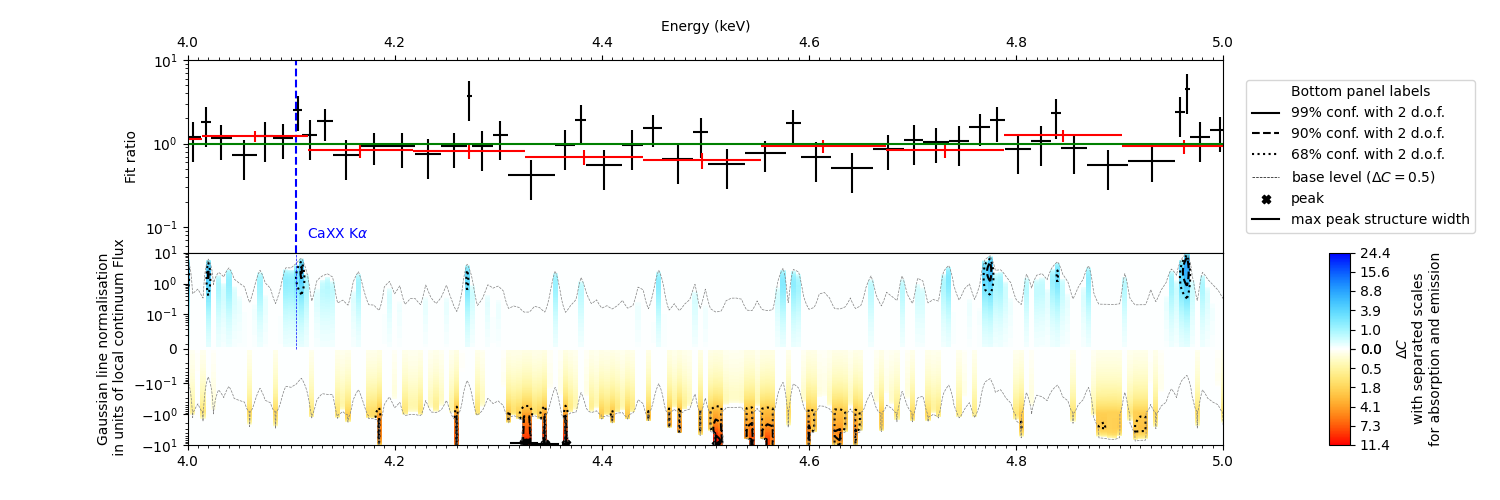}
\includegraphics[clip,trim=27.5cm 0.8cm 6.7cm 0.5cm,scale=0.315]{files_parra/blind_search_appendix/timeres/save_02_zoom_4.0_5.0.png}
\includegraphics[clip,trim=18.5cm 0.8cm 14.5cm 0.9cm,scale=0.315]{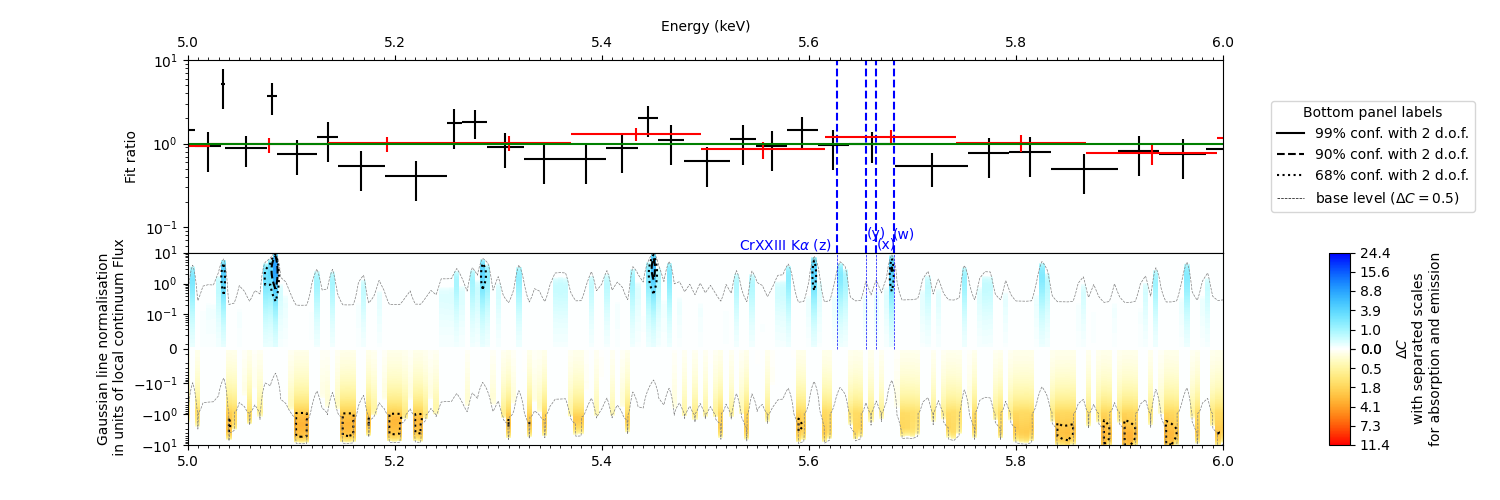}
\includegraphics[clip,trim=12.5cm 0.8cm 8.5cm 0.5cm,scale=0.315]{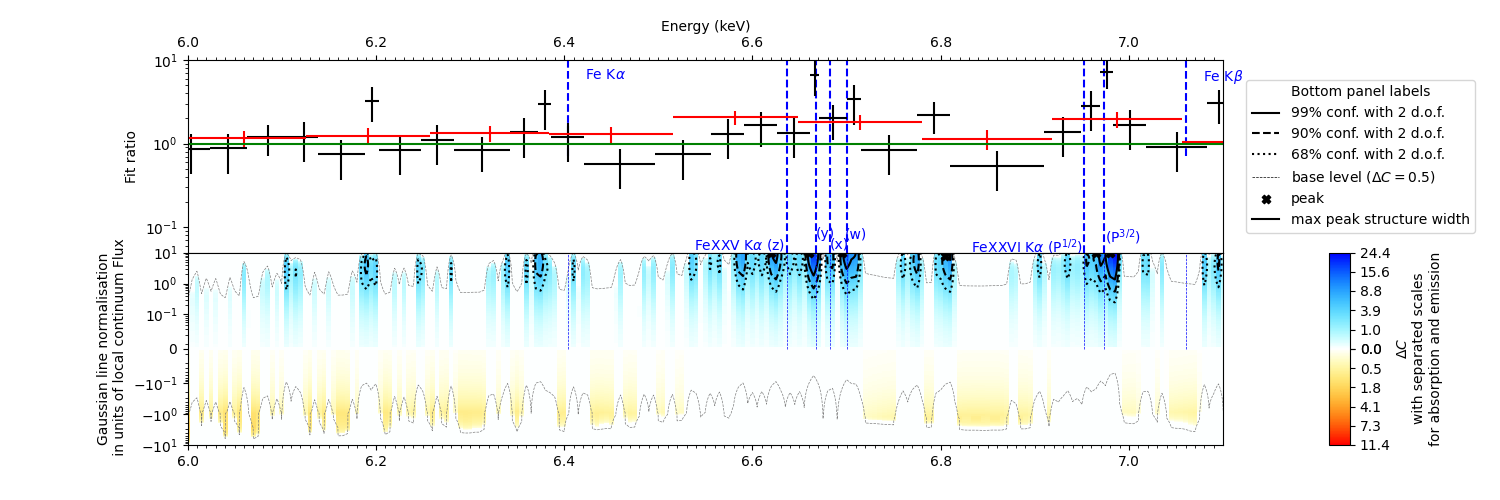}
\includegraphics[clip,trim=20.5cm 0.8cm 7.5cm 0.5cm,scale=0.315]{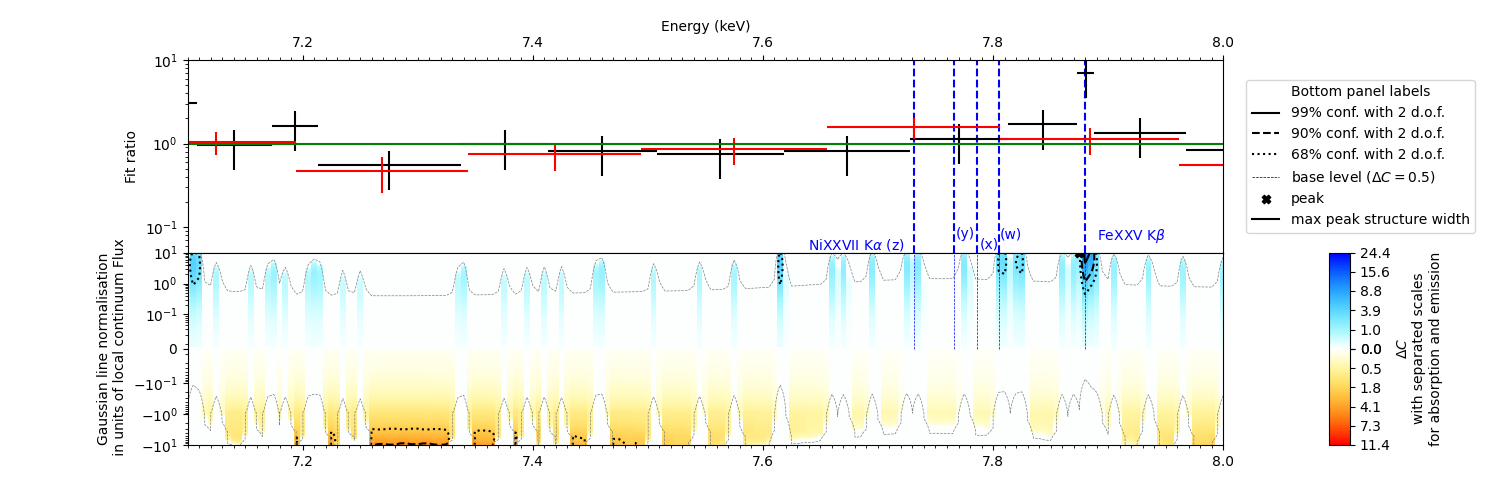}
\includegraphics[clip,trim=32.2cm 0.8cm 0.5cm 0.5cm,scale=0.315]{files_parra/blind_search_appendix/timeres/save_02_zoom_7.1_8.0.png}

\includegraphics[clip,trim=19cm 0.8cm 15cm 0.8cm,scale=0.315]{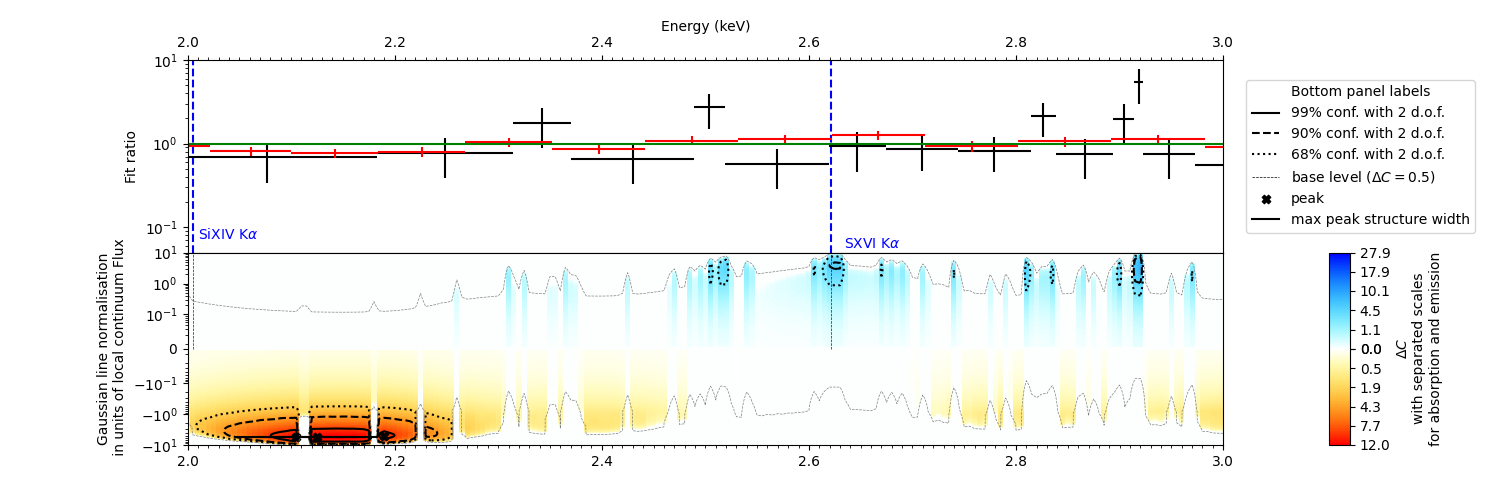}
\includegraphics[clip,trim=10.5cm 0.8cm 22.5cm 0.5cm,scale=0.315]{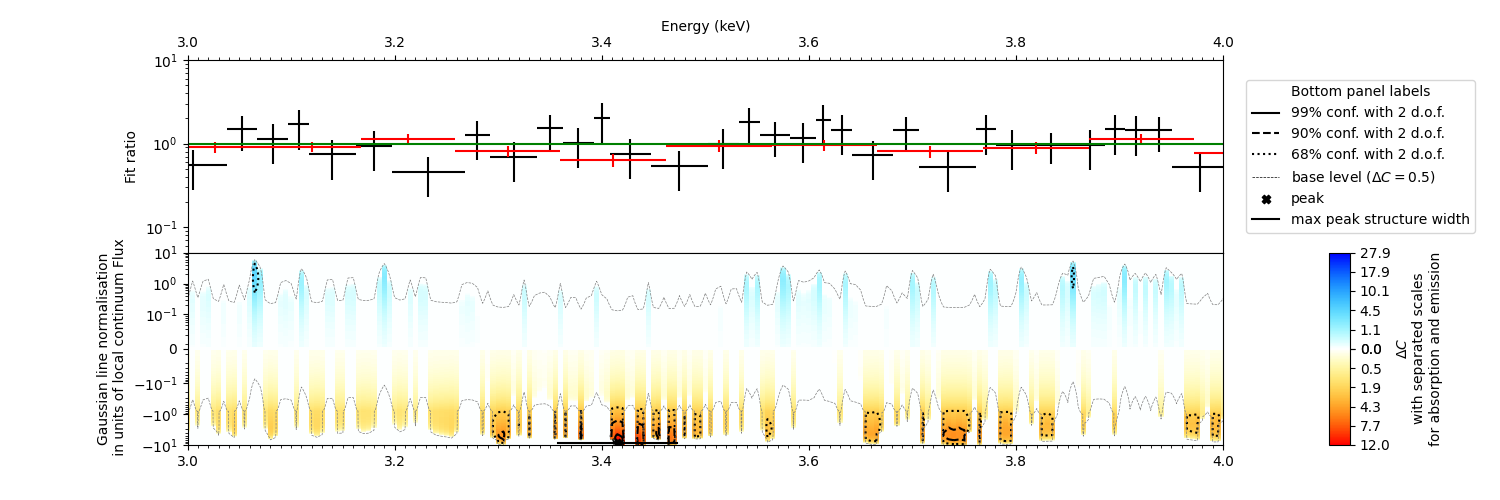}
\includegraphics[clip,trim=6.5cm 0.8cm 27.5cm 0.5cm,scale=0.315]{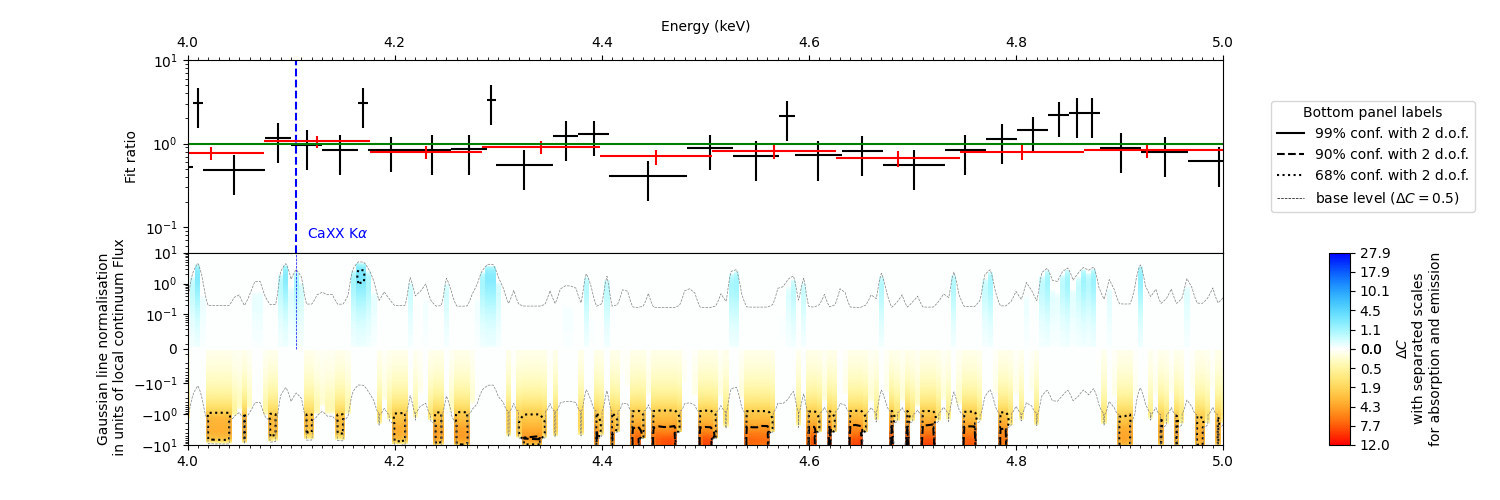}
\includegraphics[clip,trim=27.5cm 0.8cm 6.7cm 0.5cm,scale=0.315]{files_parra/blind_search_appendix/timeres/save_03_zoom_4.0_5.0.png}
\includegraphics[clip,trim=18.5cm 0.8cm 14.5cm 0.9cm,scale=0.315]{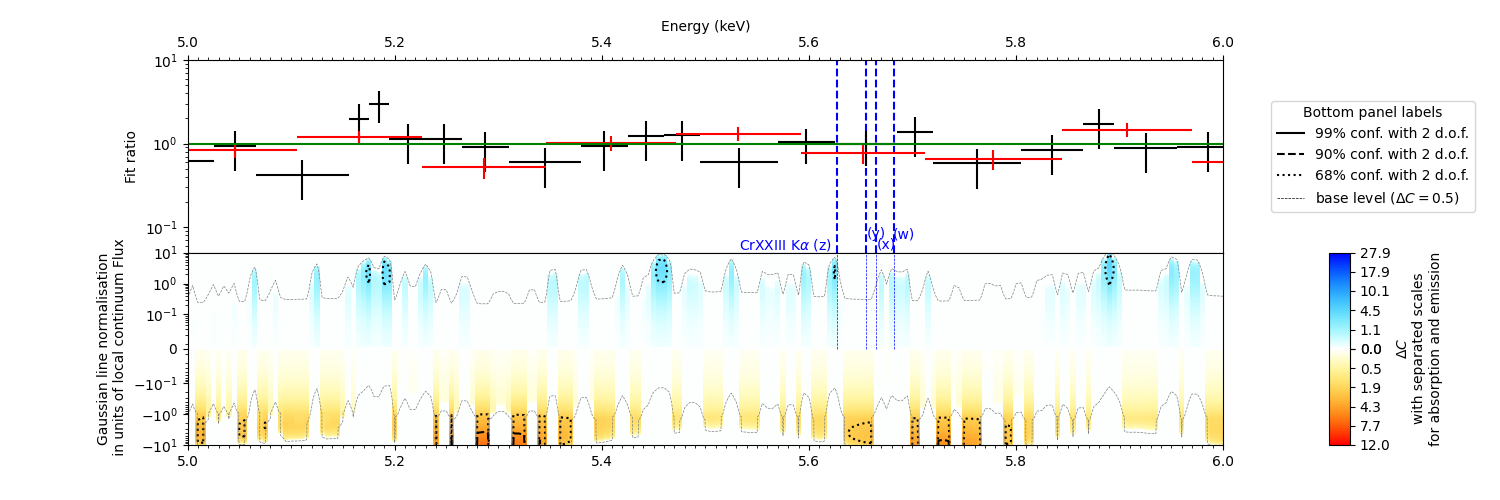}
\includegraphics[clip,trim=12.5cm 0.8cm 8.5cm 0.5cm,scale=0.315]{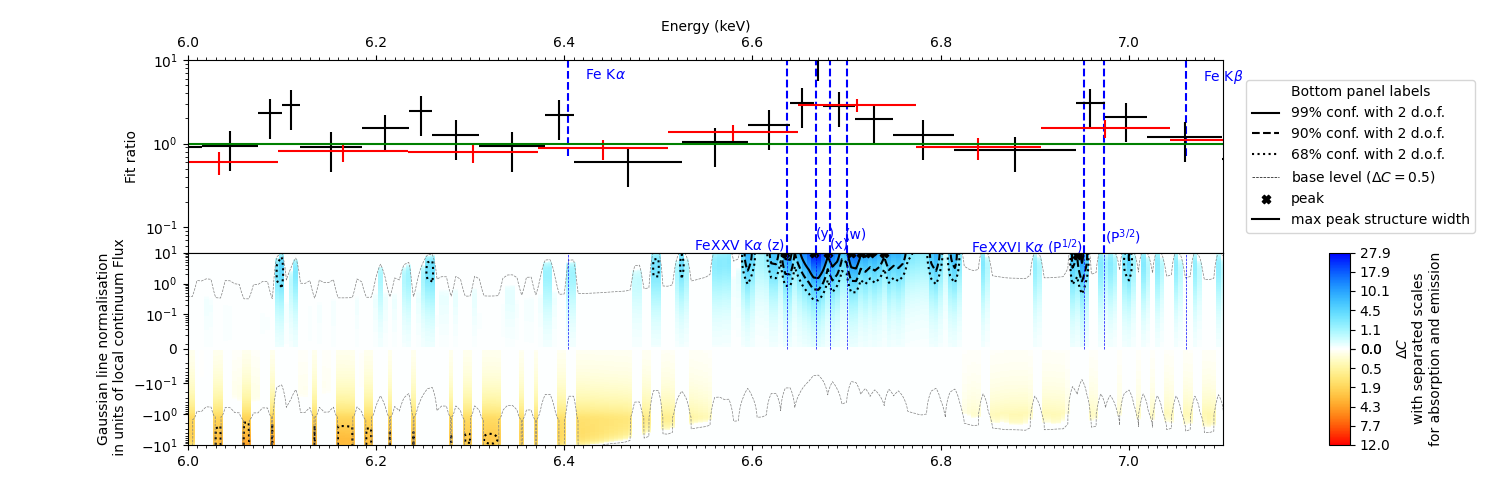}
\includegraphics[clip,trim=20.5cm 0.8cm 7.5cm 0.5cm,scale=0.315]{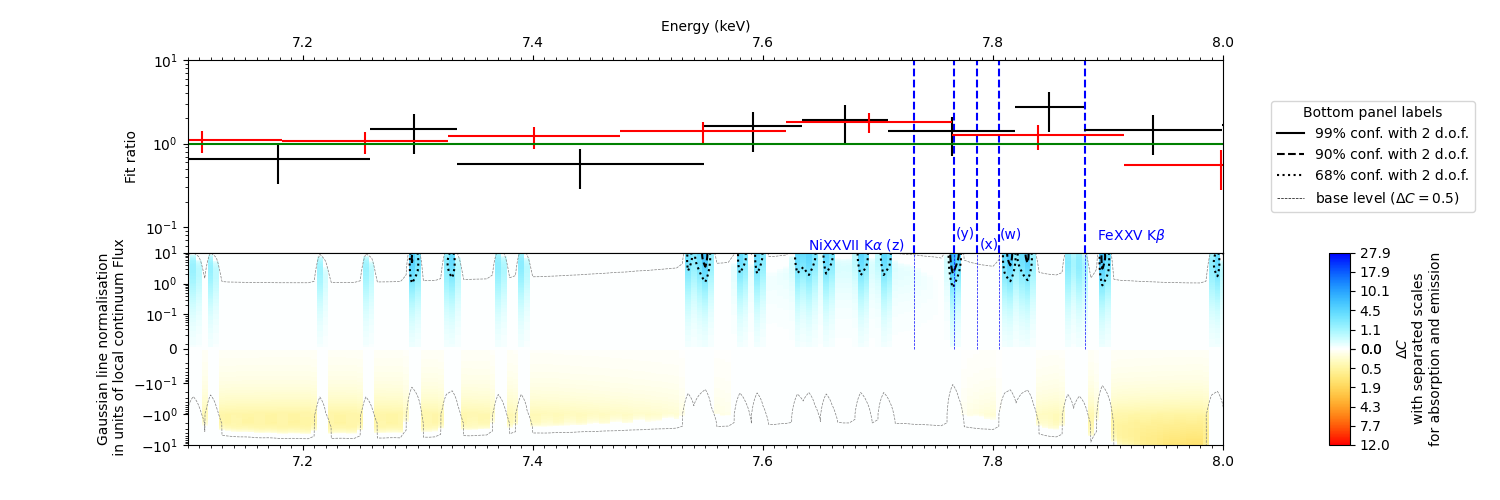}
\includegraphics[clip,trim=32.2cm 0.8cm 0.5cm 0.5cm,scale=0.315]{files_parra/blind_search_appendix/timeres/save_03_zoom_7.1_8.0.png}

\includegraphics[clip,trim=19cm 0.8cm 15cm 0.8cm,scale=0.315]{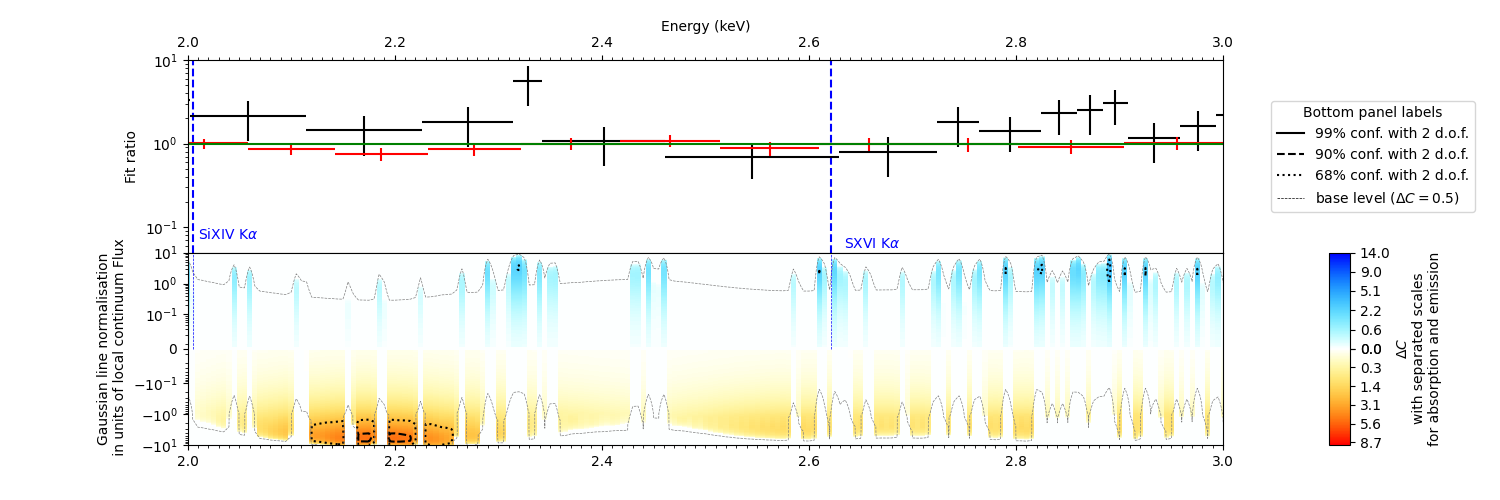}
\includegraphics[clip,trim=10.5cm 0.8cm 22.5cm 0.5cm,scale=0.315]{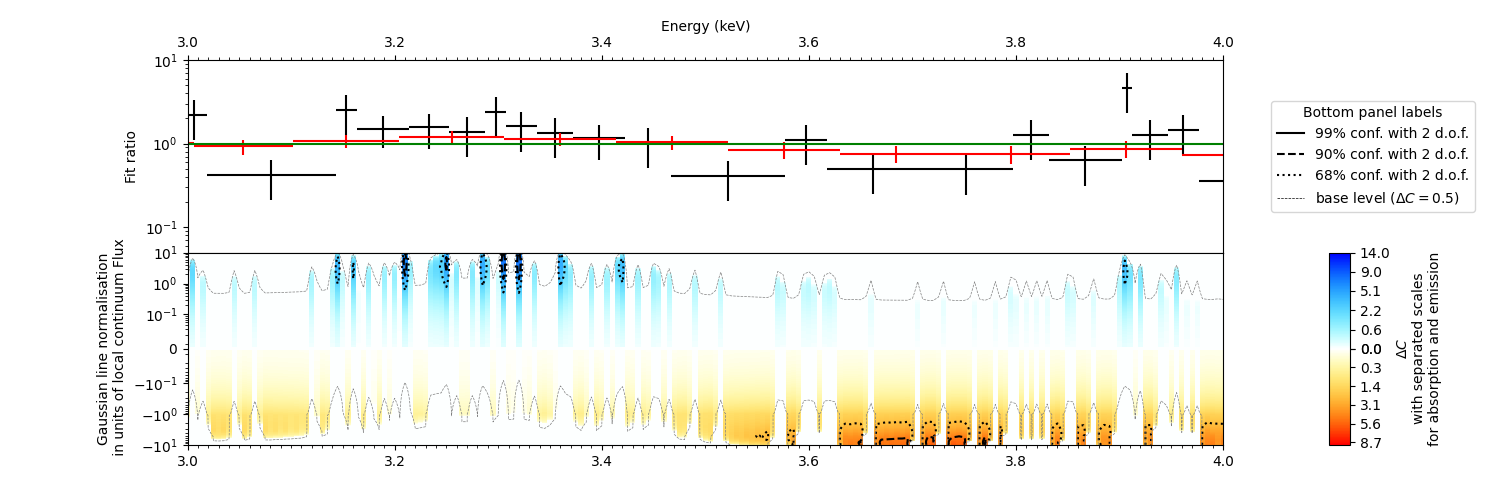}
\includegraphics[clip,trim=6.5cm 0.8cm 27.5cm 0.5cm,scale=0.315]{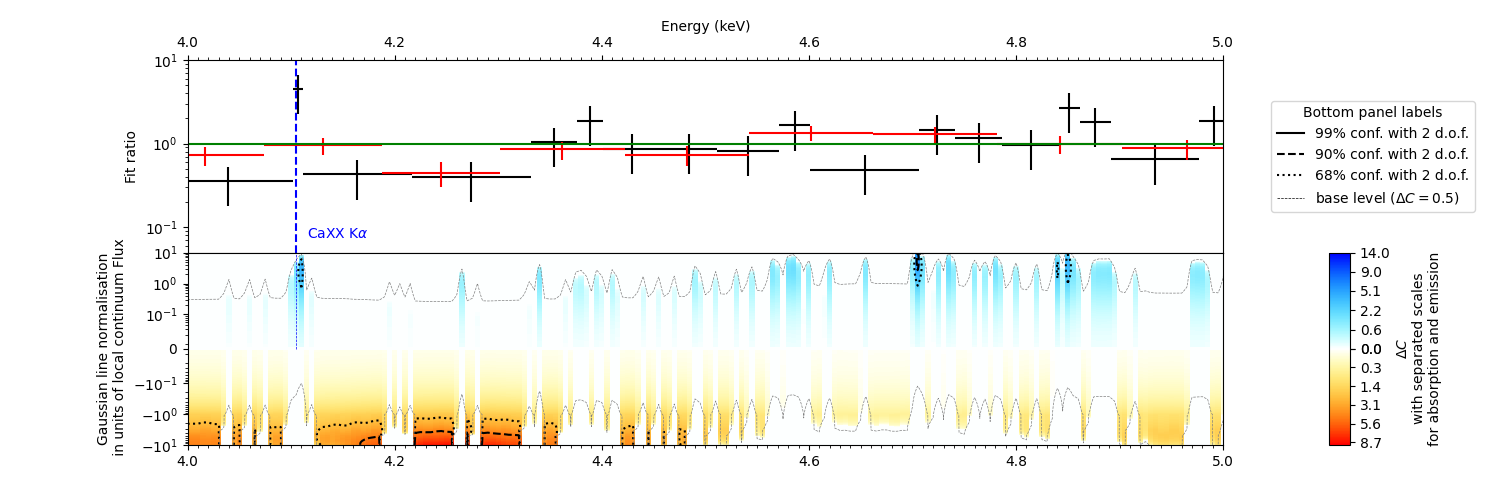}
\includegraphics[clip,trim=27.5cm 0.8cm 6.7cm 0.5cm,scale=0.315]{files_parra/blind_search_appendix/timeres/save_04_zoom_4.0_5.0.png}
\includegraphics[clip,trim=18.5cm 0.8cm 14.5cm 0.9cm,scale=0.315]{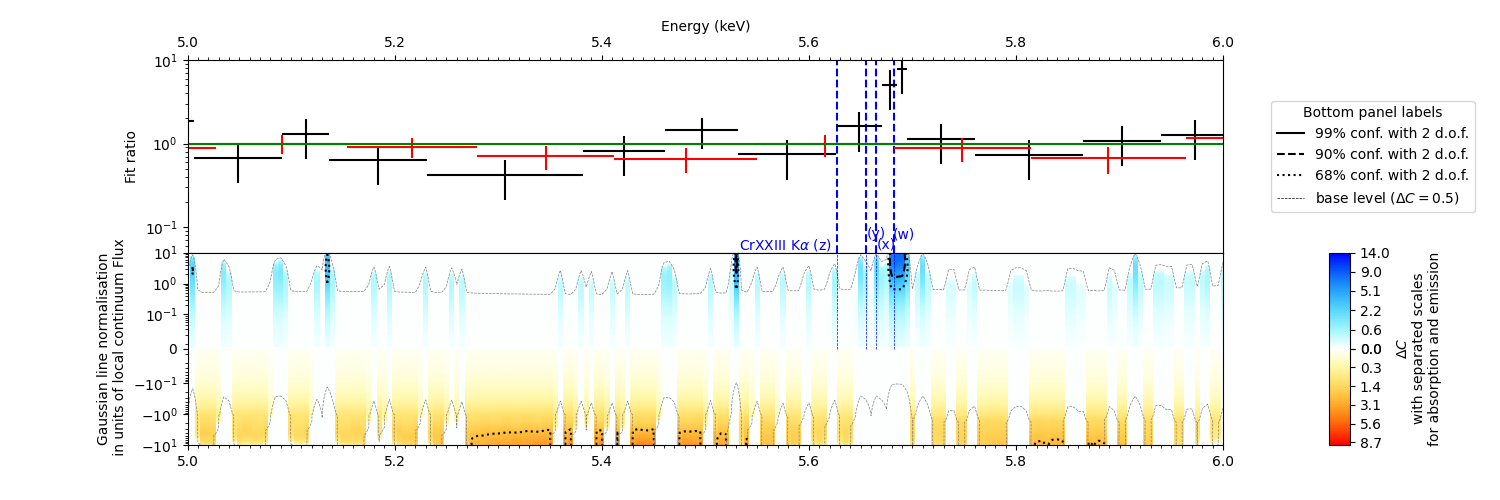}
\includegraphics[clip,trim=12.5cm 0.8cm 8.5cm 0.5cm,scale=0.315]{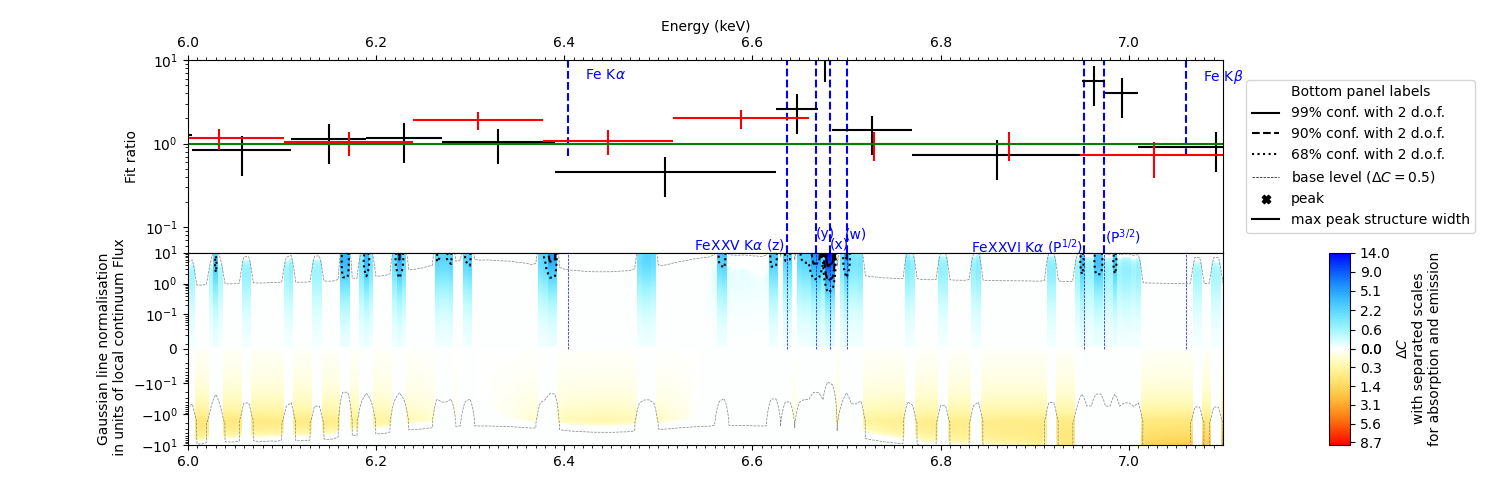}
\includegraphics[clip,trim=20.5cm 0.8cm 7.5cm 0.5cm,scale=0.315]{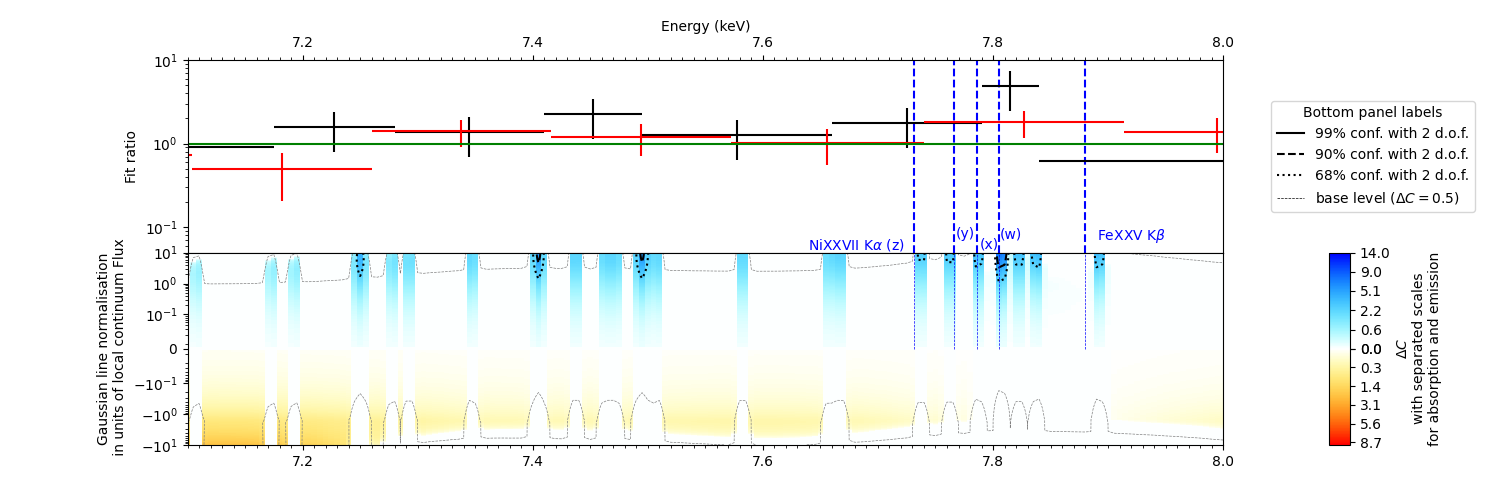}
\includegraphics[clip,trim=32.2cm 0.8cm 0.5cm 0.5cm,scale=0.315]{files_parra/blind_search_appendix/timeres/save_04_zoom_7.1_8.0.png}

\includegraphics[clip,trim=19cm 0.8cm 15cm 0.8cm,scale=0.315]{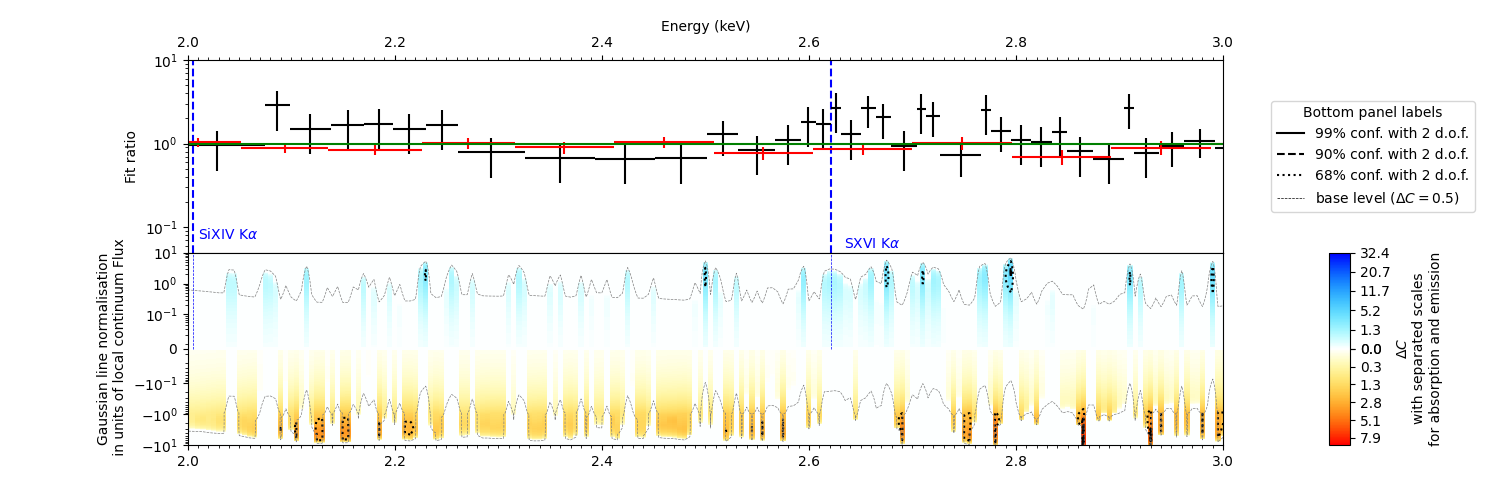}
\includegraphics[clip,trim=10.5cm 0.8cm 22.5cm 0.5cm,scale=0.315]{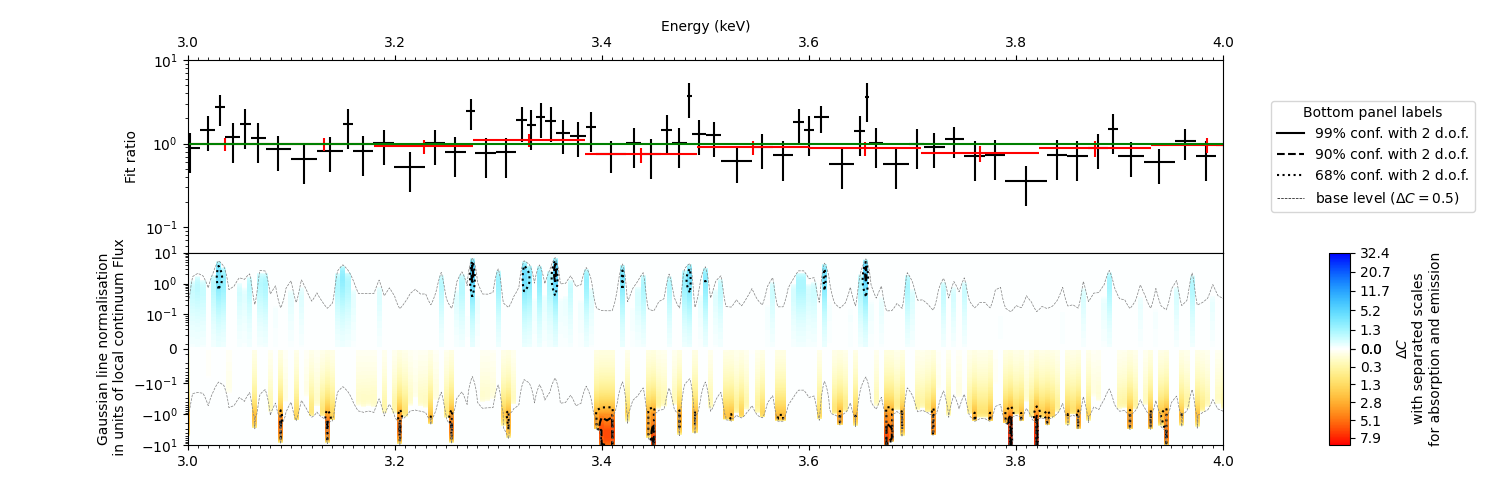}
\includegraphics[clip,trim=6.5cm 0.8cm 27.5cm 0.5cm,scale=0.315]{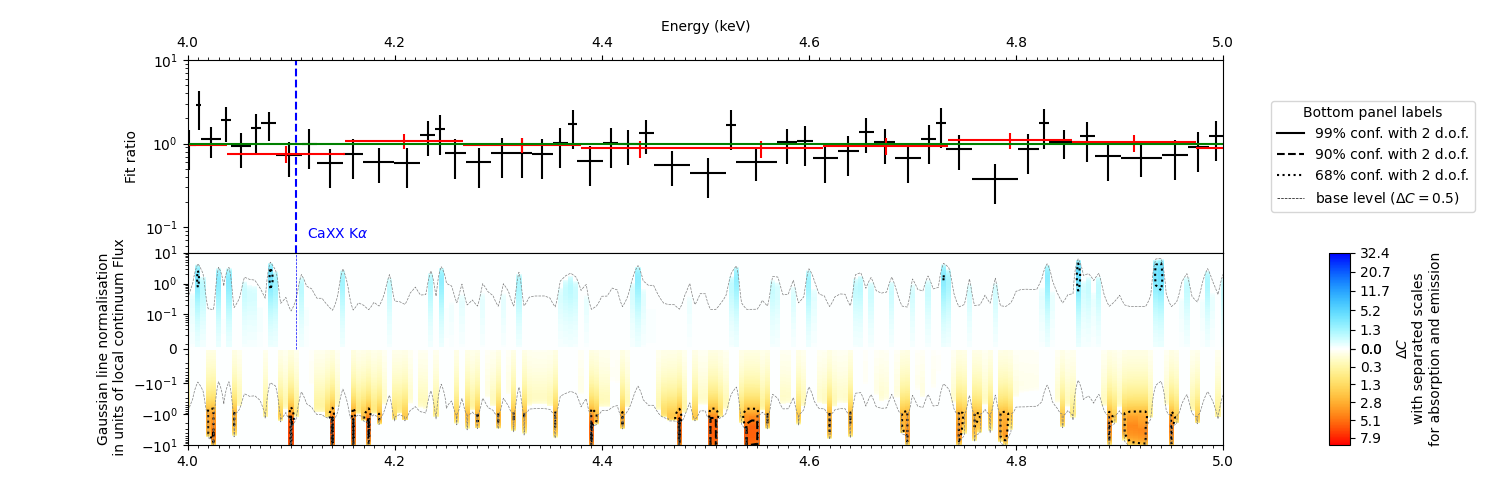}
\includegraphics[clip,trim=27.5cm 0.8cm 6.7cm 0.5cm,scale=0.315]{files_parra/blind_search_appendix/timeres/save_05_zoom_4.0_5.0.png}
\includegraphics[clip,trim=18.5cm 0.8cm 14.5cm 0.9cm,scale=0.315]{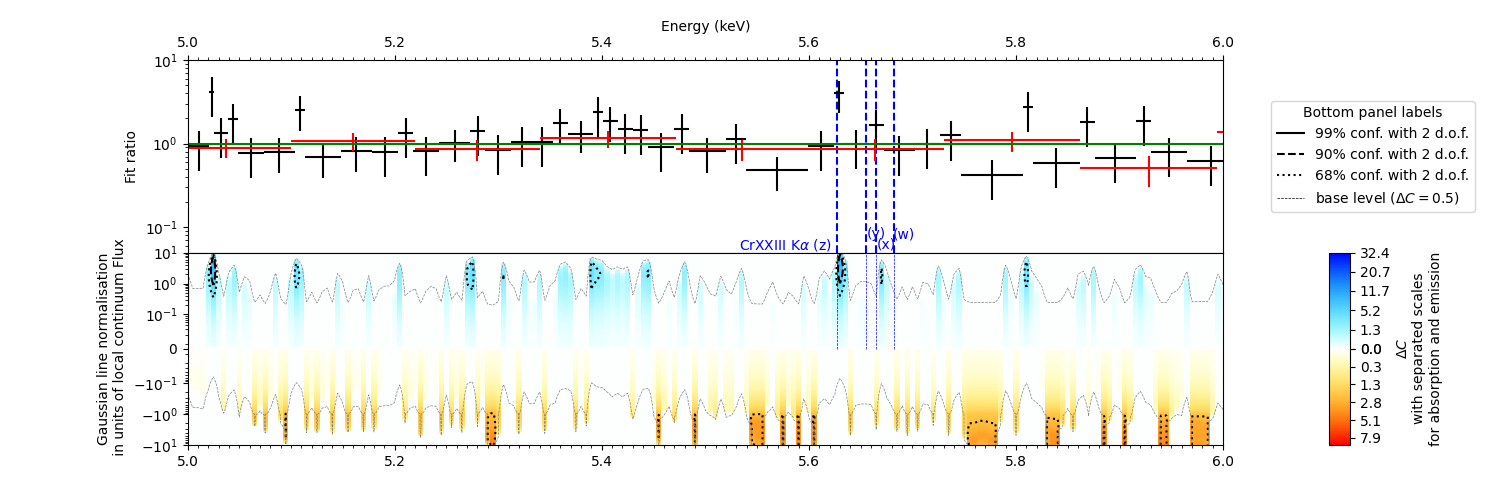}
\includegraphics[clip,trim=12.5cm 0.8cm 8.5cm 0.5cm,scale=0.315]{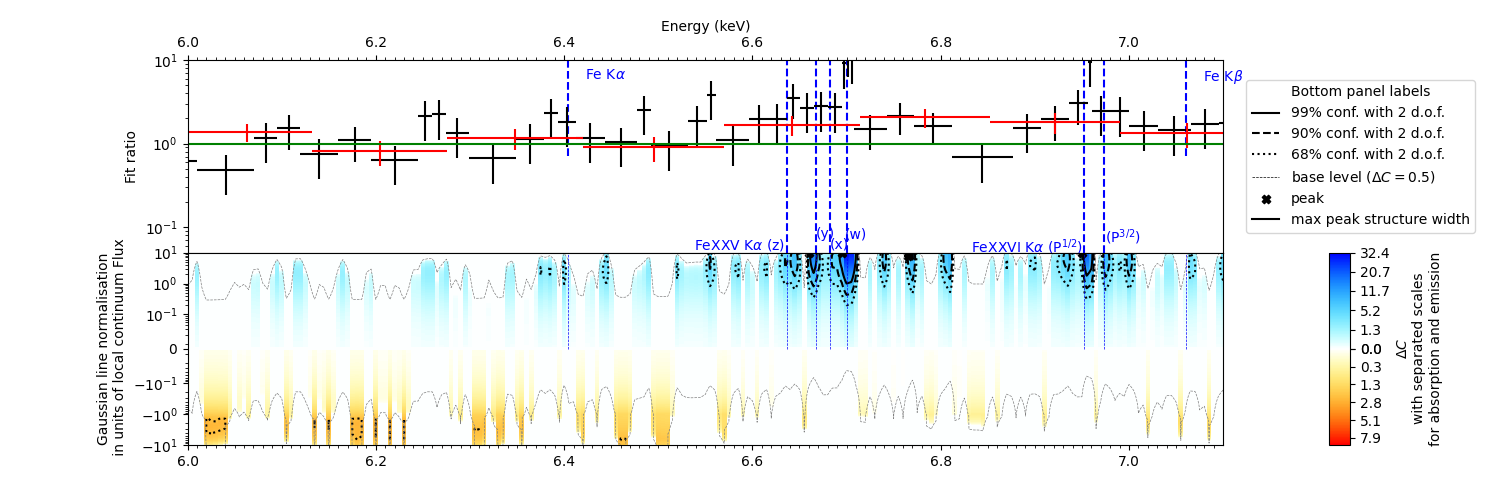}
\includegraphics[clip,trim=20.5cm 0.8cm 7.5cm 0.5cm,scale=0.315]{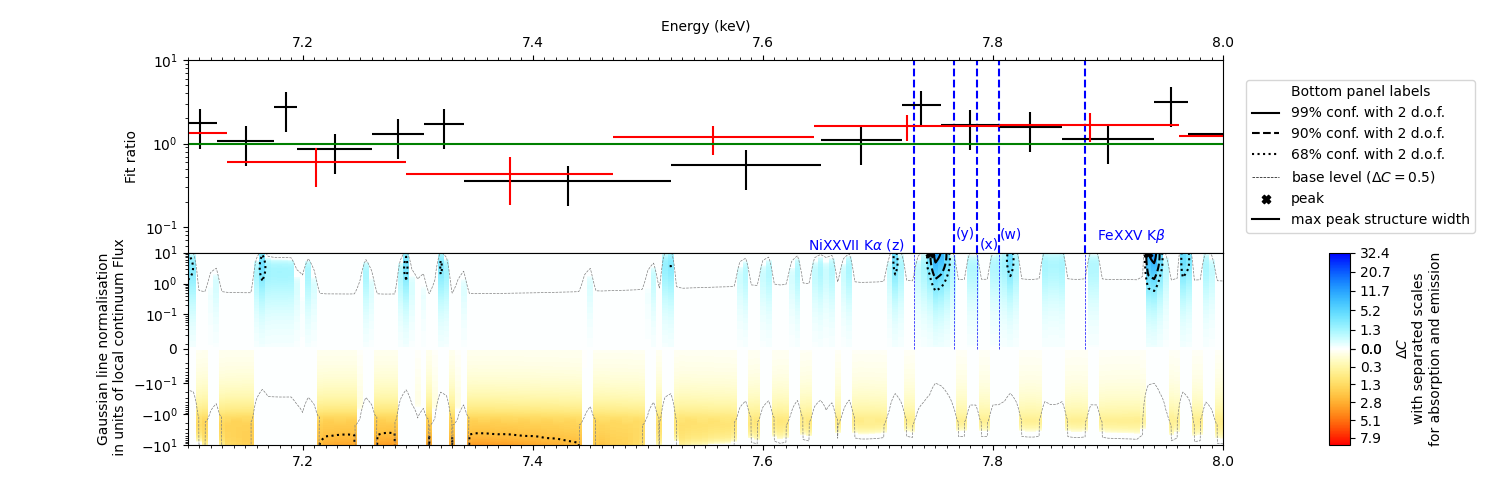}
\includegraphics[clip,trim=32.2cm 0.8cm 0.5cm 0.5cm,scale=0.315]{files_parra/blind_search_appendix/timeres/save_05_zoom_7.1_8.0.png}
   
    \caption{
    The blind search showing the significance of the residuals after a continuum fit without lines for each individual orbit spectrum (from top to bottom), zoomed to the position of the most important lines. The data are binned for visual clarity.}
    \label{fig:blind_search_timeres}
\end{figure*}

\begin{figure*}[t!]
    \centering
    \includegraphics[clip,trim=2.5cm 0.8cm 0.5cm 0.5cm,width=0.49\textwidth]{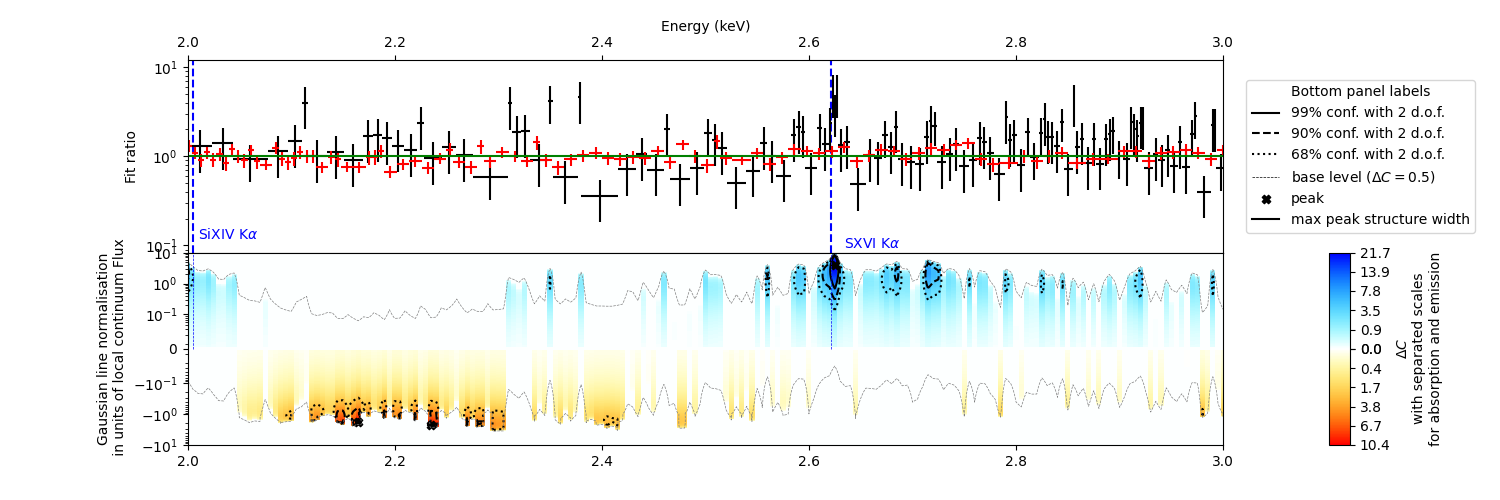}
    \includegraphics[clip,trim=2.5cm 0.8cm 0.5cm 0.5cm,width=0.49\textwidth]{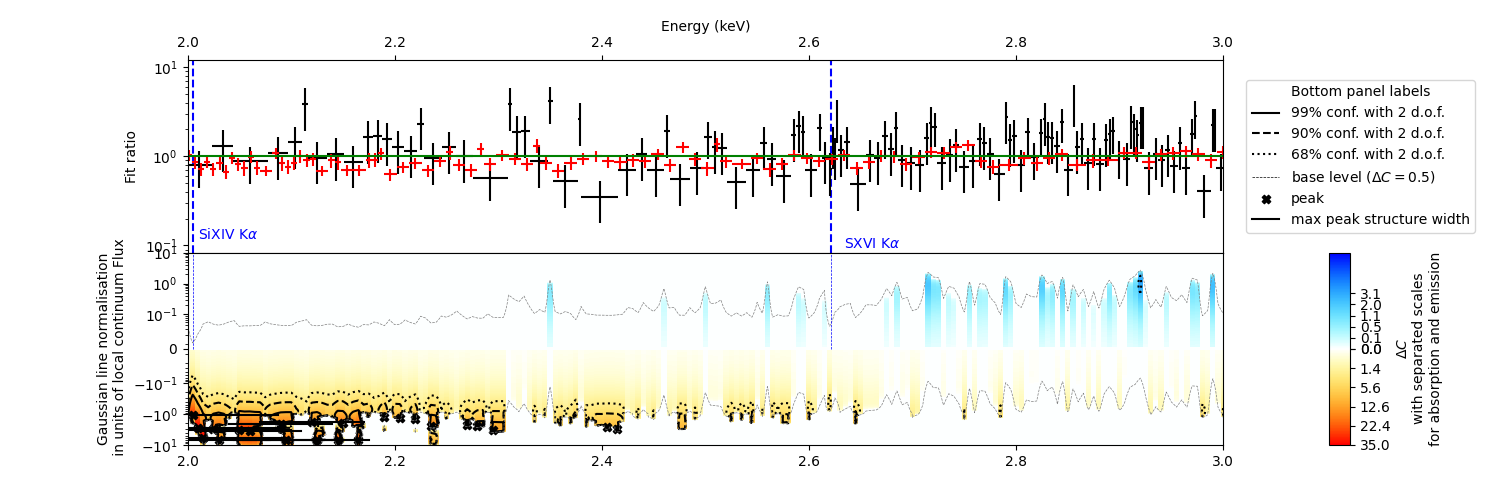}
    \includegraphics[clip,trim=2.5cm 0.8cm 0.5cm 0.5cm,width=0.49\textwidth]{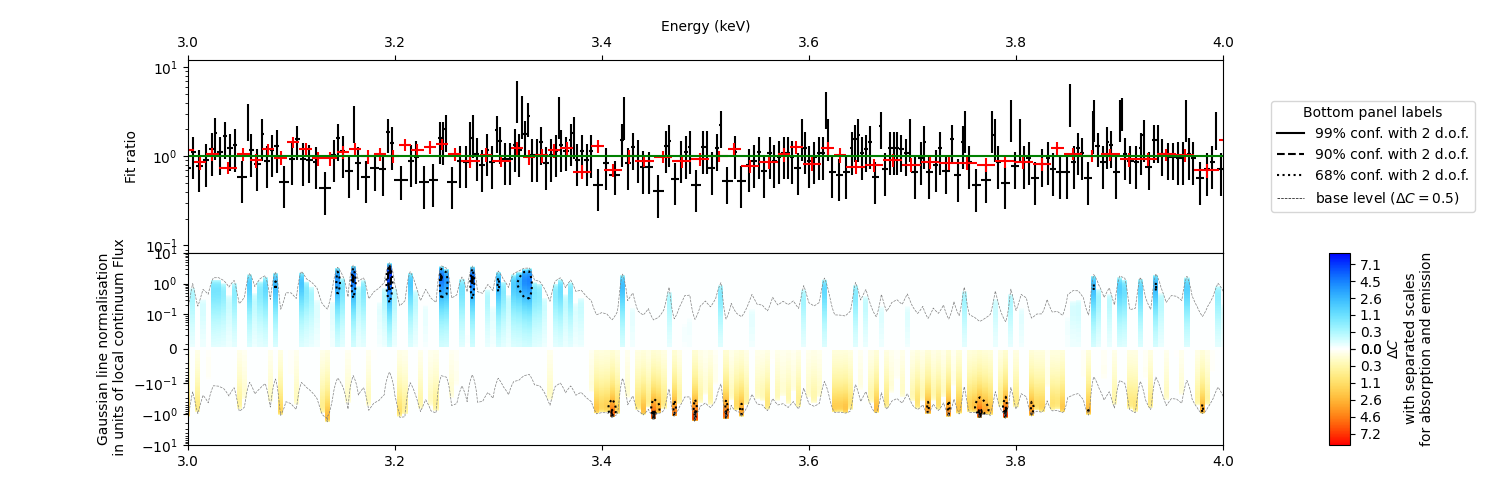}
    \includegraphics[clip,trim=2.5cm 0.8cm 0.5cm 0.5cm,width=0.49\textwidth]{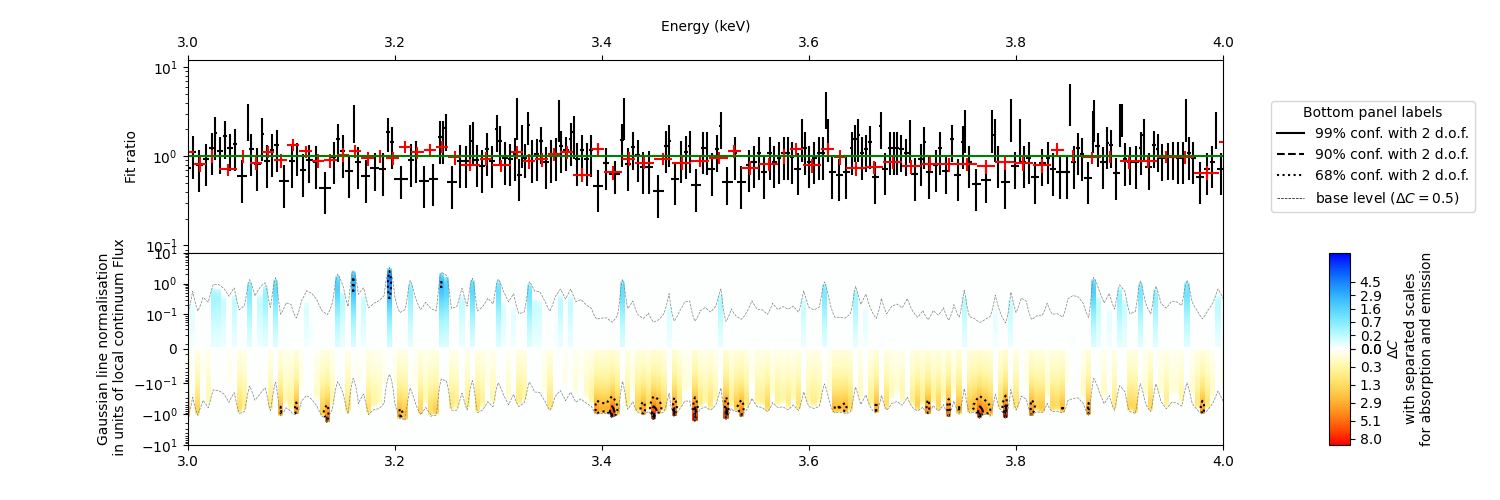}
    \includegraphics[clip,trim=2.5cm 0.8cm 0.5cm 0.5cm,width=0.49\textwidth]{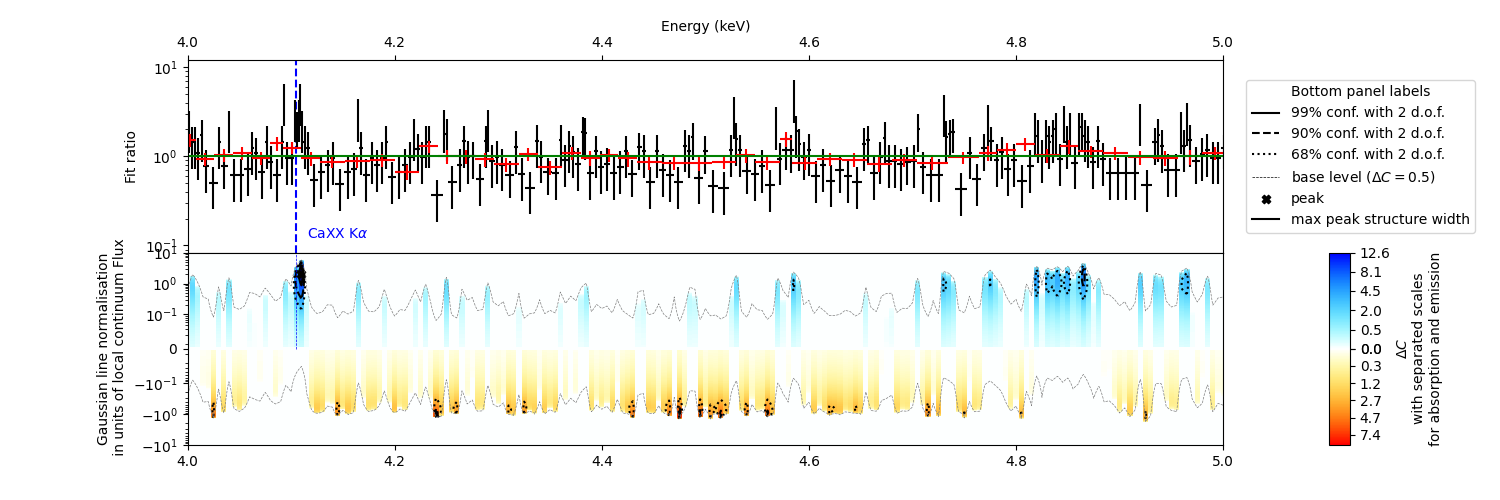}
    \includegraphics[clip,trim=2.5cm 0.8cm 0.5cm 0.5cm,width=0.49\textwidth]{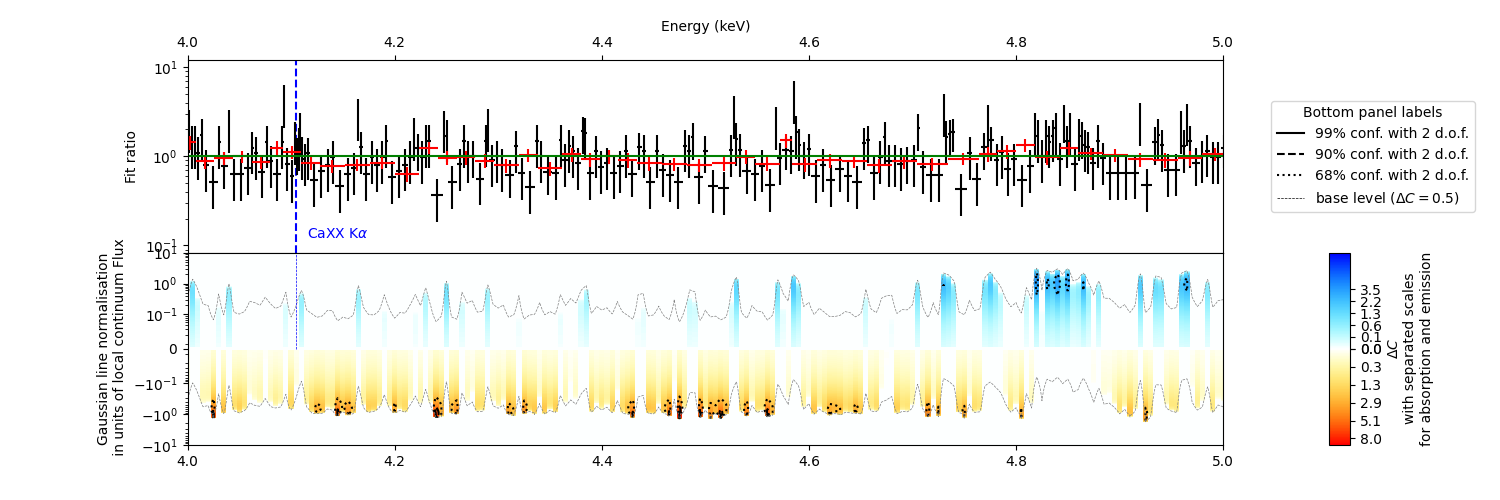}
    \includegraphics[clip,trim=2.5cm 0.8cm 0.5cm 0.5cm,width=0.49\textwidth]{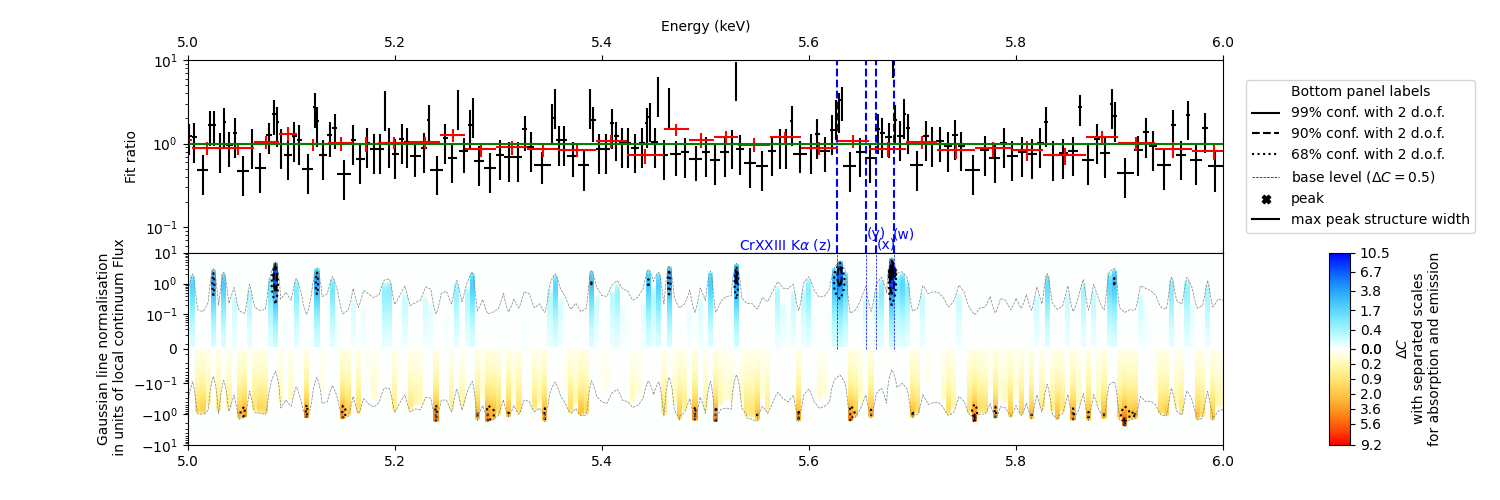}
    \includegraphics[clip,trim=2.5cm 0.8cm 0.5cm 0.5cm,width=0.49\textwidth]{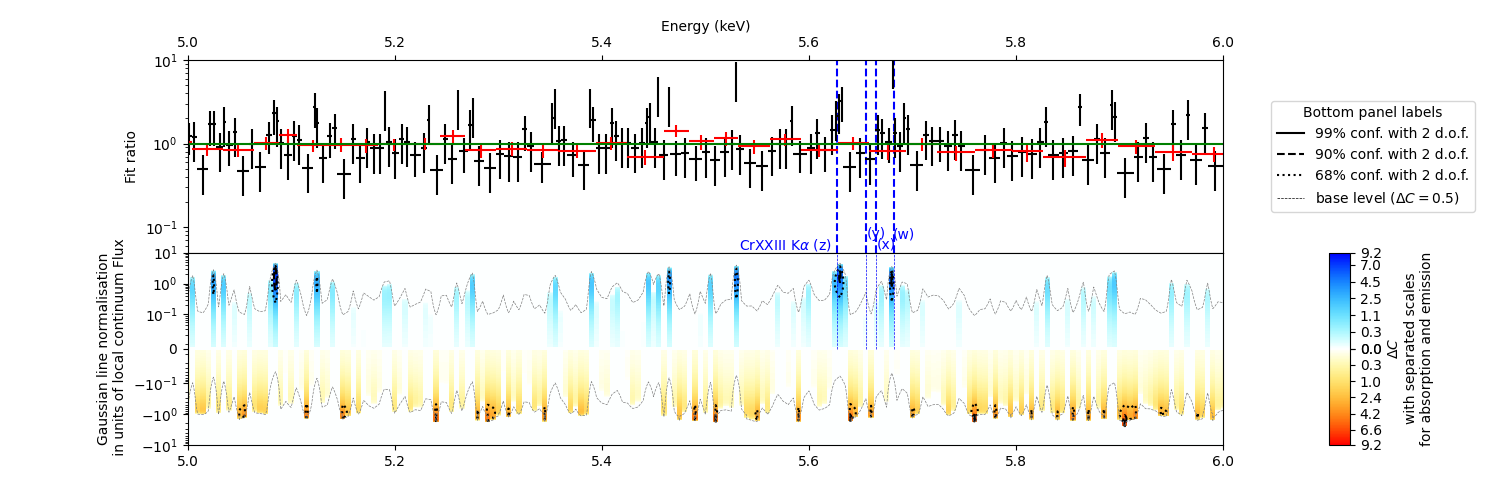}
    \includegraphics[clip,trim=2.5cm 0.8cm 0.5cm 0.5cm,width=0.49\textwidth]{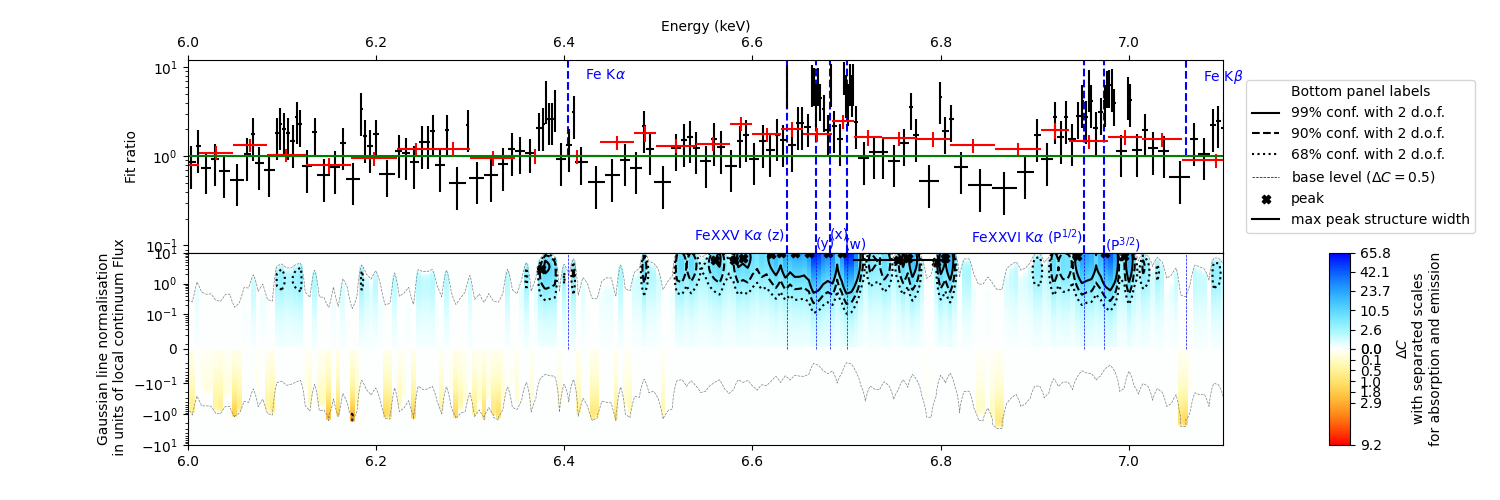}
    \includegraphics[clip,trim=2.5cm 0.8cm 0.5cm 0.5cm,width=0.49\textwidth]{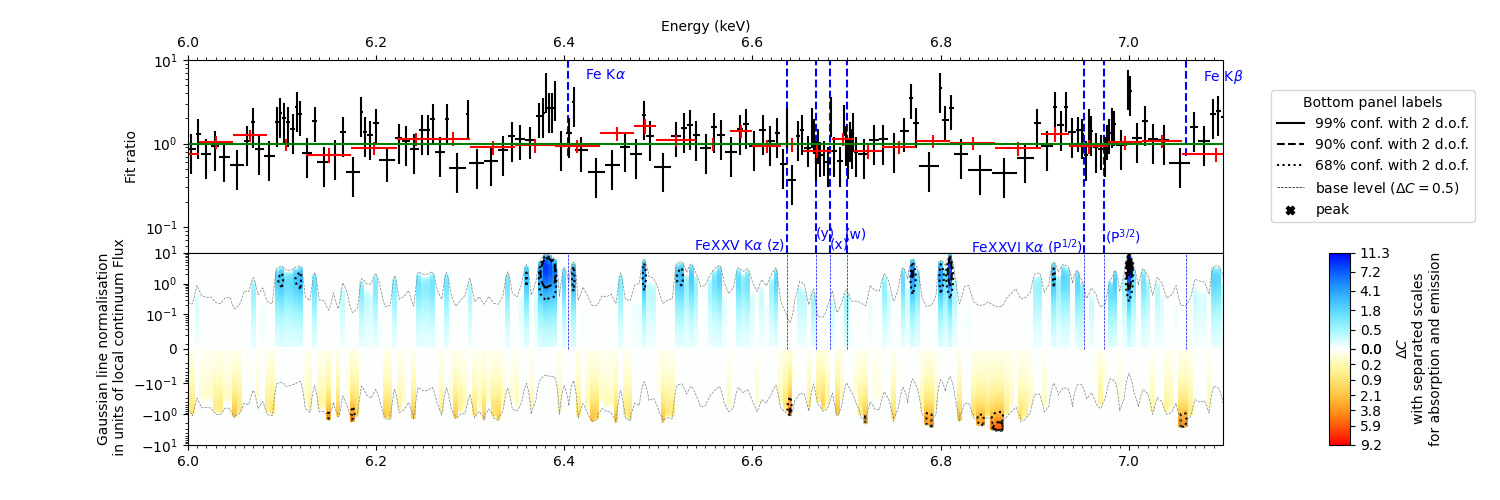}
    \includegraphics[clip,trim=2.5cm 0.8cm 0.5cm 0.5cm,width=0.49\textwidth]{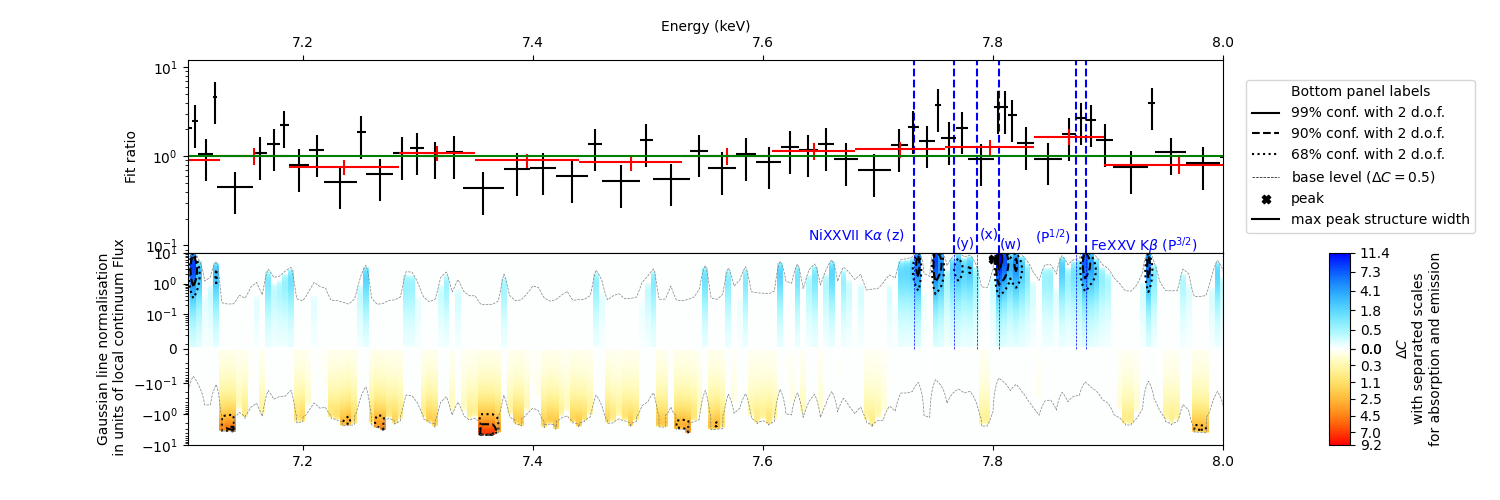}
    \includegraphics[clip,trim=2.5cm 0.8cm 0.5cm 0.5cm,width=0.49\textwidth]{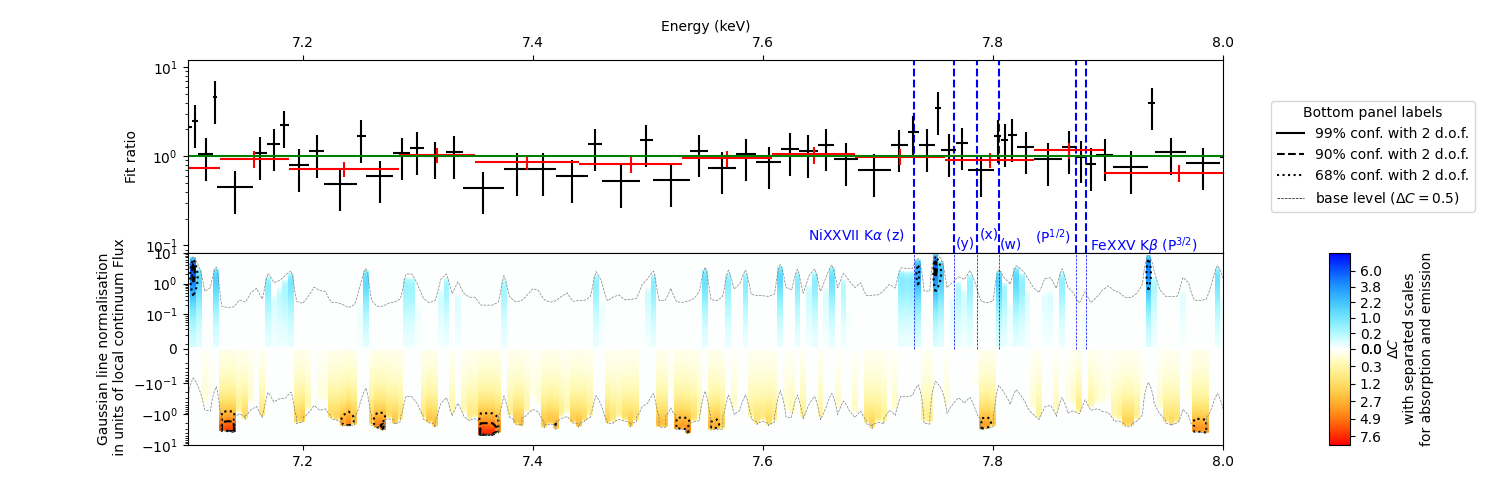}
    \includegraphics[clip,trim=2.5cm 0.8cm 0.5cm 0.5cm,width=0.49\textwidth]{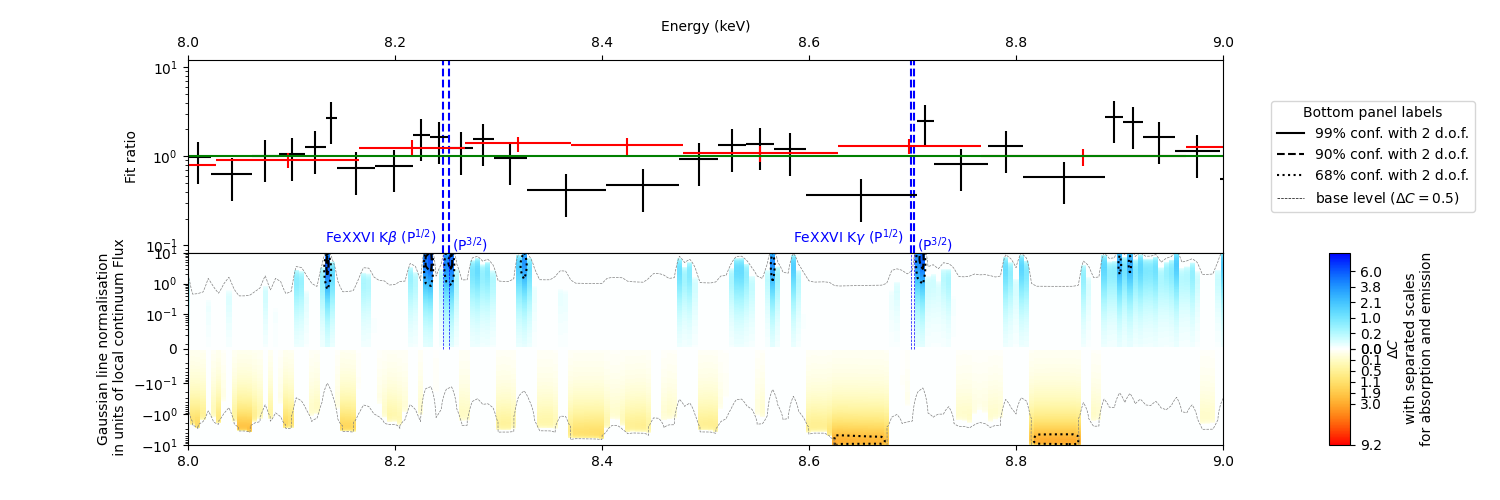}
    \includegraphics[clip,trim=2.5cm 0.8cm 0.5cm 0.5cm,width=0.49\textwidth]{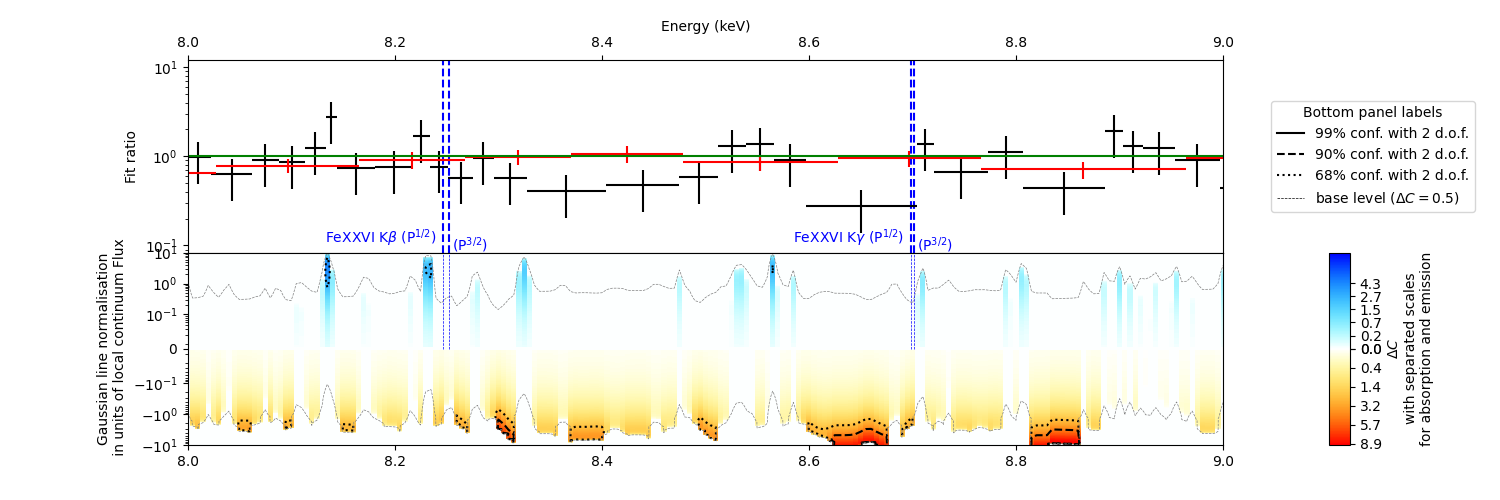}    
    \caption{The full results of the blind search showing the significance of the residuals after a continuum fit without lines \textbf{(left)} and after the pion fit \textbf{(right)}. The data from each instrument are re-binned at a 2$\sigma$ significance to increase visual clarity, and we highlight the energies and transitions of the significant and well-known features.
    }
    \label{fig:blind_search_total_phys}
\end{figure*}

\begin{figure*}[t!]
    \centering
    \includegraphics[clip,trim=0cm 0.cm 0.cm 0.cm,width=0.49\textwidth]{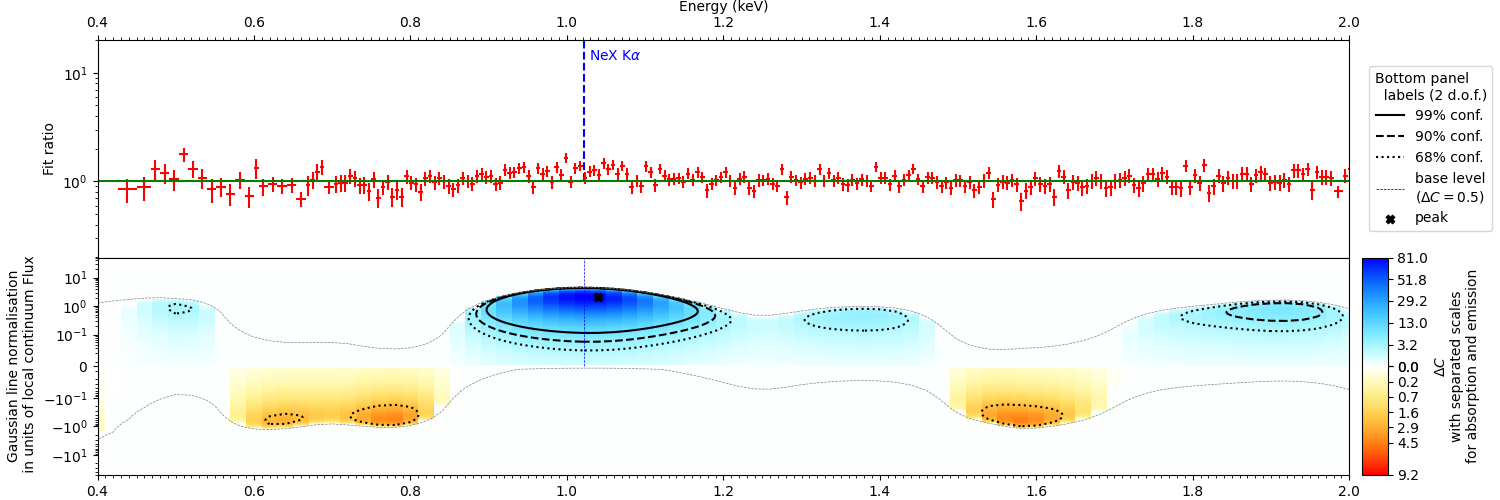}
    \includegraphics[clip,trim=0cm 0.cm 0.cm 0.cm,width=0.49\textwidth]{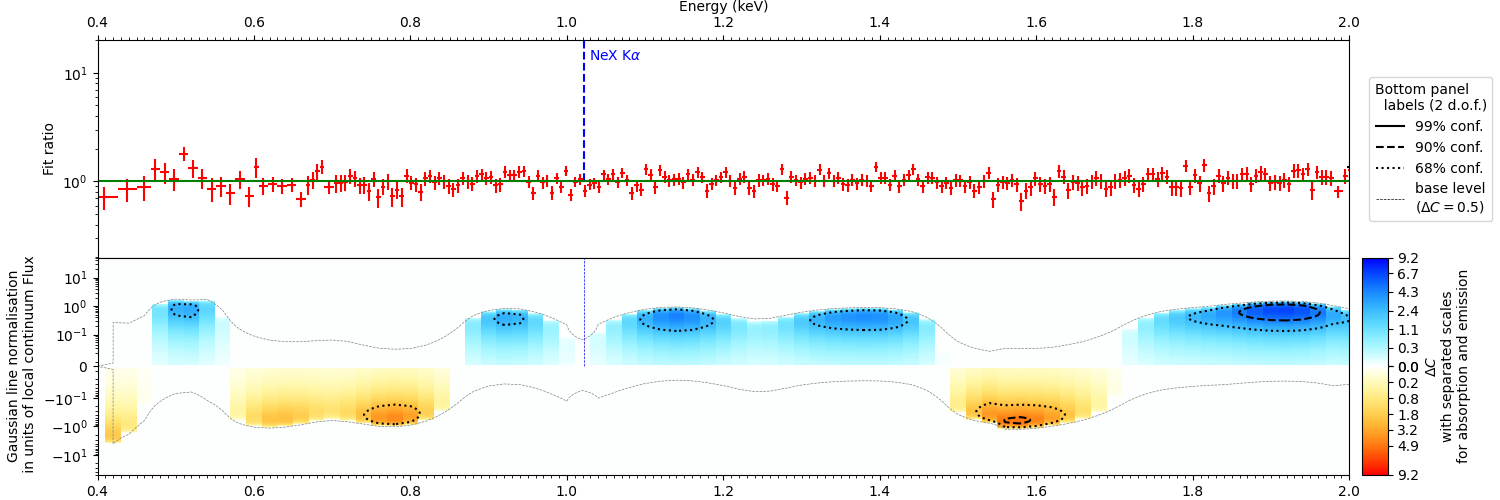}
    \includegraphics[clip,trim=0cm 0.cm 0.cm 0.cm,width=0.49\textwidth]{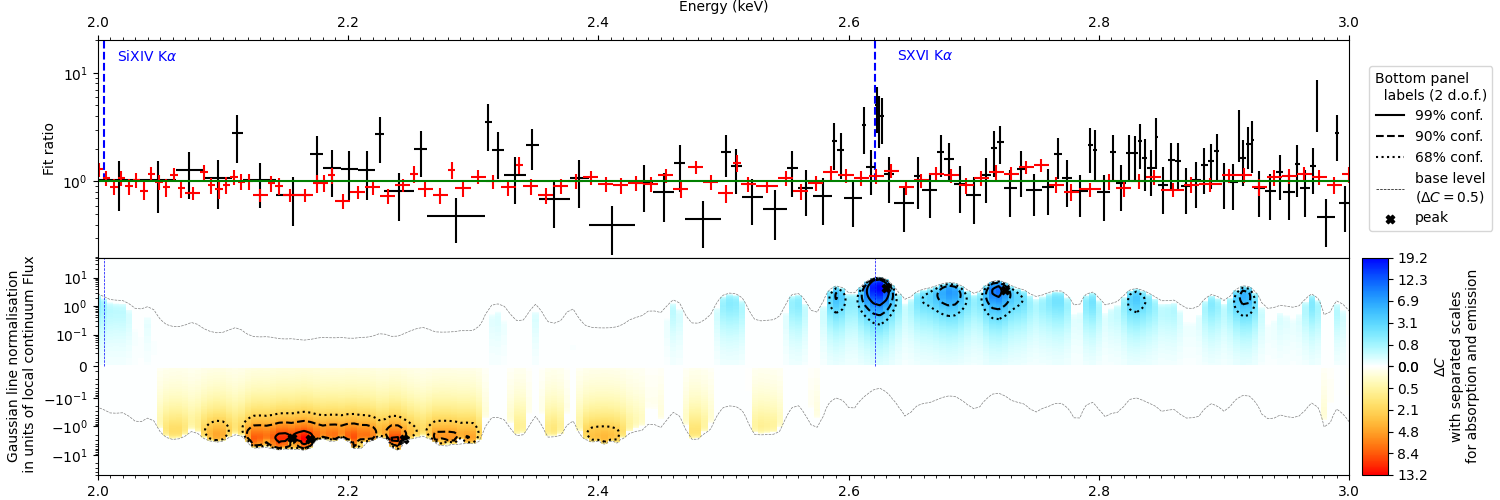}
    \includegraphics[clip,trim=0cm 0.cm 0.cm 0.cm,width=0.49\textwidth]{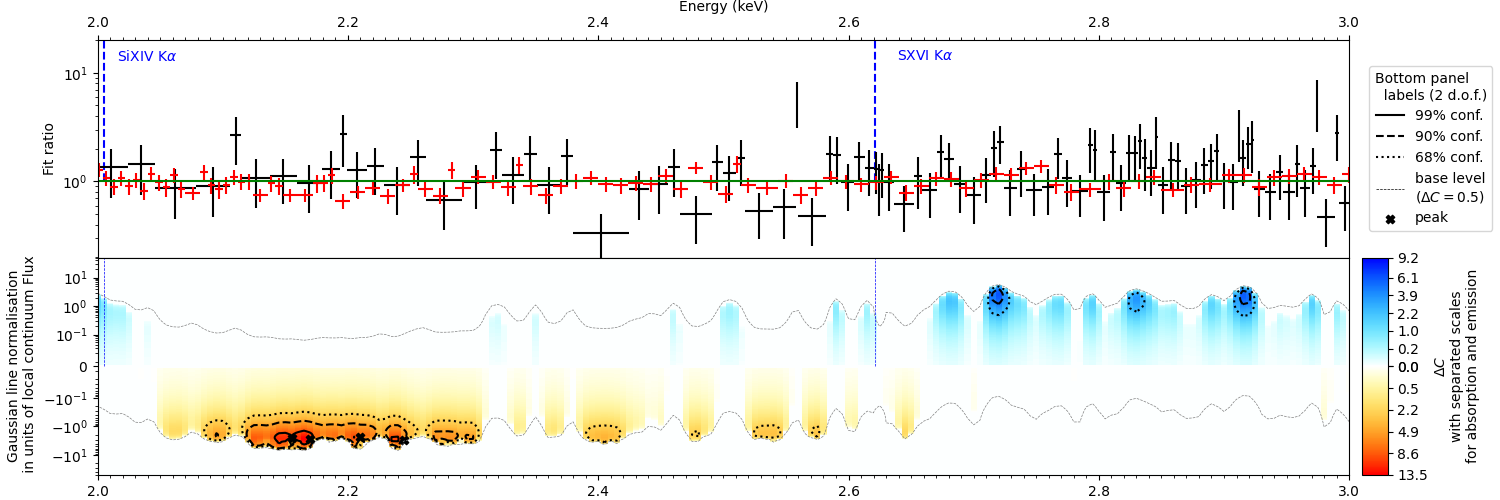}
    \includegraphics[clip,trim=0cm 0.cm 0.cm 0.cm,width=0.49\textwidth]{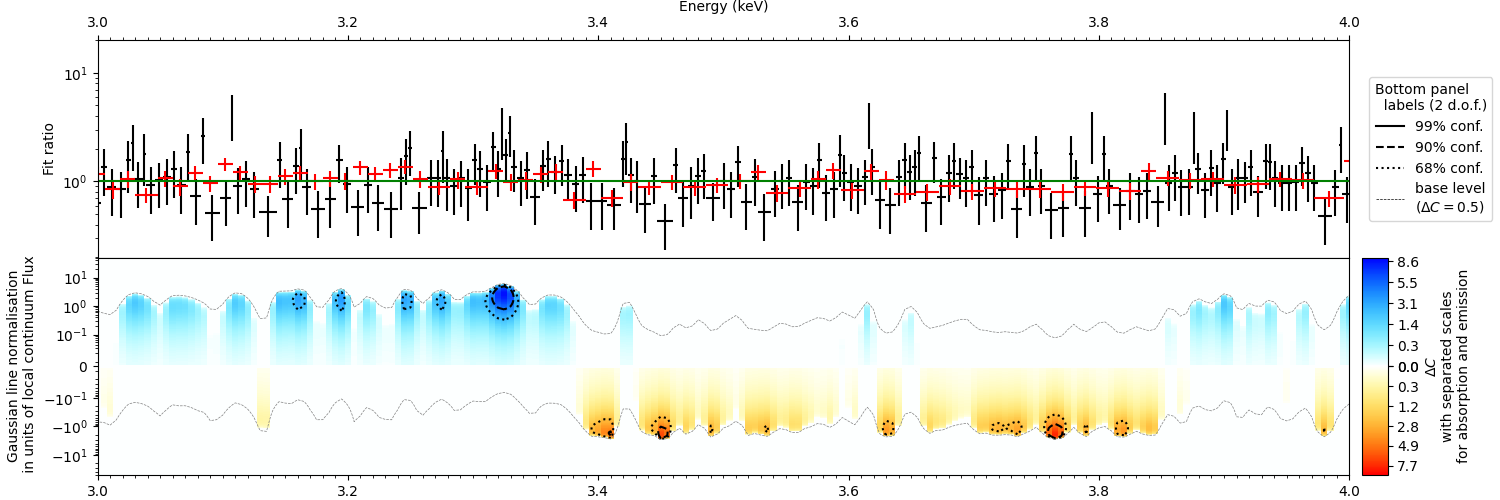}
    \includegraphics[clip,trim=0cm 0.cm 0.cm 0.cm,width=0.49\textwidth]{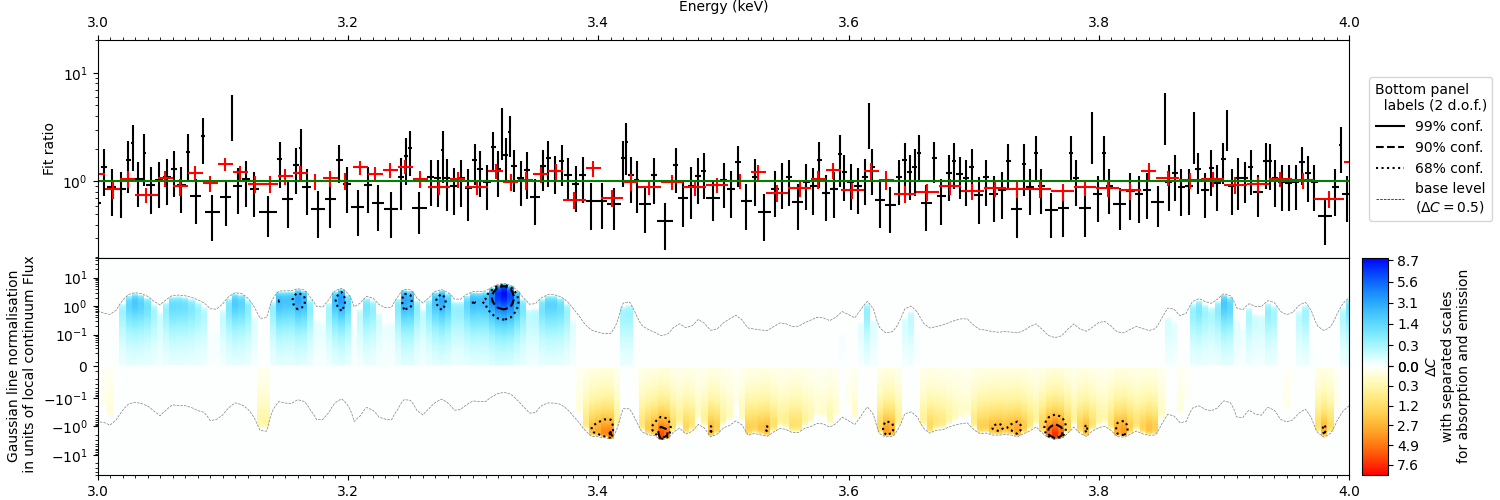}
    \includegraphics[clip,trim=0cm 0.cm 0.cm 0.cm,width=0.49\textwidth]{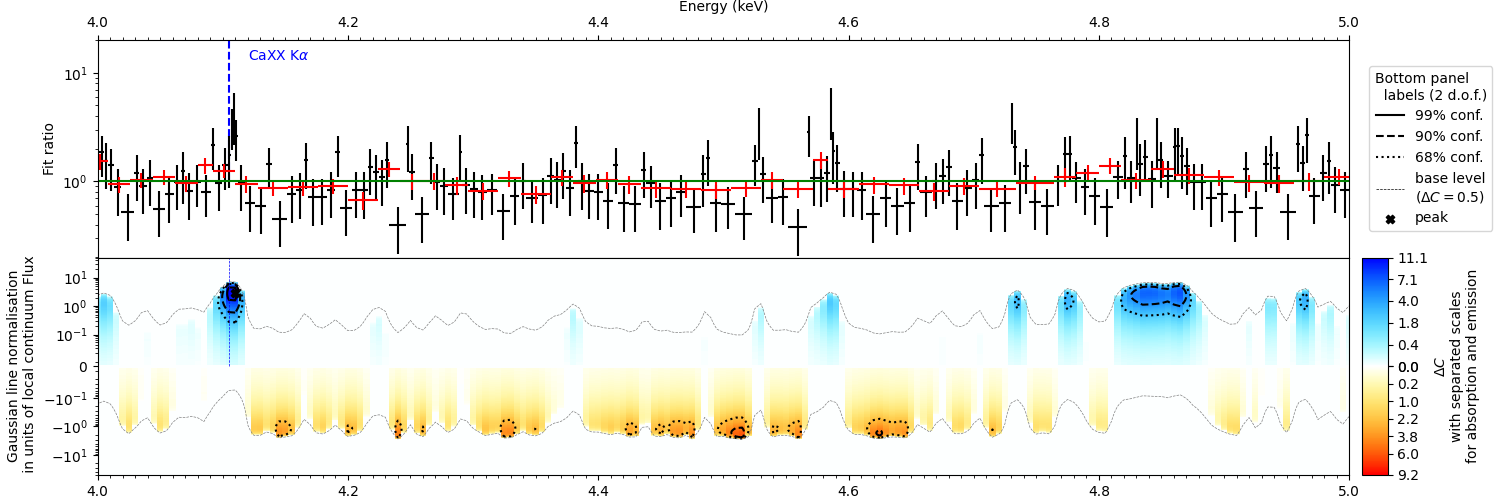}
    \includegraphics[clip,trim=0cm 0.cm 0.cm 0.cm,width=0.49\textwidth]{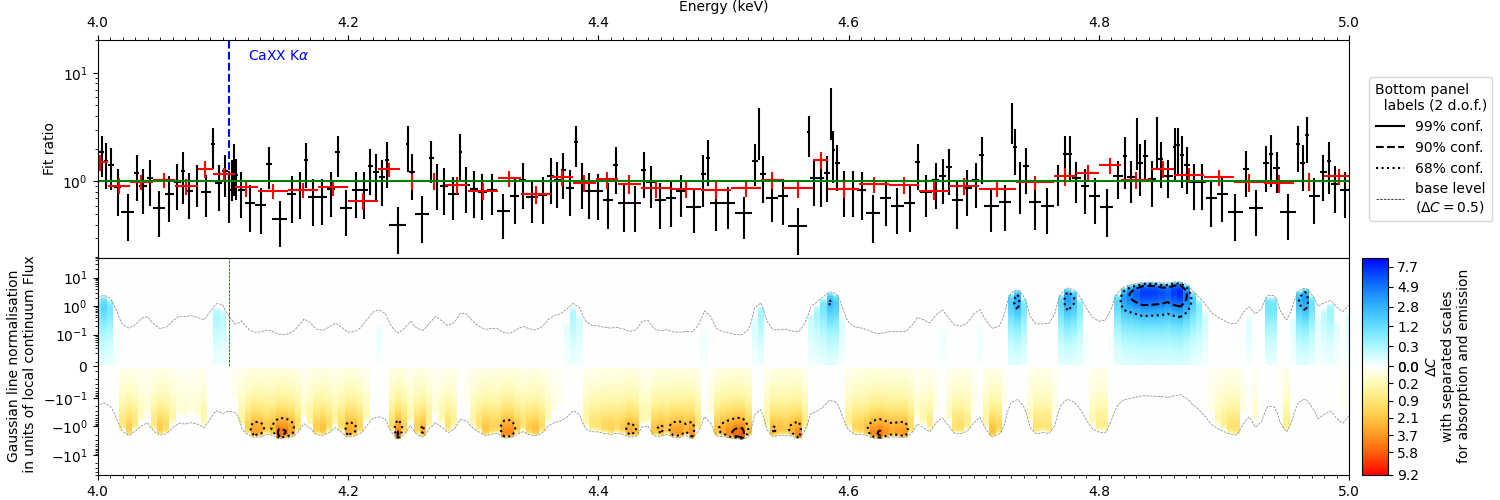}
    \includegraphics[clip,trim=0cm 0.cm 0.cm 0.cm,width=0.49\textwidth]{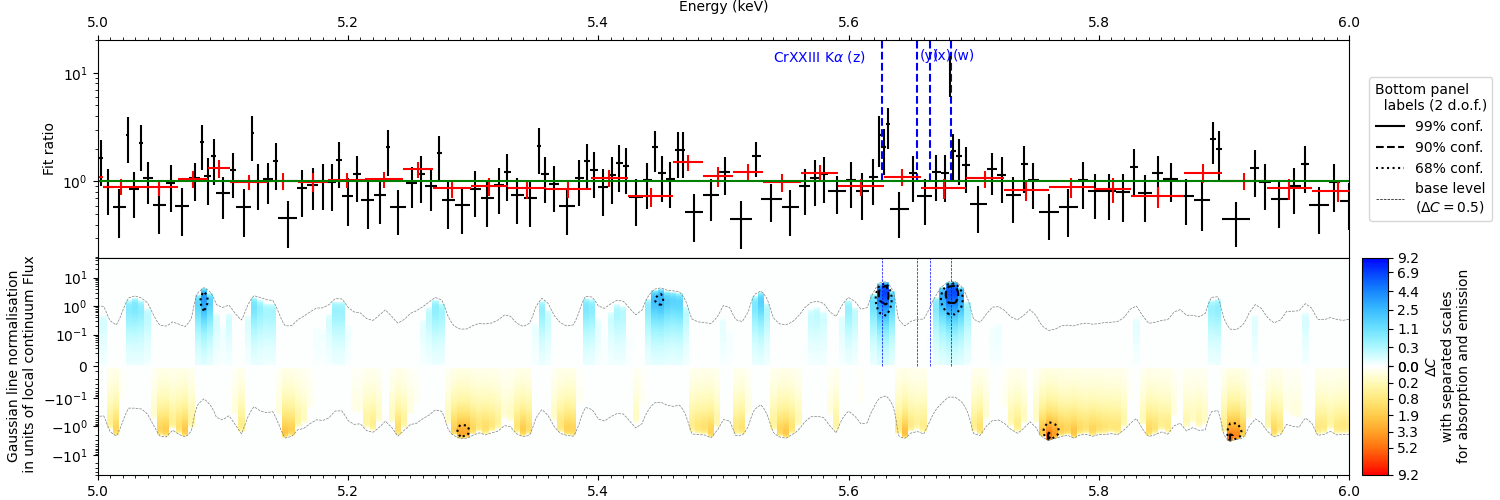}
    \includegraphics[clip,trim=0cm 0.cm 0.cm 0.cm,width=0.49\textwidth]{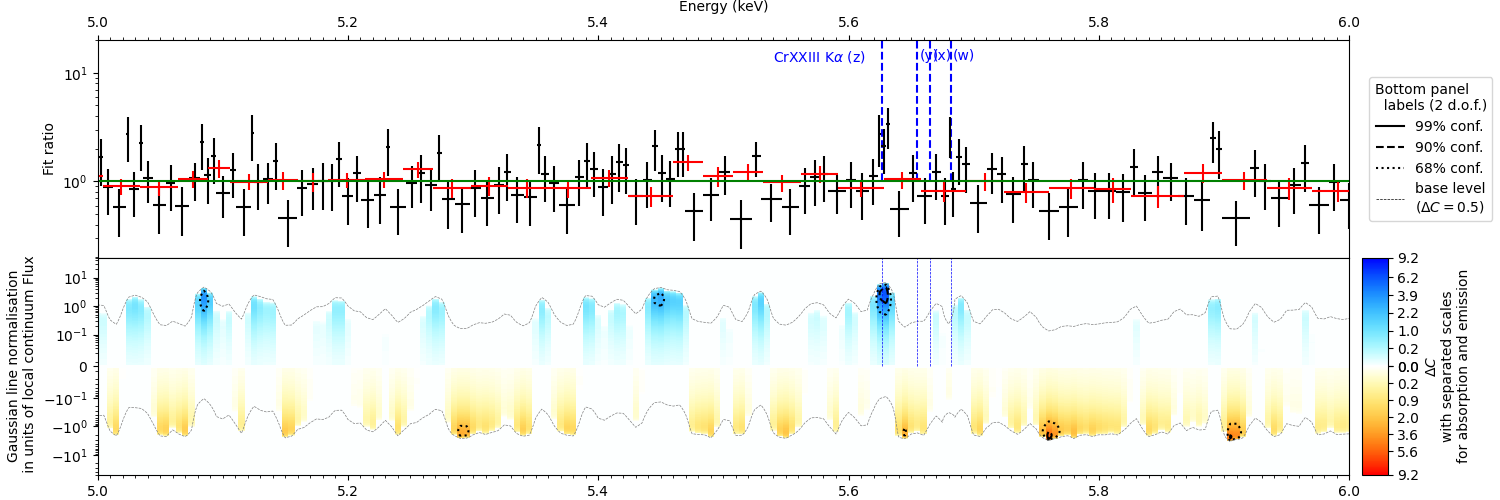}
    \includegraphics[clip,trim=0cm 0.cm 0.cm 0.cm,width=0.49\textwidth]{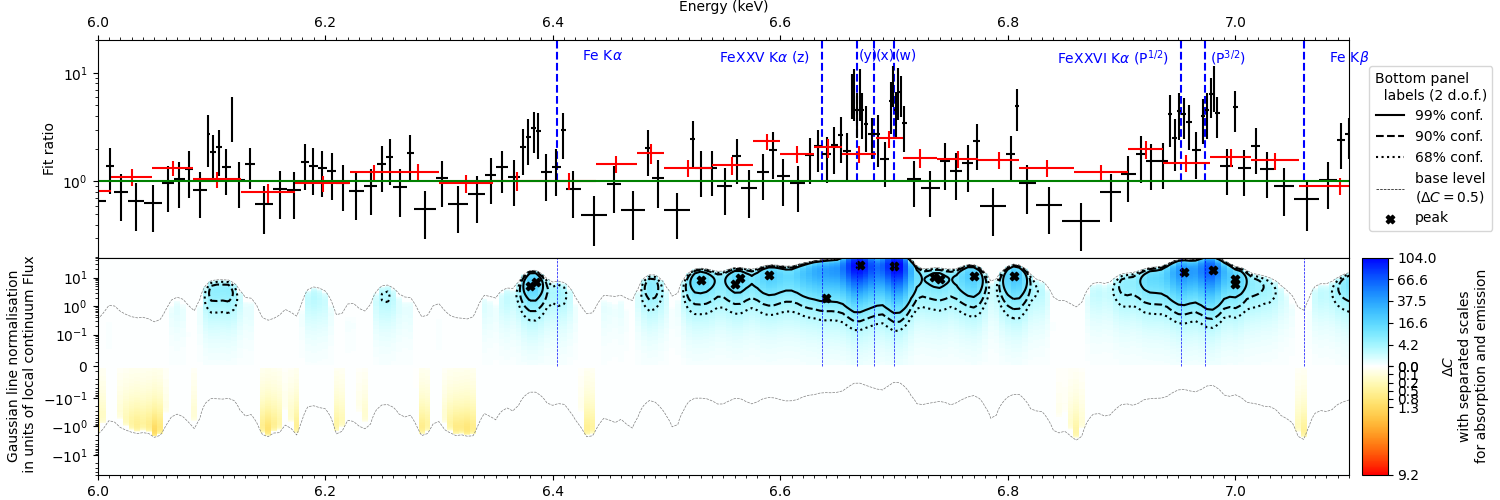}
    \includegraphics[clip,trim=0cm 0.cm 0.cm 0.cm,width=0.49\textwidth]{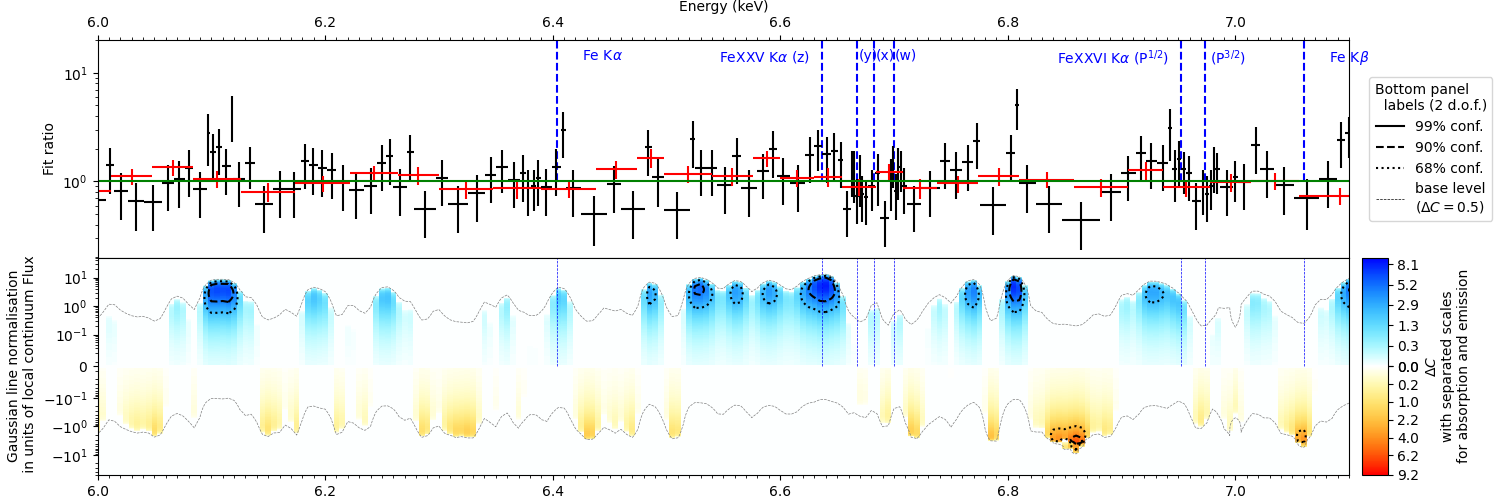}
    \includegraphics[clip,trim=0cm 0.cm 0.cm 0.cm,width=0.49\textwidth]{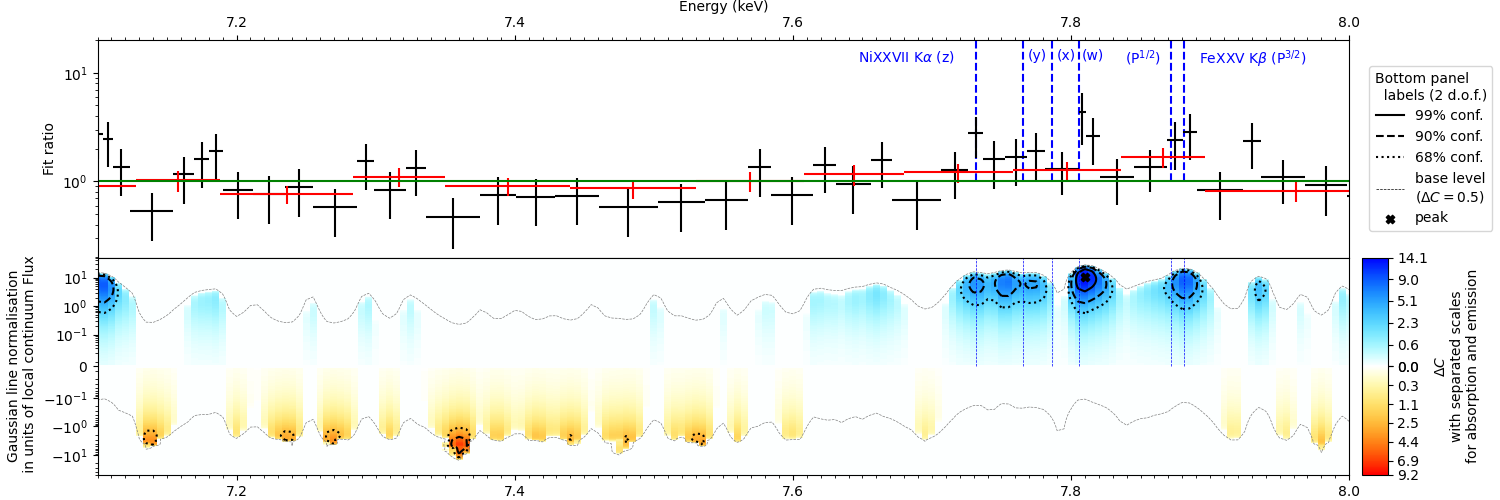}
    \includegraphics[clip,trim=0cm 0.cm 0.cm 0.cm,width=0.49\textwidth]{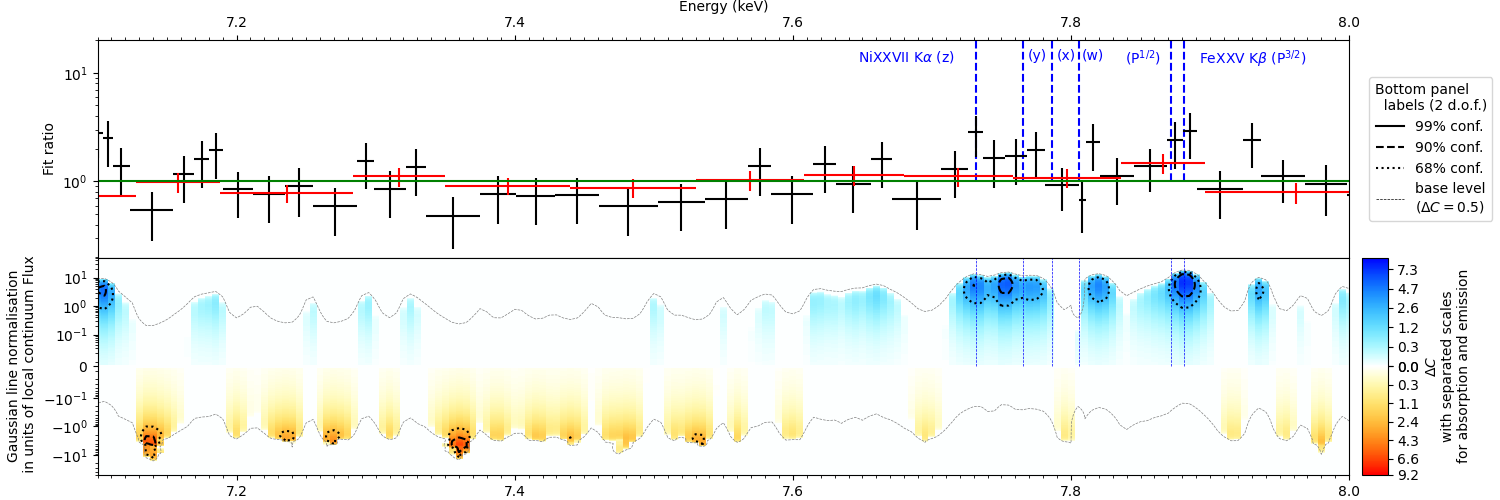}
    \includegraphics[clip,trim=0cm 0.cm 0.cm 0.cm,width=0.49\textwidth]{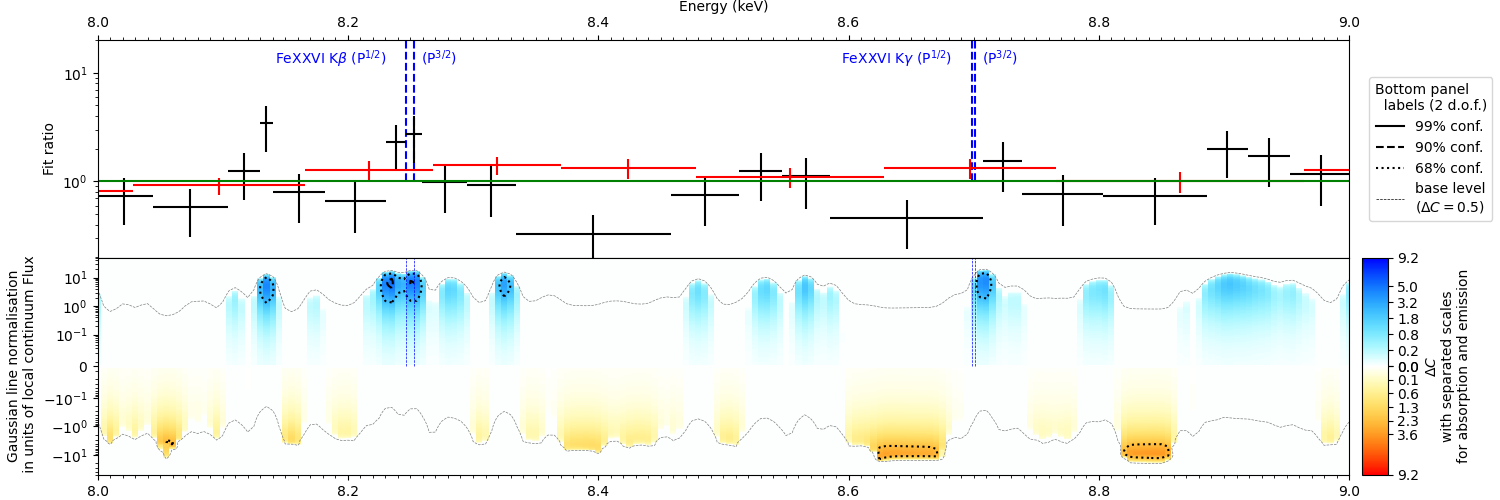}
    \includegraphics[clip,trim=0cm 0.cm 0.cm 0.cm,width=0.49\textwidth]{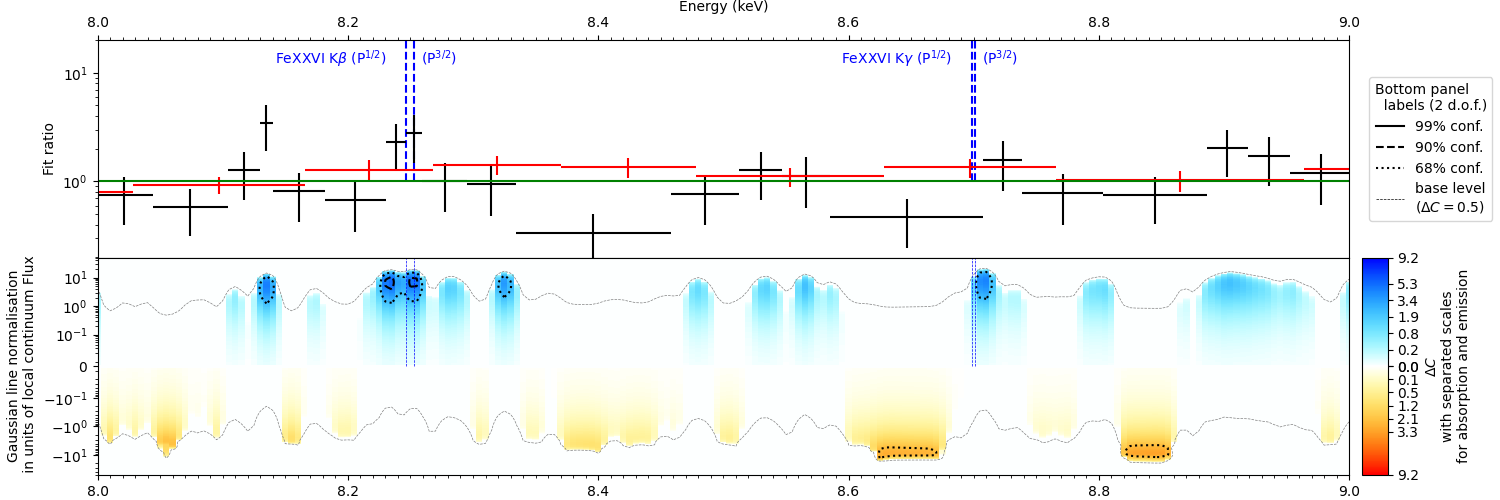}
    \caption{Same as Fig.~\ref{fig:blind_search_total_phys} but showing the significance of the residuals after a continuum fit without lines \textbf{(left)} and after the gaussian fit \textbf{(right)}.}
    \label{fig:blind_search_total_gauss}
\end{figure*}

\newpage

\begin{table*}
\centering
\caption{Evolution of spectral lines identified in the orbit-resolved Resolve spectra.}
\vspace{0.2em}
\label{tab:line_param_orbit}
\begin{tabular}{lccccccc}
\hline
 Line ID & $E_{\rm rest}$ (keV) &  
 orbit & $v_{raw}$ (km s$^{-1}$) & $\sigma$ (eV) & $\Delta v$ (km s$^{-1}$) & EW (eV) & $\Delta$ C-stat\T\B\\
 \hline
\multirow{7}{*}{\Fexxv{} Ly$\beta$} & \multirow{7}{*}{7.878} & 1 & - & - & - & $22_{-21}^{+20}$ & 2\T\B\\
& & 2 & $-150\pm220$ & - & - & $83_{-40}^{+45}$ & 12\T\B\\
& & 3 & - & - & - & $60_{-41}^{+44}$ & 6\T\B\\
& & 4 & - & - & - &  $102_{-80}^{+77}$ & 5\T\B\\
& & 5 & - & - & - & $26_{-26}^{+85}$ & 2\T\B\\ \cdashline{1-8}
%
\multirow{13}{*}{\Fexxvi{} Ly$\alpha_1$} & \multirow{13}{*}{6.973} & \multirow{2}{*}{1} & \multirow{2}{*}{$-280^{-130}_{+450}$} & \multirow{2}{*}{-} & \multirow{2}{*}{-} & $55_{-19}^{+21}$ & 18\T\B\\
\multirow{13}{*}{\Fexxvi{} Ly$\alpha_2$} & \multirow{13}{*}{6.952} &  & &  &  & $26\pm 9$ & 10\T\B\\ \cdashline{4-8}
& & \multirow{2}{*}{2} & \multirow{2}{*}{$-330_{-190}^{+230}$} & \multirow{2}{*}{$4.7_{-3.7}^{+5.2}$} & \multirow{2}{*}{$210_{-160}^{+210}$} & $106_{-26}^{+24}$ & \multirow{2}{*}{33}\T\B\\
& &  &  &  &  & $49_{-12}^{+10}$ & \T\B\\\cdashline{4-8}
& & \multirow{2}{*}{3} & \multirow{2}{*}{$-40_{-150}^{+100}$} & \multirow{2}{*}{-} & \multirow{2}{*}{-} & $57_{-29}^{+33}$ & \multirow{2}{*}{10}\T\B\\
& &  &  &  &  & $28_{-15}^{+13}$ & \T\B\\\cdashline{4-8}
& & \multirow{2}{*}{4} & \multirow{2}{*}{$-280_{-840}^{+780}$} & \multirow{2}{*}{-} & \multirow{2}{*}{-} & $71_{-39}^{+29}$ & \multirow{2}{*}{7}\T\B\\
& &  &  &  &  & $33_{-15}^{+16}$ & \T\B\\\cdashline{4-8}
& & \multirow{2}{*}{5} & \multirow{2}{*}{$790_{-110}^{+100}$} & \multirow{2}{*}{$0^{+4.1}$} & \multirow{2}{*}{$0^{+180}$} & $77_{-25}^{+23}$ & \multirow{2}{*}{27}\T\B\\
& &  &  &  &  & $36_{-11}^{+10}$ & \T\B\\\cdashline{1-8}
%
\multirow{13}{*}{\Fexxv{} He$w$} & \multirow{13}{*}{6.700} & \multirow{2}{*}{1} & \multirow{2}{*}{$-80_{-280}^{+480}$} & \multirow{2}{*}{$8.9_{-4.9}^{+1.1\dagger}$} & \multirow{2}{*}{$400_{-220}^{+50}$} & $64_{-33}^{+32}$ & \multirow{2}{*}{29}\T\B\\
\multirow{13}{*}{\Fexxv{} He$y$} & \multirow{13}{*}{6.668} &  &  &  &  & $75\pm 23$ & \T\B\\\cdashline{4-8}
& & \multirow{2}{*}{2} & \multirow{2}{*}{$90_{-250}^{+130}$} & \multirow{2}{*}{-} & \multirow{2}{*}{-} & $48_{-27}^{+28}$ & \multirow{2}{*}{30}\T\B\\
& &  &  & & & $82_{-32}^{+36}$ & \T\B\\\cdashline{4-8}
& & \multirow{2}{*}{3} & \multirow{2}{*}{$-110_{-290}^{+350}$} & \multirow{2}{*}{$9.2_{-4.0}^{+0.8\dagger}$} & \multirow{2}{*}{$410_{-180}^{+40}$} & $37_{-34}^{+44}$ & \multirow{2}{*}{52}\T\B\\
& &  &  & & & $199_{-63}^{+75}$ & \T\B\\\cdashline{4-8}
& & \multirow{2}{*}{4} &\multirow{2}{*}{ $-320_{-330}^{+260}$} & \multirow{2}{*}{$10_{-8.4}^\dagger$} & \multirow{2}{*}{$450_{-380}^\dagger$} & $40_{-40}^{+48}$ & \multirow{2}{*}{18} \T\B\\
& &  &  & &  & $180_{-80}^{+100}$ & \T\B\\\cdashline{4-8}
& & \multirow{2}{*}{5} & \multirow{2}{*}{$-10_{-109}^{+136}$} & \multirow{2}{*}{$4.6_{-2.1}^{+3.5}$} & \multirow{2}{*}{$200_{-90}^{+160}$} & $138\pm38$ & \multirow{2}{*}{57}\T\B\\
& &  &  & & & $50_{-21}^{+26}$ & \T\B\\ \cdashline{1-8}
\multirow{7}{*}{\Crxxiii{} He$w$} & \multirow{7}{*}{5.682} & 1 & - & - & - & $24_{-17}^{+18}$ & 4\T\B\\
& & 2 & - & - & - & $16_{-5}^{+14}$ & 2\T\B\\
& & 3 & - & - & - & $12_{-12}^{+14}$ & 2\T\B\\
& & 4 & $-180^{+440}_{-280}$ & 4.7$_{-2.6}^{+5.3\dagger}$ & 250$_{-140}^{+280\dagger}$ & $79_{-29}^{+32}$ & 13\T\B\\
& & 5 & - & - & - & $13\pm 13$ & 1\T\B\\\cdashline{1-8}
\multirow{7}{*}{\Nex{} Ly$\alpha$} & \multirow{7}{*}{1.022} & 1 & - & - & - & $8_{-8}^{+13}$ & 0\T\B\\
& & 2 & - & - & - & $32_{-12}^{+11}$ & 9\T\B\\
& & 3 & - & - & - & $37\pm10$ & 17\T\B\\
& & 4 & - & - & - & $77\pm18$ & 33\T\B\\
& & 5 & - & - & - & $70_{-14}^{+16}$ & 35\T\B\\
\hline
\end{tabular}
\tablefoot{To better assess the significance of the changes between orbits, we show the errors with an uncertainty of 1$\sigma$. The quoted rest wavelengths are adopted from NIST. The energies of all unresolved lines are taken as the 2/1 mean of the main transitions using the NIST values. For the two line complexes fitted with single  transitions, the line shifts and widths between the two transitions are tied. The velocities given are the measured values and do not include the correction of $-117\pm13$ km s$^{-1}$ due to the relative motion of the BH (see App.\ref{app:vel_corr}).}
\end{table*}

\end{onecolumn}
\end{document}